\documentclass[a4paper,11pt]{article}
\pdfoutput=1 % if your are submitting a pdflatex (i.e. if you have
\usepackage{jheppub} % for details on the use of the package, please
\usepackage[T1]{fontenc} % if needed
\usepackage{bm,bbm}
\usepackage{graphics}
\usepackage{subcaption}
\usepackage{amssymb}
\usepackage{hyperref}
\usepackage{slashed}
\usepackage{mathrsfs}
\usepackage{esint}
\usepackage[normalem]{ulem}
\usepackage{longtable}
\usepackage{cancel}
\hypersetup{
} 
\usepackage{xspace}
\usepackage[separate-uncertainty]{siunitx}%,multi-part-units = single]{siunitx}
\usepackage{physics}
\usepackage{mathtools}
\usepackage{graphicx}
\usepackage{verbatim}
\usepackage{multirow}
\usepackage{orcidlink}
\usepackage{etoolbox}

\newcommand{\shat}{\hat{s}}

\definecolor{purple}{rgb}{0.57, 0.36, 0.51}
\definecolor{ocorr}{RGB}{0,200,0}

\newcommand{\plus}{\! + \!}
\newcommand{\minus}{\! - \!}
\newcommand{\cdott}{\! \cdot \!}
\newcommand{\pn}{p_{\nu_\ell}}
\newcommand{\MSbar}{\overline{\text{MS}}}

\newcommand{\hata}{\hat{a}^+}
\newcommand{\bara}{\bar{a}^+}

\newcommand{\RomanNumeralCaps}[1]{\MakeUppercase{\romannumeral #1}}	
\usepackage{xcolor}

\renewcommand{\arraystretch}{1.5}
\title{
Endpoint Factorization for Semileptonic Decays of Boosted and Resonant Off-Shell Top Quarks\\ with a Large-Radius Bottom Jet
}

\preprint{
\begin{flushright}
UWThPh-2025-6
\end{flushright}
}

\author[a]{Andr\'e H. Hoang\orcidlink{0000-0002-8424-9334},}
\author[a]{Christoph Regner\orcidlink{0009-0007-7160-3844}}

\affiliation[a]{University of Vienna, Faculty of Physics, Boltzmanngasse 5, A-1090 Wien, Austria}

\emailAdd{andre.hoang@univie.ac.at}
\emailAdd{christoph.regner@univie.ac.at}

\abstract{
We derive a factorization formula for boosted double resonant top-antitop pair production in $e^+e^-$ annihilation with
a semileptonic top quark decay in the phase space region where the $b$-jet invariant mass is small. The decaying top quark state is defined through invariant mass measurements on the final states in the top and antitop hemispheres, and the $b$-jet is defined from clustering all hadrons in the top hemisphere. The factorization does not rely on the narrow width limit and accounts for the QCD off-shell and interference effects. The approach employs Soft-Collinear-Effective Theory and boosted Heavy-Quark-Effective-Theory and relies on a combination of factorization theorems known from $e^+e^-$ dijet production and inclusive semileptonic heavy meson endpoint decays. The result provides a first principles treatment of the dominant hadronization effects, which can be determined from $e^+e^-$ event shapes.
In the factorization a new distribution function arises, called the ultra-collinear-soft (ucs) function, which encodes the Fermi motion of the decaying top quark within the state defined from the invariant mass measurement. 
The ucs function is a differential generalization of the inclusive bHQET jet function and shares properties of the shape function in semileptonic heavy meson decays. In frames where the top quark is very slow, it describes the coherent soft radiation arising from top production, propagation and decay, and encodes all effects that are non-factorizable from the perspective of the NW limit. Its form and renormalization depend on two light-cone momenta related to the top-jet and $b$-jet directions and their relative angle.   
Due to the large top quark width, the ucs function can be computed perturbatively, and
we determine the QCD corrections at ${\cal O}(\alpha_s)$. The anomalous dimension is known to three loops.
}

\begin{document}
		\maketitle
\flushbottom

\section{Introduction}
\label{sec:intro}

Due to its large mass, the top quark plays an essential role in consistency checks of
the Standard Model (SM)~\cite{Cabibbo:1979ay,Alekhin:2012py,Buttazzo:2013uya,Branchina:2013jra,Baak:2014ora}, but also in models as well as in searches of physics beyond the SM.
Precise measurements of the top quark mass with control of its renormalization scheme-dependence 
and uncertainties at the level of a few hundred MeV are becoming increasingly important~\cite{Andreassen:2014gha}. The currently most precise top mass measurements, which are called ``direct measurement'' and which have reached uncertainties at the level of $300$ to $400$~MeV~\cite{CMS:2023ebf,ParticleDataGroup:2024cfk,CMS:2024irj} (see also Refs.~\cite{CMS:2015lbj,ATLAS:2018fwq,CDF:2016vzt}), are based on differential observables constructed from jets and/or charged leptons arising from the top quark decay with high kinematical sensitivity to the top quark mass. Projections indicate that at HL-LHC uncertainties as small as $200$~MeV can be reached for the direct measurements~\cite{CMS:2017gvo,CMS:2024irj}. However, the observables employed in the direct measurements have a significant sensitivity to low-energy QCD and non-perturbative as well as resummation effects, which can currently only be described using multipurpose Monte Carlo (MC) event-generator simulations. As a consequence, due to the conceptual limitations of current MC simulations (related to the precision of the parton showers, the hadronization models and their interplay), there is some conceptual ambiguity in the field theoretic interpretation of the measured top mass, which is the top mass parameter of the MC generator~\cite{Azzi:2019yne,Hoang:2020iah}.
While it is known that the MC top quark mass parameter is close to the pole mass $m_t^{\rm pole}$ or short-distance masses defined at a low renormalization scale (such as the MSR mass $m_t^{\rm MSR}(R=2~\mbox{GeV})$~\cite{Hoang:2017suc,Hoang:2017btd} that is renormalon-free and differs from the pole mass by about $250$~MeV at ${\cal O}(\alpha_s)$), the relation can currently not be made more precise for the direct LHC measurements. (See also Ref.~\cite{Kieseler:2015jzh}.)

There are concrete analytic and numerical insights in the relation from the analysis of inclusive jet mass
event-shape type observables for boosted top-antitop production in the double resonant region (where the (anti)top invariant masses are close to the top mass) in $e^+e^-$ annihilation~\cite{Fleming:2007qr,Fleming:2007xt,Butenschoen:2016lpz,Hoang:2018zrp,Dehnadi:2023msm} for which there is a rigorous understanding of (fixed-order and resummed) perturbative corrections and the structure of non-perturbative effects from factorization. Accounting for light soft drop grooming to reduce the impact of soft radiation, these results have been extended to the LHC in Refs.~\cite{Hoang:2017kmk,Hoang:2019ceu} involving top quarks with $p_T>750$~GeV.\footnote{Recently, top quark mass measurements from the invariant mass of reconstructed large radius jets from hadronically decaying boosted top quarks, and based on MC simulations, have been carried out by the CMS and the ATLAS collaborations, reaching uncertainties of $840$~MeV ~\cite{CMS:2017pcy,CMS:2019fak,CMS:2022kqg,CMS:2024irj} and $530$~MeV ($p_{T,{\rm top}}\gtrsim 2 m_t$)~\cite{Riembau:2025tlj}, respectively. These measurements are (effectively) based on reclustered identified narrow jets and supplemented by grooming/trimming procedures, which reduces the impact of soft radiation pileup contributions within the large radius top jet.} However, these insights cannot be transferred to the LHC direct measurements without making additional assumptions. These are related to the differences concerning the differential observables and cuts employed at the LHC and the universality of the results obtained for $e^+e^-$. To obtain a level of rigour for LHC observables comparable to the inclusive jet masses for $e^+e^-$ a number of conceptual advancements are mandatory. They are related to the generalization of the inclusive jet mass factorization from~\cite{Fleming:2007qr,Fleming:2007xt} towards observables sensitive to the top quark decay, based on LHC jet algorithms, and a systematic understanding of their properties concerning soft QCD and non-perturbative effects. Eventually, also a rigorous theoretical understanding of the underlying event may need to be gained.

Recently, an interesting new avenue has been opened up in feasibility studies of energy correlators for top mass measurement from the decay opening angles for boosted top quarks~\cite{Holguin:2022epo,Holguin:2023bjf,Holguin:2024tkz}, as energy correlators leads to smaller soft and non-perturbative effects. Here we adopt a different path forward by generalizing the known factorization framework for inclusive jet masses in $e^+e^-$~\cite{Fleming:2007qr,Fleming:2007xt} to differential semileptonic top quark decays. The differential observables we consider are not yet suitable for the LHC as we consider a wide large-radius $b$-jet. But our factorization treatment has the conceptual advantage that the dominant non-perturbative and off-shell effects, and non-factorizable QCD corrections, the coherent QCD radiation coming from top production and decay, can be rigourously quantified and studied analytically, at the same time when summing all large logarithms.
Apart from being an important intermediate step for analytic conceptual studies, the factorization framework we discuss here, may also provide benchmark results to test MC event generators along the lines of Refs.~\cite{Hoang:2018zrp,Hoang:2024zwl} and can be phenomenologically relevant at a future high-energy lepton collider.
The analysis of non-factorizable corrections (which become relevant when the top decay is treated differentially) is particularly interesting, since their effects should be understood systematically as well for top mass measurements with uncertainties at the level of $300$ to $400$~MeV given that the top quark width $\Gamma_t\approx 1.4$~GeV is considerably larger.

Currently, there are two main approaches for making theoretical predictions for the production and the decay of top quark. One is based on the narrow-width (NW) limit and the other 
employs full off-shell computations in fixed-order perturbation theory.
In the NW approach (see e.g. Refs.~\cite{Uhlemann:2008pm,Fuchs:2014ola}), the top quark is treated as an on-shell long-lived asymptotic particle. The approach, which simply neglects off-shell and non-factorizable effects, yields a convenient factorization of the production and decay dynamics. In the context of $e^+e^-\to t\bar t+X$, QCD corrections have been determined for the total inclusive cross section  up to ${\cal O}(\alpha_s^3)$~\cite{Hoang:2008qy,Kiyo:2009gb}, and differential $t\bar t$ production cross sections are available at ${\cal O}(\alpha_s^2)$~\cite{Gao:2014nva,Gao:2014eea,Chen:2016zbz,Bernreuther:2023jgp}.\footnote{The corresponding one-loop electroweak corrections were calculated a long time ago in Ref.~\cite{Fujimoto:1987hu,Beenakker:1991ca,Fleischer:2003kk}.} Differential calculations for the top quark semileptonic decay have been provided up to ${\cal O}(\alpha_s^2)$~\cite{Gao:2012ja,Brucherseifer:2013iv} with first results presented even at ${\cal O}(\alpha_s^3)$~\cite{Chen:2023osm}. We refer to~\cite{Chen:2022wit} (and references therein) for the two-loop QCD and the available electroweak corrections to the inclusive top quark width. The concept of the NW limit is also the basis of state-of-the-art Monte-Carlo multi-purpose event generators (supplemented by Breit-Wigner-type smearing of the top quark mass to model some of the off-shell effects)~\cite{Bierlich:2022pfr,Bewick:2023tfi,Sherpa:2024mfk} and
many experimental analyses (see e.g.\ Ref.~\cite{CMS:2024irj}).
In the off-shell fixed-order approach, full matrix element calculations for final states compatible with top decays are carried out. Here the decaying top quark is just one of all  other possible intermediate states. The method accounts for resonant and  
non-resonant effects and also for the full set of QCD corrections including the non-factorizable ones. The computations are, however, more involved when the top quark takes the role of a virtual intermediate state, so that the current precision is limited to the ${\cal O}(\alpha_s)$ level as far as QCD corrections are concerned. So far, fully differential ${\cal O}(\alpha_s)$ off-shell fixed-order calculations have been provided for leptonically decaying top quarks (i.e.\ $e^+e^-\to j_b j_{b}\ell^-{\ell'}^+\bar{\nu}_{\ell}\nu_{\ell'}$) in Refs.~\cite{Guo:2008clc,Liebler:2015ipp,ChokoufeNejad:2016qux} and for the semileptonic final state (i.e.\ $e^+e^-\to j_b j_{b} j j {\ell}^+ \nu_{\ell'}$) in Ref.~\cite{Denner:2023grl}. The corresponding NLO QCD calculations for hadron colliders have been provided in Refs.~\cite{Denner:2012yc,Cascioli:2013wga,Jezo:2016ujg}, with first steps towards NNLO precision discussed very recently in Ref.~\cite{Buonocore:2025fqs}.

The factorization formula we derive in this article refers to boosted $t\bar t$ production ($E_{cm}=Q\gg m_t$) in the double resonant region with a semileptonically decaying top quark 
into a low mass $b$-jet, a neutrino $\nu_\ell$ and a charged lepton $\ell^+$. The formula
does not rely on the NW limit and combines the factorization of production and decay stage radiation, when they cannot interfere for kinematic reasons, with an additional factorization of radiation arising coherently from the production or decay stage, which contains non-factorizable effects when both types of radiation occupy the same phase space region and interfere. 
The factorization also provides the means for the summation of all large logarithms and a first principles description of the dominant hadronization effects which in particular includes those which are linearly sensitive to $\Lambda_{\rm QCD}$. The low mass $b$-jet is defined as all the hadronic radiation in the top quark hemisphere defined with respect to the thrust axis of the entire event, and the factorization allows for the description 
of differential distributions constructed from the momenta of the (large-radius hemisphere) $b$-jet and the leptons. The antitop jet is treated fully inclusively. Our factorization applies to the process where $W$ from the top decays into a lepton pair and the antitop decays fully hadronically, and requites that the neutrino momentum is reconstructed.

In this situation we have the following hierarchies of scales: $\Gamma_t\lesssim  M_{t,\bar t}-m_t = \Gamma \ll (m_t \Gamma)^{1/2}\ll m_t\ll Q$, where $M_{t}$ and $M_{\bar t}$ are the top and antitop hemisphere invariant masses (defined with respect to the plane perpenticular to the thrust axis and including the lepton momenta for the top hemisphere) and $(m_t \Gamma)^{1/2}$ is the typical $b$-jet invariant mass. There are four distinct QCD radiation modes that are separated kinematically and can be mutually factorized: (i) collinear-soft QCD radiation associated to the boosted top (which is soft in the top rest frame\footnote{We refer to frames where the off-shell resonant top quark is very slowly moving.} and called "top-ultra-collinear"), (ii) collinear-soft QCD radiation associated to the $\bar t$ (which is soft in the antitop rest frame and called "antitop-ultra-collinear"), (iii) large-angle soft radiation (in the $e^+e^-$ c.m.\ frame) that is sensitive to definition of the top and antitop jet hemisphere boundaries and (iv) hard-collinear radiation associated to the $b$-jet production within the top quark hemisphere. A graphical illustration of these modes is provided in Fig.~\ref{fig:topdecay_modes}. By definition, the hard-collinear $b$-jet radiation, the top-ultra-collinear radiation and the large-angle soft radiation within the top hemisphere all contribute to the $b$-jet momentum. 
The restriction that the $b$-jet has small invariant mass implies that the factorization is derived from the phase space region where the $W$-boson and the $b$-jet are produced back-to-back in frames where the top quark is slowly moving, which is the kinematics of the tree-level top quark decay.
For example, for the  $b$-jet lepton invariant mass $M_{j_b\ell}$ this refers to the endpoint regions where $M_{j_b\ell}$ is either small or large and which are sensitive to the masses of the $b$-jet and the top quark.

The resulting factorization formula for the multi-differential cross section
\begin{align}
\label{eq:diffcrossmulti}
\frac{\mathrm{d}^{3}\sigma}{\mathrm{d} M_t^2 \, \mathrm{d} M_{\bar{t}}^2 \, \mathrm{d} X}\,,
\end{align}
where $X$ stands for an observable constructed from the $b$-jet and the lepton momenta, 
represents a merging of previously established factorizations for the double hemisphere inclusive invariant mass distribution for boosted top pair production in the double resonant region in Refs.~\cite{Fleming:2007qr,Fleming:2007xt} and the inclusive\footnote{In the context of double hemisphere invariant mass distribution the term 'inclusive' refers to the treatment that only the invariant masses of the top and antitop hemispheres are measured.	
In the context of semileptonic heavy meson decays the term 'inclusive' refers to the treatment that the entire hadronic final state is clustered into a jet.} semileptonic heavy meson decay into a light quark jet, a charged lepton and a neutrino~\cite{Bauer:2001yt,Bosch:2004th}. 
The formulation of the factorization involves boosted heavy quark effective theory (bHQET)~\cite{Fleming:2007qr,Fleming:2007xt} and soft-collinear effective theory (SCET)~\cite{Bauer:2000yr,Bauer:2001ct,Bauer:2001yt}. We find that the QCD modes (ii), (iii) and (iv) remain to be strictly associated to either production or decay. However, the top-ultra-collinear QCD radiation of type (i), which refers to soft radiation with momenta $k^\mu\sim \Gamma\ll m_t$ in the rest frame of the boosted top quark, is sensitive to top production, propagation and decay, and encodes all the effects which are non-factorizable with respect to the NW approach. These coherent top-ultra-collinear effects are described by a new factorization function, which we call {\it ultra-collinear-soft (ucs) function} and which is expressed in terms non-local matrix elements that involve a time-ordered product of top production and decay operators. The usc function describes the light-cone momentum distribution (with respect to the $b$-jet direction) of the coherent top-ultra-collinear radiation and thus carries a similar physical interpretation as the shape function for the $\bar B$ meson semileptonic decay. However, here the decaying physical state is not a hadron, but the top quark jet defined through the hemisphere mass $M_t$ measurement. The ucs function may thus be associated with the ``Fermi motion'' of the decaying top quark within the measured large top quark jet, in analogy to the terminology frequently used for semileptonic heavy meson decays~\cite{Bigi:1993ex,Neubert:1993mb,Manohar:2000dt}. 
Due to the finite top quark width, the ucs function can be computed perturbatively, and we determine its QCD corrections at 
${\cal O}(\alpha_s)$. We also find that the resummation of large logarithms related to the scale hierarchies above, which is described by the renormalization group evolution of the Wilson coefficients and factorization functions, remains factorized and can still be associated to either the production or the decay stage. Since all Wilson coefficients and the other factoriation functions appearing in factorization formula are already known to ${\cal O}(\alpha_s^2)$ or even ${\cal O}(\alpha_s^3)$, and since all anomalous dimensions are known to at least NNLL order precision, our results allows for a NNLL or NLL' resummed description of the differential cross section.
Our result demonstrates that factorized and consistently QCD resummed predictions for top quark semileptonic decay observables, without relying on the NW limit can be carried out with full analytic control of the non-factorizable and hadronization effects. 

In this article we report on the details of the derivation and consistency of our factorization formula and the computation of the ucs function up to ${\cal O}(\alpha_s)$. We leave the resummation of all large logarithms as well as further phenomenological and conceptual applications of these results to upcoming future work.  
The content of the article is a follows:
In Sec.~\ref{sec:reviewfactorization} we start by reviewing the derivation of the factorization formulae for the double hemisphere inclusive invariant mass distribution $\mathrm{d}^2\sigma/\mathrm{d} M_t^2 \mathrm{d} M_{\bar{t}}^2$ for boosted top-antitop production and the differential semileptonic decay rate $\mathrm{d}\Gamma/\mathrm{d} X$ of the $\bar B$ meson in the low inclusive jet mass endpoint region. The discussion of the inclusive double hemisphere invariant mass distribution is the starting point for the treatment of the top decay in Sec.~\ref{sec:factorizationtopdecay} and provides information (not provided in Refs.~\cite{Fleming:2007qr,Fleming:2007xt}) how the notations and the choices for the bHQET and SCET label momenta impact the definition of the reference frame used for the description of the top quark decay. The discussion on the $\bar B$ decay factorization is carried out for an arbitrary reference frame as this generalization is also mandatory for the factorization of the boosted top quark decay.
In Sec.~\ref{sec:jetfctfactorization} we then proceed with the derivation of the factorization formula for Eq.~(\ref{eq:diffcrossmulti}) by further factorizing the inclusive bHQET top quark jet function appearing in the factorization formula for $\mathrm{d}^2\sigma/\mathrm{d} M_t^2 \mathrm{d} M_{\bar{t}}^2$. In Sec.~\ref{sec:inclusivefactorization} we show that a differential treatment of the top quark decay requires an additional (non-local) insertion of a top quark decay operator by rederiving a known dispersion relation that relates the inclusive bHQET jet function for an unstable top quark to the one for a stable top quark (originally obtained using analyticity and the optical theorem in Ref.~\cite{Fleming:2007qr,Fleming:2007xt}) from an explicit treatment of the top decay final states. In Sec.~\ref{sec:factorizationtopdecay} we then proceed with the derivation of the remaining factorization for the top quark decay and the definition of the ucs function. Some general properties of the ucs function are discussed in 
Sec.~\ref{sec:Sucsproperties}. Here we also elaborate on how the non-factorizable character of the ucs function is imprinted in the form of the convolutions appearing in the factorization formula and how the ucs function needs to be defined such that the non-factorizable effects remain confined to it in such a way that all RG evolution remains factorized to either the production or the decay stages. 
In Sec.~\ref{sec:renormalizationtopdecay} we then derive the renormalization of the usc function, which involves an unconventional two-dimensional $Z$-factor convolution from the RG consistency of the factorization formula, and in Sec.~\ref{sec:ucsoftcomputations}
we compute the ucs function to ${\cal O}(\alpha_s)$. In Sec.~\ref{sec:pheno} we provide a brief NLO fixed-order phenomenological analysis of
the factorization theorem for $b$-jet lepton invariant mass, $X=M_{j_b\ell}$. We keep this analysis qualitative and brief since the summation of large-logarithms and non-perturbative effects have to be accounted for to carry out fully a realistic phenomenological study, which will be carried out in future work. Section~\ref{sec:conclusions} contains our conclusions and an outlook.
We also provide some appendices. In App.~\ref{app:Wilson} we collect the definitions of all Wilson lines that appear in the factorization formulas we discuss, and in App.~\ref{app:apprenor} we provide our notations concerning the renormalization of all Wilson coefficients and factorization functions relevant for this article. In App.~\ref{app:NLO_corrections} we provide the results for all renormalized factorization functions and their $Z$-factors at ${\cal O}(\alpha_s)$, and in App.~\ref{eq:udsfctIandII} we provide individual results for the ${\cal O}(\alpha_s)$ corrections of the ucs function from different classes of diagrams. In App.~\ref{app:phasespace} we specify some details concerning the 3-body phase space integrations relevant for our phenomenological discussion.

\section{Review of Factorization Theorems}
\label{sec:reviewfactorization}

\subsection{Double Hemisphere Jet Mass Distribution in the Top Resonance Region}
\label{sec:hemispherefactorization_new}

In the following we revisit the derivation of the factorization theorem for the double-hemisphere invariant mass distribution in the resonance region for boosted top jet production in $e^+e^-$-collisions carried out in Ref.~\cite{Fleming:2007qr} in the SCET and bHQET frameworks.
In addition to the discussion in Ref.~\cite{Fleming:2007qr}, which is adequate for inclusive jet masses, our derivation specifies more concretely the choice of the bHQET top quark velocity labels, as this is important for the frame dependence in the description of the top quark decay in Sec.~\ref{sec:jetfctfactorization}.
We note that we use the SCET/bHQET notation as employed in Refs.~\cite{Bauer:2000yr,Bauer:2001ct,Bauer:2001yt,Fleming:2007qr}. Moreover, to keep expressions simple we drop the dependence on the renormalization scales in the derivation of the factorization formulae in this and the following sections. 
We also note that in the following discussions there are quantities defined in $6$-flavor QCD and in $5$-flavor QCD without the top quark. As the case considered will be clear from the context, we do not indicate this via additional indices on the quantities. Here and throughout the rest of this work, we use the notation that Latin indices (except for $i, j, j'$) refer to color. Greek indices are used for Dirac or Lorentz quantities. 
Finally, we mention that for the top-antitop direction we use the light-cone momentum decomposition with respect to a light-like reference vector 
$n^\mu=(1,\vec{n})$ and the auxiliary vector $\bar{n} = (1, - \vec{n})$ in the form $p^\mu=(p^+, p^-, p^\perp) = (n \cdott p , \bar{n} \cdott p, p^\perp )$.

We start from the generic QCD cross section for boosted $t\bar t$ production in $e^+e^-$ annihilation which has the form
\begin{equation} 
	\label{eq:QCD_cross_section_new}
	\mathrm{d} \sigma = \sum_X^{\mathrm{res.}} (2\pi)^4 \delta^{(4)} (q_{ee} \minus P_X)  \sum_{i= v,a} L_{\mu \nu}^i \bra{0} \mathcal{J}_{t\bar t,i}^{\mu \dagger} (0) \ket{X} \bra{X} \mathcal{J}_{t\bar t,i}^\nu (0) \ket{0} \,,
\end{equation}
where the superscript `res.' in the sum over the final states $X$ indicates that $X$ is restricted to the dijet limit in the double resonant region where we have a back-to-back $t\bar t$ pair plus ultra-collinear (along the top and antitop quarks) and soft radiation in the $e^+e^-$ c.m.\ frame. This restriction implies that the invariant masses $M_{t,\bar t}$ of the two hemispheres, which are defined through the plane perpendicular to the thrust axis, are close to the top quark mass (in a renormalization scheme close to the pole mass), i.e.\ $Q\gg m_t\gg M_{t,\bar t}-m_t=\Gamma\gtrsim \Gamma_t$.  The vector ($i=v$) and axial-vector ($i=a$) QCD $t\bar t$ production currents are defined as $\mathcal{J}^\mu_{t\bar t,v} = \bar{t} \gamma^\mu t$ and $\mathcal{J}^\mu_{t\bar t,a} = \bar{t} \gamma^\mu \gamma_5 t$, respectively, and $L_{\mu \nu}^i$ denotes the angular averaged leptonic tensor for both contributions including photon as well as $Z$ boson exchanges, as we sum over the directions of the $t\bar t$ dijet axis. More specifically, we have 
\begin{align}
L^{i,\mu \nu} = F^i(Q^2) \bigg( \frac{q^\mu_{ee} q^\nu_{ee}}{Q^2} \minus   g^{\mu \nu} \bigg) \,,
\end{align}
where
\begin{align}
&F^v(Q^2) = \frac{8\pi^2 \alpha^2}{3 Q^4} \bigg[ e_t^2 \minus \frac{2Q^2 v_e v_t e_t}{Q^2 \minus M_Z^2} \plus \frac{Q^4 (v_e^2 \plus a_e^2) v_t^2}{(Q^2 \minus M_Z^2)^2} \bigg] \ , \\
&F^a(Q^2) = \frac{8\pi^2 \alpha^2}{3 Q^4} \bigg[  \frac{Q^4 (v_e^2 \plus a_e^2) a_t^2}{(Q^2 \minus M_Z^2)^2} \bigg] \ , \notag 
\end{align}
with the total momentum $q^\mu_{ee} = p_{e^-}^\mu + p_{e^+}^\mu$ and $Q^2 \equiv q^2_{ee}$. To be concrete, we formulate the cross section in the $e^+e^-$ c.m.\ frame where we have $q^\mu_{ee}=(Q,\vec{0})$. In addition, $e_t$ denotes the top quark charge and
\begin{align}
	v_f = \frac{T_3^f \minus 2 Q_f \sin^2 \theta_W}{2 \sin \theta_W \cos \theta_W} \ , \qquad \qquad a_f = \frac{T_3^f}{2 \sin \theta_W \cos \theta_W} \ ,
\end{align}
with the third component of weak isospin, $T_3^f$, and the weak mixing angle, $\theta_W$.

In order to describe the dijet-like final state configuration in the top resonance region,
we proceed in a two-stage process.\footnote{In Ref.~\cite{Fleming:2007qr} first a factorization formula for
$M_{t,\bar t}-m_t\sim m_t$ was derived in SCET, which was then matched at the factorization level to bHQET where $M_{t,\bar t}-m_t\ll m_t$. Here we derive the bHQET factorization directly and also use a more direct treatment of the large SCET/bHQET field labels to simplify the notations.} We start from the QCD cross section in Eq.~(\ref{eq:QCD_cross_section_new}) and first match the QCD $t\bar t$ currents onto SCET $t\bar t$ currents at the hard production scale $Q$. The corresponding matching relation can be formulated in the form ($i = v, a$)
\begin{align} 
\label{eq:SCET_currents-3}
& \mathcal{J}_{t\bar t,i}^\mu (0)  =  C_Q(Q) \, \mathcal{J}_{i,{\rm SCET}}^\mu(0) \, + \, {\cal O}(m_t/Q) \,,\\
& \mathcal{J}_{i,{\rm SCET}}^\mu(0) =
 [\bar{\chi}_{n} \, S_{n,+}^\dagger\, \Gamma^\mu_i \, S_{\bar{n},-} \, \chi_{\bar{n}}] (0) \,, \notag
\end{align}
where we use the compact notation $\Gamma_i^\mu \in \{ \gamma^\mu, \gamma^\mu \gamma_5 \}$ for later convenience. All quantities of the SCET current in the second line of Eq.~(\ref{eq:SCET_currents-3}) are defined in 6-flavor QCD involving the top quark and the five lighter quarks, which are all treated as massless for simplicity. The Wilson coefficient $C_Q (Q)$ contains the hard contributions for the production of the two top jets. Moreover, we stress that the SCET currents are already expressed in terms of soft decoupled (anti)top quark (hard) collinear jet fields $\bar{\chi}_n = \bar{\xi}_n  W_{n}$ and $\chi_{\bar{n}} = W_{\bar{n}}^\dagger \xi_{\bar{n}}$. 
The collinear $W$-Wilson lines generate partons with momenta $p^\mu\sim(m_t^2/Q,Q,m_t)$ and $p^\mu\sim(Q,m_t^2/Q,m_t)$ for the $n$ and $\bar n$ sectors, respectively, and the large-angle soft $S$-Wilson lines generate partons with momenta  $k^\mu\sim m_t$ in the $e^+e^-$ c.m.\ frame. We use the back-to-back light-like reference vectors $n^\mu=(1,\vec{n})$ and $\bar n^\mu=(1,-\vec{n})$, where the 3-unit-vector $\vec{n}$ ($-\vec{n}$) is the spatial direction of the total momentum in the top (antitop) quark hemisphere $M_t^\mu$ ($M_{\bar t}^\mu$). These directions do in general not agree with the top or the antitop quark directions.
The jet fields satisfy the projection relations 
$\bar{\chi}_n=\bar{\chi}_n\frac{\slashed{\bar n}\slashed{n}}{4}$ and 
$\chi_{\bar{n}}=\frac{\slashed{\bar n}\slashed{n}}{4}\chi_{\bar{n}}$ and we adopt  
\begin{align} 
\label{eq:SCET-ttbar-largelabels}
& \tilde{P}_{X_n}^\mu =  (\tilde{P}_{X_n}^+,\tilde{P}_{X_n}^-,\tilde{P}_{X_n}^\perp)=(0,Q,0)  \,,  &          &
\tilde{P}_{X_{\bar{n}}}^\mu =  (\tilde{P}_{X_{\bar{n}}}^+,\tilde{P}_{X_{\bar{n}}}^-,\tilde{P}_{X_{\bar{n}}}^\perp)=(Q,0,0) \,,
\end{align}
as their respective large label momenta. It is the large label momentum light cone component $\tilde{P}_{X_n}^-=\tilde{P}_{X_{\bar{n}}}^+=Q$ that determines the scale of SCET current Wilson coefficient in Eq.~(\ref{eq:SCET_currents-3}).
We note that the formulation of the matching relation in Eq.~(\ref{eq:SCET_currents-3}) provides an alternative approach to the QCD-SCET matching condition used in Ref.~\cite{Fleming:2007qr}, where the jet fields of the SCET currents are defined with delta functions involving the label operators $\mathcal{P}$ and $\mathcal{P}^\dagger$ that specify the large labels of the jet fields. Here, we use that fact that the label momenta of the jet fields can be specified right away from physical considerations as this streamlines some of the following steps of the factorization derivation. 

After matching to the SCET currents and integrating out the local hard production scale $Q$ effects, the hemisphere masses can still fluctuate by contributions of order the top quark mass, i.e.\  $Q^2\gg M_{t,\bar t}^2\gtrsim M_{t,\bar t}^2-m_t^2 \sim m_t^2$. In order to further integrate out local hard effects at the top mass scale, which provides the adequate description for the double resonant region where  $Q^2\gg M_{t,\bar t}^2\gg M_{t,\bar t}^2-m_t^2\gtrsim m_t \Gamma_t$, we proceed by carrying out a second matching from the SCET currents onto bHQET currents. Since in the resonance region the production of additional $t\bar t$ pairs beyond the primary dijet top-antitop pair in the two collinear sectors is not allowed, we can match the $6$-flavor SCET current on the $5$-flavor bHQET current at the scale $m_t$, 
\begin{align} 
\label{eq:SCET-bHQET-current-match-2}
& \mathcal{J}_{i,{\rm SCET}}^\mu(0)  =  C_m\Big( m_t, \frac{Q}{m_t} \Big) \, \mathcal{J}_{i,{\rm bHQET}}^\mu(0) \, + \, {\cal O}(\Gamma_t/m_t) \,,\\
& \mathcal{J}_{i,{\rm bHQET}}^\mu(0) =
[\bar h_{v_n} W_{n,-}^{\rm uc} \, Y_{n,+}^\dagger\, \Gamma^\mu_i \, Y_{\bar{n},-} \, W_{\bar{n},+}^{\rm uc \,\dagger}  h_{v_{\bar{n}}}] (0) \,.  \notag
\end{align} 
Here $\bar h_{v_n}$ and $h_{v_{\bar{n}}}$ are bHQET top and antitop quark fields with the velocity labels $v_n^\mu$ and $v_{\bar{n}}^\mu$ that we define more specifically below. They satisfy the relation $P_v h_v =h_v$ for the projectors $P_v = (1 + \slashed{v})/2$.
These and the Wilson lines 
$W_{n,-}^{\rm uc}$ and $W_{\bar{n},+}^{\rm uc \,\dagger}$ (see Eqs.~(\ref{eq:ucWilsonlines})) form bHQET jet fields and
describe ultra-collinear radiation off the resonant (anti)top, which refers to soft momenta $m_t\gg k_{n,\bar n}^\mu\gtrsim\Gamma_t$ in the boosted (anti)top rest frame. This refers to total momenta $p_{n,\bar n}=m_t v_{n,\bar n}+k_{n,\bar n}$ so that we can now drop the restriction on the final state sums mentioned above. We note that the Wilson lines $W_{n,-}^{\rm uc}$ and $W_{\bar{n},+}^{\rm uc \,\dagger}$ are light-like since the soft radiation in let's say the boosted top rest frame sees the boosted antitop quark and its decay products as a light-like color source at leading order in the power counting~\cite{Fleming:2007qr,Fleming:2007xt}.
We also recall that at leading order in the power counting we can assign the $n$ and $\bar n$ ultra-collinear momenta to contribute exclusively to the top and antitop hemispheres, respectively, as let's say the radiation of the top-ultra-collinear gluons into the antitop hemisphere is power suppressed due to the boosted kinematics. The $Y$-Wilson lines (without the superscript uc) describe radiation that is ultra-soft in the $e^+e^-$ c.m.\ frame with momenta $m_t\gg k_s^\mu\gtrsim m_t\Gamma_t/Q$. The definition of all Wilson lines we discuss in this article are collected in App.~\ref{app:Wilson}. Explicit ${\cal O}(\alpha_s)$ results for all Wilson coefficients and factorization functions that arise in this work and references for the available expressions at higher order are collected in App.~\ref{app:NLO_corrections}. 

Employing the matching relations in Eqs.~(\ref{eq:SCET_currents-3}) and (\ref{eq:SCET-bHQET-current-match-2}), the cross section in bHQET can be written in the form
\begin{align} 
\label{eq:SCET_cross_section-2}
\mathrm{d} \sigma =& \sum_{\vec{n}} \sum_{X_n, X_{\bar{n}}, X_s} \sum_{i} L_{\mu \nu}^i  \, H_Q(Q) \, H_m\Big( m_t , \frac{Q}{m_t} \Big)  \, (2\pi)^4 \delta^{(4)} (q_{ee} \minus P_{X_n} \minus P_{X_{\bar{n}}} \minus P_{X_s})  \notag \\
&\times  \bra{0} \bar{T} \{  [\bar h_{v_{\bar{n}}} W_{\bar{n},-}^{\rm uc} \, Y_{\bar{n},+}^\dagger\, \overline{\Gamma}^\mu_i \, Y_{n,-} \, W_{n,+}^{\rm uc \,\dagger}  h_{v_{n}}] (0) \} \ket{X_n X_{\bar{n}} X_s}   \notag\\
&\times  \bra{X_n X_{\bar{n}} X_s} T \{  [\bar h_{v_n} W_{n,-}^{\rm uc} \, Y_{n,+}^\dagger\, \Gamma^\nu_i \, Y_{\bar{n},-} \, W_{\bar{n},+}^{\rm uc \,\dagger}  h_{v_{\bar{n}}}] (0) \} \ket{0} \,, 
\end{align}
where we used $\overline{\Gamma}_i^\mu = \gamma^0 \Gamma^{\mu \dagger}_i \gamma^0$ and introduced the hard matching factors $H_Q(Q) \equiv |C_Q(Q)|^2$ and $H_m\Big( m_t,\frac{Q}{m_t} \Big) \equiv \Big|C_m\Big( m_t, \frac{Q}{m_t} \Big) \Big|^2$. We note that the hard factor $H_Q$ also appears in the dijet factorization theorems for $e^+e^-$ event-shapes. The restriction on the final states $X$ has been rewritten by pulling out an explicit sum over the jet directions $\vec{n}$ for the collinear sectors and by splitting the final states into two ultra-collinear sectors $X_n$, $X_{\bar{n}}$ for the respective jets and a soft sector $X_s$ for large-angle soft radiation. The decoupling of the final states in SCET into collinear and soft sectors allows us to decompose the momentum $P_X$ into a sum of collinear and soft momenta, i.e. $P_X = P_{X_n} + P_{X_{\bar{n}}} + P_{X_s}$. In addition, the momenta of the observed hadrons in the detector can also be assigned to either a specific collinear sector or the soft sector, which motivates to factorize the final states $\ket{X}$ as a tensor product, $\ket{X} = \ket{X_n X_{\bar{n}} X_s}=\ket{X_n} \ket{X_{\bar{n}}} \ket{X_s}$. We remind the reader that due to our restriction on the hemisphere invariant masses in the resonance region, the soft final states $\ket{X_s}$ contain only light quarks (which here includes the bottom quarks) and gluons, but no additional $t\bar t$ pairs.

\begin{table}[]
\begin{tabular}{|c|l|l|l|}
\hline
\multicolumn{1}{|c|}
{\begin{tabular}[c]{@{}c@{}}
$\frac{m_t}{Q} \ll 1 \,, \quad  \frac{\Gamma}{m_t} \ll 1$\\[-2mm] $s^*_{t,\bar{t}} \sim \Gamma \,, \quad  \Delta \sim \frac{m_t}{Q} \Gamma$
\end{tabular}} 
& \multicolumn{1}{c|}{$e^+ e^-$ frame}  
& \multicolumn{1}{c|}{top rest frame}    
& \multicolumn{1}{c|}{virtuality}     \\ \hline 
\begin{tabular}[c]{@{}c@{}}$n$-ultra-collinear (top)  \\[-3mm] ($h_{v_n},A_n^{\mathrm{uc},\mu}$) \end{tabular}  
& $k_n^\mu \sim \Gamma \Big( \frac{m_t}{Q}, \frac{Q}{m_t}, 1 \Big)$         
& \begin{tabular}[c]{@{}l@{}}$k_n^\mu \sim \Gamma (1,1,1)$\\ $ \quad \, \, \sim \Gamma(1,1,1)'$\end{tabular}                                                                                                             
& $\mu_{B_t} \sim \Gamma$               \\ \hline
\begin{tabular}[c]{@{}c@{}} $\bar{n}$-ultra-collinear (antitop)  \\[-3mm] ($h_{v_{\bar{n}}}, A_{\bar{n}}^{\mathrm{uc},\mu}$) \end{tabular}  
& $k_{\bar{n}}^\mu \sim \Gamma \Big( \frac{Q}{m_t}, \frac{m_t}{Q}, 1 \Big)$ 
&$k_{\bar{n}}^\mu\sim\Gamma\Big(\frac{Q^2}{m_t^2},\frac{m_t^2}{Q^2},1\Big)$  
& $\mu_{B_{\bar t}} \sim \Gamma$       \\ \hline
\begin{tabular}[c]{@{}c@{}} $n'$-collinear ($b$-jet)  \\[-3mm] ($\xi_{n'},A_{n'}^\mu$)  \end{tabular}  
&                                                                             
& $p_{n'}^\mu \sim m_t\Big( \frac{\Gamma}{m_t}, 1, \sqrt{\frac{\Gamma}{m_t}}\Big)'$                                                                                                                        
& $\mu_J \sim \sqrt{m_t \Gamma}$    \\ \hline
%
% hemi. soft ($q_s,A_s^\mu$) 
\begin{tabular}[c]{@{}c@{}} large-ang. hemisphere soft \\[-3mm] ($q_s,A_s^\mu$)  \end{tabular}  
& $k_s^\mu \sim \Gamma \frac{m_t}{Q } ( 1,1, 1)$         
& \begin{tabular}[c]{@{}l@{}}$k_s^\mu \sim \Gamma \Big( 1 , \frac{m_t^2}{Q^2}, \frac{m_t}{Q} \Big) $\\ 
$ \quad \, \, \sim m_t \Big( \frac{\Gamma}{m_t}, \frac{\Gamma m_t}{Q^2}, \frac{\Gamma}{Q} \Big)'$
\end{tabular} 
& $\mu_S \sim \frac{m_t}{Q} \Gamma$ 
\\ \hline
\end{tabular}
\caption{Summary of the fields required in bHQET and SCET for the description of the semileptonic top quark decay in the endpoint region with additional measurements of the invariant masses of the top and antitop hemispheres in the resonance region. The first field in parentheses is a quark and the second is a gluon. The scaling of the light-cone momentum components is given for $(p^+, p^-, p^\perp) = (n\cdot p, \bar{n} \cdot p, p^\perp)$ w.r.\ to the light-like vector $n^\mu$ in the top hemisphere direction, or $(\hat{p}^+, \hat{p}^-, \hat{p}^\perp)' = (n'\cdot p, \bar{n}' \cdot p, \hat{p}^\perp)'$ w.r.\ to the light-like vector $n'^\mu$ in the $b$-jet direction. $\Delta$ is the scale of the large-angle hemisphere soft modes, which yield the dominant sensitivity to non-perturbative effects. The scale $\Gamma$ is determined  by the off-shellness $s^*_{t, \bar{t}}$ of the hemispheres, $\Delta \frac{Q}{m_t}$, and is bounded from below by the top quark width $\Gamma_t$. In the resonance region, when also non-perturbative effects from large-angle hemisphere soft radiation are accounted for, $\Gamma$ is in the range between $5$ and $10$~GeV~\cite{Dehnadi:2016snl,Bachu:2020nqn}. For top-antitop production, the finite bottom mass yields leading power effects for the ultra-collinear and the large-angle soft radiation arising from diagrams at ${\cal O}(\alpha_s^2)$. It is straightforward to account for these effects~\cite{Gritschacher:2013tza,Pietrulewicz:2014qza,Hoang:2019fze,Bris:2024bcq}, but they are not discussed in this article. For the semileptonic $\bar{B}$ decay only the modes analogous to the $n$-ultra-collinear and $n'$-collinear modes are relevant, and the scale $\Gamma$ is close to the hadronization scale $\Lambda_{\text{QCD}}$.}
\label{tab:modetable}
\end{table}

\begin{figure}[!t]%!htbp
	\centering
	\includegraphics[width=0.6\linewidth]{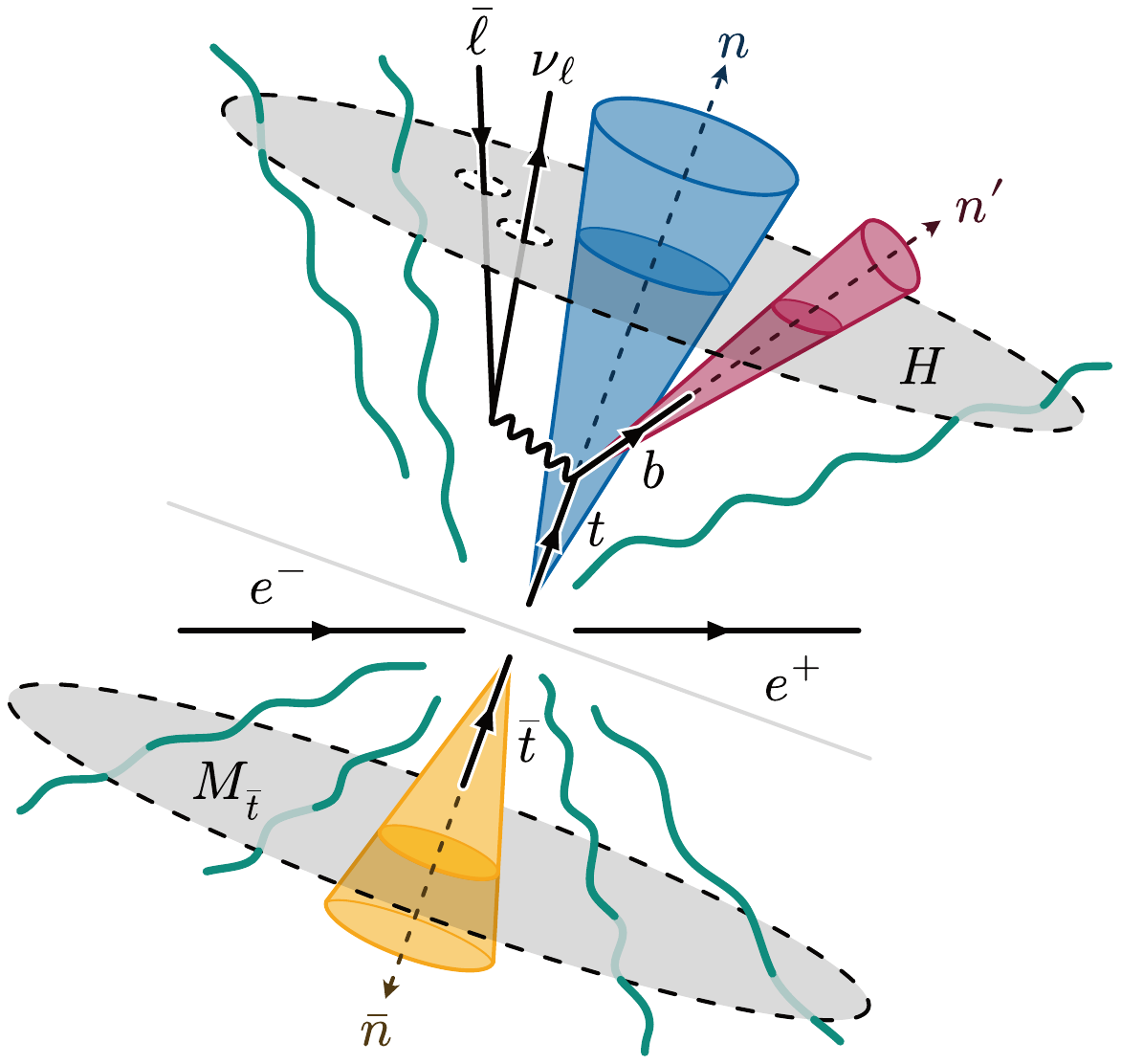}
	\caption{Graphical illustration of all distinct radiation modes relevant for the leading power factorization formula for boosted double resonant top-antitop pair production in $e^+e^-$ annihilation with a semileptonic top quark decay in the phase space region where the $b$-jet invariant mass is small. The illustrated modes are large-angle hemisphere soft (green), $n$-ultra-collinear (blue), $\bar n$-ultra-collinear (orange) and $n'$-hard-collinear (red). The total hadronic momentum $H^\mu$ in the top hemisphere, which excludes the leptonic momentum $q^\mu = p_\ell^\mu + \pn^\mu$, is indicated by the upper gray ellipse. The total top hemisphere momentum is $M_t^\mu=H^\mu+q^\mu$. The total antitop hemisphere momentum $M_{\bar t}^\mu$ is indicated by the lower gray ellipse.}
	\label{fig:topdecay_modes}
\end{figure}

In order to disentangle and separate the fields in the matrix elements into the distinct sectors, we proceed by employing the HQET Fierz identity \cite{Lee:2004ja} 
\begin{align} \label{eq:HQET_Fierz_relation_new}
(h_v)_\alpha (\bar h_v)_\beta = \mathrm{tr} \Big[ h_v \bar h_v\Big] \, \Big( \! \frac{P_v}{2} \! \Big)_{\alpha \beta} - \mathrm{tr} \big[ h_v \bar h_v  \, s^\mu \big]\, \Big( \! \frac{s_\mu}{2} \! \Big)_{\alpha \beta} \,,
\end{align}
where Greek indices (except for $\mu, \nu$) denote spin. Here, we use $s^\mu = \gamma_T^\mu \gamma_5$ with $\gamma_T^\mu = \gamma^\mu - \slashed{v} v^\mu$ and introduce our convention that $\mathrm{tr} [ \dots ]$ denotes exclusively a trace over spin, while $\mathrm{Tr} [\dots ]$ represents a trace over both color and spin. In addition, we also introduce $\mathrm{tr}_c[\dots]$, which denotes a pure color trace. By applying this Fierz relation twice, first between the heavy quark fields $h_{v_n}$ and $\bar{h}_{v_n}$, and then between the heavy quark fields $h_{v_{\bar{n}}}$ and $\bar{h}_{v_{\bar{n}}}$, we can rearrange the matrix elements in the cross section (\ref{eq:SCET_cross_section-2}) and separate the jet fields into the different ultra-collinear sectors pertaining to the top and antitop jets. We stress that only the first term in the Fierz relation (\ref{eq:HQET_Fierz_relation_new}) leads to non-vanishing matrix elements in the collinear sectors at leading order in the power counting. 
We note that at this point the two ultra-collinear sectors, encoded in the trace factor $\mathrm{tr}[ h_v \bar h_v]$, have also been rendered separately Lorentz-invariant, which later allows us to independently pick any frame for them. This is important for the description of the top quark decay in Sec.~\ref{sec:jetfctfactorization}.
Using the fact that the resulting ultra-collinear matrix elements are color-diagonal (due to ultra-collinear gauge invariance~\cite{Bauer:2000yr}), we can also identify
{ %
\allowdisplaybreaks
\begin{align} 
\label{eq:color_rearrangement_collinear_matrix_elements-2}
&\bra{0} (\bar{h}_{v_{\bar{n}}}  W_{\bar{n},-}^{\rm uc})^a_\alpha \ket{X_{\bar{n}}} \bra{X_{\bar{n}}} (W_{\bar{n},+}^{\rm uc \,\dagger} h_{v_{\bar{n}}})^{b}_\beta \ket{0} =\frac{\delta^{a b}}{N_c}\,  \bra{0} (\bar{h}_{v_{\bar{n}}} W_{\bar{n},-}^{\rm uc})^c_\alpha \ket{X_{\bar{n}}} \bra{X_{\bar{n}}} (W_{\bar{n},+}^{\rm uc \,\dagger} h_{v_{\bar{n}}})^{c}_\beta \ket{0} , \notag \\
&  \bra{0} (W_{n,+}^{\rm uc \,\dagger} h_{v_n})^d_\sigma \ket{X_{n}} \bra{X_{n}} (\bar{h}_{v_n} W_{n,-}^{\rm uc})^{e}_\rho \ket{0} = \frac{\delta^{d e}}{N_c}\, \bra{0} (W_{n,+}^{\rm uc \,\dagger} h_{v_n})^f_\sigma \ket{X_{n}} \bra{X_{n}} (\bar{h}_{v_n} W_{n,-}^{\rm uc})^{f}_\rho \ket{0} \ ,
\end{align}%
}%
to rearrange the color indices. These modifications lead to 
{ %
\allowdisplaybreaks
\begin{align} \label{eq:bHQET_cross_section_new-1}
\mathrm{d} \sigma =& \sum_i L_{\mu \nu}^i \, \mathrm{tr}\left[ \frac{P_{v_n}}{2} \Gamma_i^\nu \frac{P_{v_{\bar n}}}{2} \bar{\Gamma}_i^\mu \right] \, H_Q(Q) \, H_m\Big( m_t , \frac{Q}{m_t} \Big)  \\
&\times \sum_{\vec{n}} \sum_{X_n, X_{\bar{n}}, X_s}  (2\pi)^4 \delta^{(4)} (q_{ee} \minus P_{X_n} \minus P_{X_{\bar{n}}} \minus P_{X_s})  \notag \\
&\times \frac{1}{N_c} \, \mathrm{Tr} \big[ \bra{0} [W_{n,+}^{\rm uc \,\dagger} h_{v_n}] (0) \ket{X_{n}} \bra{X_{n}} [\bar{h}_{v_n} W_{n,-}^{\rm uc}] (0) \ket{0} \big] \notag \\
&\times \frac{1}{N_c} \, \mathrm{Tr} \big[ \bra{0} [\bar{h}_{v_{\bar{n}}} W_{\bar{n},-}^{\rm uc}] (0) \ket{X_{\bar{n}}} \bra{X_{\bar{n}}} [W_{\bar{n},+}^{\rm uc \,\dagger} h_{v_{\bar{n}}}] (0) \ket{0}\big]  \notag \\
&\times \mathrm{tr}_c \big[ \bra{0} [(\overline{Y}_{\bar{n},-})^\intercal \, Y_{n,-} ](0) \ket{X_s} \bra{X_s} [Y_{n,+}^\dagger  \, (\overline{Y}_{\bar{n},+}^{\dagger})^\intercal ](0) \ket{0} \big] \,, \notag
\end{align}%
}%
where all color and spin indices are contracted among the fields within each sector. For the large-angle soft matrix elements in the last line we have used the identities $T \big[ Y_{\bar{n},-} \big]^{ab} = \big[(\overline{Y}_{\bar{n},+}^\dagger)^\intercal \big]^{ab}$ and
$\overline{T} \big[ Y_{\bar{n},+}^\dagger \big]^{ab}=\big[ (\overline{Y}_{\bar{n},-} )^\intercal \big]^{ab}$ for the two Wilson lines where the $T$ ($\overline{T}$) operators overwrite the Wilson lines' path-ordering prescription, see Eq.~(\ref{eq:overwriteP}). This introduces $\overline Y$-Wilson lines of soft gluons fields in the $\bar{3}$ representation. Path-ordering in each matrix element then implies the correct time-ordering so that we can drop the $T$ and $\overline{T}$ operators.\footnote{This manipulation is not mandatory, if one keeps the $T$ ($\overline{T}$) operators.} Also note that we explicitly use the notation Tr for both collinear sectors for clarity, even though it is not strictly needed for the $\bar{n}$-collinear antitop sector, since the ordering of the heavy quark fields already implies a trace over color and spin.

For the velocity labels of the top ($n$) and antitop ($\bar n$) heavy quark fields we adopt the particular expressions
\begin{align}
\label{eq:vlabeldef-2}
v^\mu_{n/\bar n} = \frac{Q}{2m_t} \Big(1, \pm \vec{n} \sqrt{1-\frac{4 m_t^2}{Q^2}} \Big)\,,
\end{align}
so that the large ultra-collinear bHQET label momenta have the form
\begin{align} \label{eq:collinear_label_momenta_bHQET}
& \tilde{P}_{X_n} = m_t v_n \,,  &          & \tilde{P}_{X_{\bar n}} = m_t v_{\bar n}\,,
\end{align}
where $m_t$ refers to the pole mass. If a mass scheme $m_t(R)$ other than the pole mass is adopted, an additional residual mass term arises, which needs to be carried through the manipulations.
Our choice of velocity labels implies that the sum of all perp momenta in each hemisphere vanishes and that the sum of the large collinear labels gives the c.m.\ momentum, i.e.\ $q_{ee}^\mu =\tilde{P}_{X_n}^\mu + \tilde{P}_{X_{\bar{n}}}^\mu$. The traces over the (axial-)vector Dirac structures in the first line of Eq.~(\ref{eq:bHQET_cross_section_new-1}) then give ($g^{\mu\nu}_\perp = g^{\mu\nu} - (n^\mu \bar n^\nu + n^\nu \bar n^\mu)/2$)
\begin{align}
	\mathrm{tr}\left[ \frac{P_{v_n}}{2} \Gamma_{1,2}^\nu \frac{P_{v_{\bar n}}}{2} \bar{\Gamma}_{1,2}^\mu \right]
	=
	\frac{1}{4}\Big(v_{n}^\mu v_{\bar n}^\nu+v_{n}^\nu v_{\bar n}^\mu \pm g^{\mu\nu}-(v_n\cdott v_{\bar n})g^{\mu\nu} \Big)
	\approx
	-2 \Big( \frac{Q}{4m_t} \Big)^2 g^{\mu\nu}_\perp \,.
\end{align}

We now decompose all physical momenta into large label ($\tilde P$) and small residual dynamical ($k$) parts,
\begin{align}\label{eq:variabledecompo-2}
&P_{X_n} = \tilde{P}_{X_n} + K_{X_n},  &          &P_{X_{\bar{n}}} = \tilde{P}_{X_{\bar{n}}}+ K_{X_{\bar{n}}} , &   &P_{X_{s,a}} = K_{X_{s,a}} , 
&   &P_{X_{s,b}} = K_{X_{s,b}}  \,, 
\end{align}
where we distinguish between the momentum of the soft particles in the top ($a$) and the antitop ($b$) hemispheres and introduce the indentity
\begin{align} 
\label{eq:introduction_of_integration_variables-new}
1 = & \int \mathrm{d}^4 k_n \, \delta^{(4)} (k_n \minus K_{X_n}) \int \mathrm{d}^4 k_{\bar{n}} \, \delta^{(4)} (k_{\bar{n}} \minus K_{X_{\bar{n}}})  \notag \\ & \times
\int \mathrm{d}^4 k_{s,a} \, \delta^{(4)} (k_{s,a} \minus K_{X_s,a}) 
\int \mathrm{d}^4 k_{s,b} \, \delta^{(4)} (k_{s,b} \minus K_{X_s,b}) \,.
\end{align} 
This insertion introduces the residual total momentum variables $k_n^\mu$, $k_{\bar{n}}^\mu$, $k_{s,a}^\mu$ and  $k_{s,b}^\mu$ that we will use for the formulation of the double hemisphere mass measurement function below. 

We can now write the cross section as
\begin{align}
\label{eq:bHQET_cross_section_new-2}
\mathrm{d} \sigma =&\, K_0\,  H_Q(Q) \, H_{m} \Big( m_t, \frac{Q}{m_t} \Big)  \sum_{\vec{n}} \\
&\times  \sum_{X_n} \int \mathrm{d}^4 k_n \, \delta^{(4)} (k_n \minus K_{X_n}) \, \frac{Q}{4 N_c m_t} \, \mathrm{Tr} \big[ \bra{0} [W_{n,+}^{\mathrm{uc},\dagger} h_{v_n} ](0) \ket{X_n} \bra{X_n} [\bar{h}_{v_n} W_{n,-}^{\mathrm{uc}}] (0) \ket{0} \big] \notag \\
&\times \sum_{X_{\bar{n}}} \int \mathrm{d}^4 k_{\bar{n}} \, \delta^{(4)} (k_{\bar{n}} \minus K_{X_{\bar{n}}}) \, \frac{Q}{4 N_c m_t}  \, \mathrm{Tr} \big[ \bra{0} [\bar{h}_{v_{\bar{n}}} W_{\bar{n}, -}^{\mathrm{uc}}] (0) \ket{X_{\bar{n}}} \bra{X_{\bar{n}}} [W_{\bar{n}, +}^{\mathrm{uc}, \dagger} h_{v_{\bar{n}}}] (0) \ket{0} \big]  \notag \\
&\times  \sum_{X_{s}}  \int \mathrm{d}^4 k_{s,a}  \int \mathrm{d}^4 k_{s,b} \, \delta^{(4)} (k_{s,a} \minus K_{X_{s,a}})\, \delta^{(4)} (k_{s,b} \minus K_{X_{s,b}}) \notag \\
&\times  \mathrm{tr}_c \big[ \bra{0} [(\overline{Y}_{\bar{n},-})^\intercal \, Y_{n,-} ](0) \ket{X_s} \bra{X_s} [Y_{n,+}^\dagger  \, (\overline{Y}_{\bar{n},+}^{\dagger})^\intercal ](0) \ket{0} \big]  \notag \\[1mm]
&\times (2\pi)^4 \, \delta^{(4)} (k_n \plus k_{\bar{n}} \plus k_{s,a}+ k_{s,b})  \,, \notag
\end{align}
where we exploited the identity $q_{ee}^\mu -\tilde{P}_{X_n}^\mu - \tilde{P}_{X_{\bar{n}}}^\mu =0$, mentioned above, to remove all large labels from the overall energy-momentum conserving $\delta$-function in the last line, and we defined
\begin{align}
K_0 &\equiv -2 g^{\mu \nu}_\perp \sum_i L_{\mu \nu}^i  = 4 (F^v(Q^2)+F^a(Q^2)) \\
& = \frac{32 \pi^2 \alpha^2}{3Q^4} \, \Big[ e_t^2 \minus \frac{2Q^2 v_e v_t e_t}{Q^2 \minus M_Z^2} \plus \frac{Q^4 (v_e^2 \plus a_e^2) (v_t^2 \plus a_t^2)}{(Q^2 \minus M_Z^2)^2} \Big]    \notag \,.
\end{align}
The physical momentum of the two hemispheres can now be written down as
\begin{align} 
\label{eq:hemisphere_momenta}
M_t  = \, & m_t v_{n}+k_t^*\,, 
& M_{\bar t}  = \, &  m_t v_{\bar n}+k_{\bar t}^* \,,
\end{align}
where the residual momenta $k_{t}^{*}$ and $k_{\bar t}^{*}$ parametrize the {\it off-shellness} of the jets constituted by the hemispheres w.r.\ to an on-shell top quark. 
The off-shellness of the two hemispheres is then defined by
\begin{align} 
\label{eq:hemisphere_offshellness_bHQET}
& s_t^* \equiv %\frac{s_t}{m_t}=
\frac{M_{t}^2-m_t^2}{m_t} 
= 2 v_{n} \cdot k_t^* = v_{n}^- k_t^{*+} + v_{n}^+ k_t^{*-} 
\,, \\
& s_{\bar{t}}^* \equiv %\frac{s_{\bar{t}}}{m_t}=
\frac{M_{\bar t}^2-m_t^2}{m_t} 
= 2 v_{\bar n} \cdot k_{\bar t}^* = v_{\bar n}^- k_{\bar t}^{*+} + v_{\bar n}^+ k_{\bar t}^{*-} \,, \notag 
\end{align}
where at leading order in the power counting we can neglect contributions that are quadratic in the residual momenta.
The fact that for our concrete bHQET large label choice there is no residual perp momentum (i.e.\ $k_t^{*\perp}=k_{\bar t}^{*\perp}=0$), will be important for the manipulations we carry out when considering the top quark decay in Sec.~\ref{sec:jetfctfactorization}.
The star as a superscript has been introduced to provide a better distinction of these physical off-shellness momenta from the ultra-collinear and soft residual momentum variables introduced above which arise from the factorization $\ket{X} = \ket{X_n} \ket{X_{\bar{n}}} \ket{X_s}$. The state $X_{n}$ ($X_{\bar n}$) contains the (anti)top quark decay products with ultra-collinear residual momentum and ultra-collinear gluon as well as light quark final states, which render it gauge-invariant under ultra-collinear gauge transformations.~\cite{Bauer:2003mga}.
Furthermore, we can now distinguish between the $e^+e^-$ c.m.\ scaling of the velocity labels and the residual momenta ($v_n^-=v_{\bar n}^+\approx Q/m_t$, 
$v_n^+=v_{\bar n}^-\approx m_t/Q$, 
$k_n^-\sim k_t^{*-}\sim k_{\bar n}^+\sim k_{\bar t}^{*+}\sim \Gamma Q/m_t$, \
$k_n^+\sim k_t^{*+}\sim k_{\bar n}^-\sim k_{\bar t}^{*-}\sim \Gamma m_t/Q$) and the respective homogeneous top/antitop rest frame scaling (
$v_{n,r}^\pm=v_{{\bar n},r}^\pm=1$, 
$k_{n,r}^\pm \sim k_{{\bar n},r}^\pm  \sim k_{t,r}^{*\pm} \sim k_{{\bar t},r}^{*\pm} \sim\Gamma$). It is the information on the two rest frames defined by the velocity labels $v_{n,\bar n}^\mu$ which is the reason why in bHQET two light-cone momenta need to be specified for the masses of each of the two hemispheres. 
The choices $v_{n}^- k_t^{*+}=v_{n}^+ k_t^{*-}$ and $ v_{\bar n}^- k_{\bar t}^{*+} = v_{\bar n}^+ k_{\bar t}^{*-} $ implement that the residual momenta $k_t^*$ and $k_{\bar t}^*$ are proportional to $v_{n}$ and $v_{\bar n}$, respectively, so that in reference frames where $v_{n}^\mu$ or $v_{\bar n}^\mu$ have the static form $(1,\vec{0})$, we are in a hemisphere rest frame where $\vec{M}_{t}=0$ (or $\vec{M}_{\bar t}=0$). This information is not relevant when the top and antitop hemispheres are treated inclusively, but it is important for the definition of the 3-body final state momenta for the semileptonic top quark decay.

To implement the measurement function we recall that the hemisphere masses are given by 
$M_t^2 = ({P}_{X_n} + P_{X_{s,a}})^2 = (m_t v_n + k_n + k_{s,a})^2$ and 
$M_{\bar{t}}^2 = ({P}_{X_{\bar n}} + P_{X_{s,b}})^2 = (m_t v_{\bar n} + k_{\bar{n}} + k_{s,b})^2$. Using the physical off-shellness momenta introduced in Eq.~(\ref{eq:hemisphere_offshellness_bHQET}), the measurement function in the $e^+ e^-$ c.m. frame at leading order in the power counting is given by
\begin{align} \label{eq:measurement_function_bHQET}
\frac{1}{m_t^2}\delta\Big( 2 v_{n} \cdot k_{n} + v_{n}^- k_{s,a}^{+} - 2 v_{n} \cdot k_t^* \Big) \, 
\delta \Big( 2 v_{\bar n} \cdot k_{\bar n}  + v_{\bar n}^+ k_{s,b}^{-} -2 v_{\bar n} \cdot k_{\bar t}^* \Big) \,,
\end{align}
where for the large-angle soft momenta only the $+$ ($-$) momenta contribute in the top (antitop) hemispheres, since 
\begin{align}
v_n^-=v_{\bar n}^+\approx \frac{Q}{m_t}
\end{align} 
are the dominant light-cone velocity components in the $e^+e^-$ c.m.\ frame. 
In addition, we rewrite the sums over the ultra-collinear $X_n$ and $X_{\bar n}$ states in the cross section (\ref{eq:bHQET_cross_section_new-2}) in terms of the inclusive bHQET jet functions
\begin{align}
\label{eq:bHQETjetfcts-2}
& J_{B_{t}}^{\Gamma_t}  ( 2 v_{n} \cdot k_{n} ) =  \sum_{X_n}  \delta^{(4)} (k_n \minus K_{X_n})  \frac{(2\pi)^3}{4 N_c m_t} \, \mathrm{Tr} \big[ \bra{0} [W_{n,+}^{\mathrm{uc},\dagger} h_{v_n} ](0) \ket{X_n} \bra{X_n} [\bar{h}_{v_n} W_{n,-}^{\mathrm{uc}}] (0) \ket{0} \big] , \notag  \\
& J_{B_{\bar{t}}}^{\Gamma_t}  (2v_{\bar n} \cdot k_{\bar n} ) =\sum_{X_{\bar{n}}}  \delta^{(4)} (k_{\bar{n}} \minus K_{X_{\bar{n}}}) \, \frac{(2\pi)^3}{4 N_c m_t} \, \mathrm{Tr} \big[ \bra{0} [\bar{h}_{v_{\bar{n}}} W_{\bar{n}, -}^{\mathrm{uc}}] (0) \ket{X_{\bar{n}}} \bra{X_{\bar{n}}} [W_{\bar{n}, +}^{\mathrm{uc}, \dagger} h_{v_{\bar{n}}}] (0) \ket{0} \big]  \,.
\end{align}
That the bHQET functions are only functions of $2v_{n}.k_{n}$ or  $2v_{\bar n}.k_{\bar n}$ is discussed in Sec.~\ref{sec:inclusivefactorization}.

We continue by noting that the ultra-collinear and large-angle soft momentum scalings and the measurement function in Eq.~(\ref{eq:measurement_function_bHQET}) imply that we can replace the residual momentum conserving $\delta$-function in the $e^+e^-$ c.m. frame by\footnote{\label{foot:kperp} Due to our choice of the light-like reference vector direction $\vec{n}$ the sum of the ultra-collinear and large-angle soft perp momenta vanish individually in each hemisphere.
This forces both ultra-collinear perp momenta $k_n^\perp$ and $k_{\bar{n}}^\perp$ to scale with the power-suppressed large-angle soft momenta,	
so that we may as well replace the perp $\delta$-function by $\delta^{(2)} ( k_{n}^\perp )$.}
\begin{align}
\label{eq:deltareplace}
\delta^{(4)} (k_n \plus k_{\bar{n}} \plus k_{s,a} \plus k_{s,b}) \to 
2 \, \delta ( k_{\bar{n}}^+ ) \, \delta (k_n^- ) \, \delta^{(2)} (k_n^\perp \plus k_{\bar{n}}^\perp \plus k_{s,a}^\perp \plus k_{s,b}^\perp) \,.
\end{align}
Using also that $v^\perp_{n,\bar n}=0$, we thus obtain the following expression for the double hemisphere invariant mass distribution with all integration light-cone momenta defined in the $e^+e^-$ c.m. frame
\begin{align}  
\label{eq:bHQET_cross_section_new-3}
\frac{\mathrm{d}^2\sigma}{\mathrm{d} M_t^2 \mathrm{d} M_{\bar{t}}^2} =&\, \frac{2}{\pi} \, \frac{\sigma_0}{(2\pi)^4} \, H_Q(Q) \, H_{m} \Big(m_t,\frac{Q}{m_t}\Big)  \sum_{\vec{n}} \int \mathrm{d}^4 k_n \, \mathrm{d}^4 k_{\bar{n}} \,   J_{B_t}^{\Gamma_t} \bigg(  v_n^- k_n^+ \bigg) \, J_{B_{\bar{t}}}^{\Gamma_t} \bigg( v_{\bar n}^+  k_{\bar{n}}^-\bigg) \notag \\
&\times \sum_{X_{s}}\int \mathrm{d}^4 k_{s,a}  \int \mathrm{d}^4 k_{s,b} \, \delta^{(4)} (k_{s,a} \minus K_{X_{s,a}})\, \delta^{(4)} (k_{s,b} \minus K_{X_{s,b}}) \\ 
&\times \frac{1}{N_c} \,  \mathrm{tr}_c \big[ \bra{0} [(\overline{Y}_{\bar{n},-})^\intercal \, Y_{n,-} ](0) \ket{X_s} \bra{X_s} [Y_{n,+}^\dagger  \, (\overline{Y}_{\bar{n},+}^{\dagger})^\intercal ](0) \ket{0} \big] \notag \\
&\times 2\, (2\pi)^4 \,\delta ( k_{\bar{n}}^+ ) \, \delta (k_n^- ) \, \delta^{(2)} (k_{n}^\perp )  \notag \\
& \times
\frac{1}{m_t^2}\delta\Big( 2 v_{n} \cdot k_{n} + v_{n}^- k_{s,a}^{+} - 2 v_{n} \cdot k_t^* \Big) \, 
\delta \Big( 2 v_{\bar n} \cdot k_{\bar n}  + v_{\bar n}^+ k_{s,b}^{-} -2 v_{\bar n} \cdot k_{\bar t}^* \Big) \notag
\end{align}
where $\sigma_0 = N_c K_0 Q^2/(8\pi)$ denotes the Born total cross section. We can now proceed by trivially carrying out the integration over $k_n^-$, $k_{\bar{n}}^+$ and $k_{n}^\perp$ using the remainder of the residual momentum conserving $\delta$-function in the second to last line. In addition, we can also trivially perform the integration over $k_{s,a}^-$, $k_{s,a}^\perp$, $k_{s,b}^+$ and $k_{s,b}^\perp$ which do not enter the measurement function. Finally, we combine the integration  $\mathrm{d}^2 k_{\bar{n}}^\perp$ with the sum over the jet axis directions $\vec{n}$ which yields the integration over the square of the collinear 3-momentum  $|\vec{P}^2_{X_{\bar{n}}}|\approx m_t^2 |\vec{v}_{\bar{n}}|^2 = Q^2/4-m_t^2\approx Q^2/4$ in hemisphere $b$ over 
all possible dijet directions, $\sum_{\vec{n}} \int \mathrm{d}^2 k_{\bar{n}}^\perp \approx Q^2/4  \int \mathrm{d}\Omega = \pi Q^2\approx \pi m_t^2 v_{n}^- v_{\bar n}^+$. By relabelling $k_{s,a}^+=\ell^+$ and $k_{s,b}^-=\ell^-$ for better readability, we finally obtain
\begin{align}
\label{eq:bHQET_cross_section_new-4}
\frac{\mathrm{d}^2\sigma}{\mathrm{d} M_t^2 \mathrm{d} M_{\bar{t}}^2} =&\, \sigma_0 \, H_Q(Q) \, H_{m} \Big(m_t,\frac{Q}{m_t}\Big)  \, v_{n}^- v_{\bar n}^+ \int \mathrm{d} k_n^+ \, \mathrm{d} k_{\bar{n}}^- \int \mathrm{d} \ell^+ \,  \mathrm{d} \ell^-  \\
& \times \delta \Big(v_n^- k_n^+ \plus v_n^- \ell^+ \minus v_n^- k_t^{*+} \minus v_n^+ k_t^{*-} \Big) \, \delta \Big(  v_{\bar n}^+ k_{\bar{n}}^- \plus  v_{\bar n}^+ \ell^- \minus  v_{\bar n}^+ k_{\bar{t}}^{*-} \minus  v_{\bar n}^- k_{\bar{t}}^{*+} \Big) \notag \\
&\times  J_{B_t}^{\Gamma_t} \Big(v_n^-  k_n^+ \Big) \, J_{B_{\bar{t}}}^{\Gamma_t} \Big(  v_{\bar n}^+  k_{\bar{n}}^-\Big) \, S_{\mathrm{hemi}} (\ell^+, \ell^-) \,,  \notag 
\end{align}
where the $5$-flavor double hemisphere soft function is defined by
\begin{align} 
\label{eq:nf5hemisphere_soft_function_new}
	&S_{\mathrm{hemi}} (\ell^+, \ell^-) = \\
&=  \sum_{X_{s}} \delta (\ell^+ \minus K_{X_{s,a}}^{+})  \delta (\ell^- \minus K_{X_{s,b}}^{-})  \frac{1}{N_c} \mathrm{tr}_c \big[\! \bra{0} [(\overline{Y}_{\bar{n},-})^\intercal \, Y_{n,-} ](0) \ket{X_s} \bra{X_s} [Y_{n,+}^\dagger  \, (\overline{Y}_{\bar{n},+}^{\dagger})^\intercal ](0) \ket{0}\! \big] \notag \\
&= \frac{1}{N_c} \,  \mathrm{tr}_c \big[ \bra{0} [(\overline{Y}_{\bar{n},-})^\intercal \, Y_{n,-} ](0) \, \delta (\ell^+ \minus n \cdott \hat{P}_a) \,    \delta (\ell^- \minus \bar{n} \cdott \hat{P}_b)                     [Y_{n,+}^\dagger  \, (\overline{Y}_{\bar{n},+}^{\dagger})^\intercal ](0) \ket{0} \big] \,. \notag
\end{align}
Here, $\hat{P}_{a,b}$ denote the soft momentum projection operators in the hemispheres $a$ and $b$. This 5-flavor double hemisphere soft function also appears in the dijet factorization for $Z$-pole invariant mass based event-shape distributions such as thrust~\cite{Abbate:2010xh,Benitez:2024nav}, $C$ parameter~\cite{Hoang:2014wka} or heavy jet mass~\cite{Hoang:2025uaa}.
For the purpose of making predictions for the double hemisphere mass distribution in the resonance region we can now carry out the $k_n^+$ and $k_{\bar n}^-$ integrations, yielding
\begin{align}
\label{eq:bHQET_cross_section_new-5}
\frac{\mathrm{d}^2\sigma}{\mathrm{d} M_t^2 \mathrm{d} M_{\bar{t}}^2} 
=&\, \sigma_0 \, H_Q(Q) \, H_{m} \Big(m_t,\frac{Q}{m_t}\Big)  \int \mathrm{d} \ell^+ \,  \mathrm{d} \ell^- \\
&\times  J_{B_t}^{\Gamma_t} \Big( s_t^* \minus v_n^-\ell^+  \Big) \, J_{B_{\bar{t}}}^{\Gamma_t} \Big( s_{\bar t}^*-v_{\bar n}^+ \ell^-  \Big) \, S_{\mathrm{hemi}} (\ell^+, \ell^-) \,.  \notag 
\end{align}
This is the form of the factorized bHQET double hemisphere invariant mass distribution originally derived in Refs.~\cite{Fleming:2007qr,Fleming:2007xt}. It contains all contributions that are leading order in the double resonant region, reflecting the power counting of all distinct relevant modes summarized in Tab.~\ref{tab:modetable}.  A graphical illustration of all modes covering also the case of the semileptonic top decay in the endpoint region is shown in Fig.~\ref{fig:topdecay_modes}.
Its evaluation at N$^2$LL$+{\cal O}(\alpha_s)$ has been used to the top mass calibration analyses of Refs.~\cite{Butenschoen:2016lpz,Dehnadi:2023msm} and a N$^3$LL$+{\cal O}(\alpha_s^2)$ 
analysis was provided in Ref.~\cite{Bachu:2020nqn}.

Let us comment on the non-perturbative effects that enter the leading factorization formulae~(\ref{eq:bHQET_cross_section_new-5}) in the resonance region. In principle,
all factorization (and matching) functions have a sensitivity to small momenta and thus non-perturbative contributions. For inclusive vacuum matrix element matching functions and Wilson coefficients this sensitivity to small momenta is quadratic or higher, while for factorization functions that involve cuts or restrictions on infrared momenta, also linear sensitivities can arise. 
For the matching coefficients $H_Q$ and $H_m$ the associated non-perturbative effects are ${\cal O}(\Lambda_{\rm QCD}^2/Q^2)$ and ${\cal O}(\Lambda_{\rm QCD}^2/m_t^2)$, respectively, and thus power suppressed and not part of the factorization theorem. For the inclusive bHQET jet functions
$J_{B_t}^{\Gamma_t}$ and $J_{B_{\bar{t}}}^{\Gamma_t}$, which describe the inclusive $n$ ($\bar{n}$) ultra-collinear radiation off the boosted top (antitop) quarks, it is also at most quadratic, when 
a suitable short-distance scheme is employed for the top quark mass~\cite{Fleming:2007qr,Fleming:2007xt}.\footnote{The bHQET function has a linear sensitivity to small momenta in the top quark pole mass scheme, which, however, is absent when switching to a short-distance mass scheme~\cite{Fleming:2007qr,Hoang:2018zrp}. The linear sensitivity to small momenta related to the heavy quark mass is not related to any non-perturtative effects for the top quark due to its large width.}
Due to the finite top width, these non-perturbative effects are suppressed by $\Lambda_{\rm QCD}^2/\Gamma^2$ in the double resonant region and thus also beyond leading power. However, leading power  
non-perturbative effects arise from the double hemisphere soft function as it involves a restriction of the soft radiation through its assignment to either the top or the antitop hemisphere. These non-perturbative effects are associated to powers of $\Lambda_{\rm QCD}Q/(M_{t,\bar t}^2-m_t^2)\sim \Lambda_{\rm QCD}Q/(m_t\Gamma)$  in the resonance region that must be summed to all orders. 
They can be factored from the partonic soft function 
$\hat{S}_{\mathrm{hemi}}$ in the form of a non-perturbative soft distribution function $F$~\cite{Fleming:2007qr,Hoang:2025uaa},
\begin{align}
\label{eq:nonpertF}
S_{\rm hemi}(\ell^+,\ell^-) = \int\!\! \mathrm{d} k^+\, \mathrm{d} k^-\,\hat S_{\rm hemi}(\ell^+-k^+,\ell^--k^-)F(k^+,k^-)\,.
\end{align}
The distribution function $F$ is normalized to unity, $\int \mathrm{d} \ell'^+ \,  \mathrm{d} \ell'^- F (\ell'^+, \ell'^-) =1$, has positive support for both light cone momenta, peaks at $\ell'^+= \ell'^-\sim \Lambda_{\rm QCD}$ and falls off exponentially.
In the resonance region it provides a smearing of the partonic contributions in Eq.~(\ref{eq:bHQET_cross_section_new-5}) that shifts the partonic
resonance positions to larger masses by an amount of order $v_n^-=v_{\bar n}^+=Q/m_t$ times $\Lambda_{\rm QCD}$.  The essential aspect is that $F$ also arises in the leading power factorization formulae of $e^+e^-$ event-shape distributions such as thrust~\cite{Abbate:2010xh,Benitez:2024nav}, $C$ parameter~\cite{Hoang:2014wka} or heavy jet mass~\cite{Hoang:2025uaa}. It can thus be determined from experimental data unrelated to top quark production.   
The form of the convolution in Eq.~(\ref{eq:nonpertF}) can also be used to associate the distribution function $F$ to the parton-to-hadron migration matrix describing the effect of hadronization in multi-purpose Monte-Carlo event generators~\cite{Hoang:2024zwl}. For hemisphere masses $M_{t,\bar t}$ above the resonance (for $\Gamma_t\ll M_{t,\bar t}-m_t\ll m_t$), the effects of $F$ can be expanded in terms of an operator product expansion where the leading non-perturbative correction is suppressed by $\Lambda_{\rm QCD}Q/(M_{t,\bar t}^2-m_t^2)$ and effectively yields a shift in the partonic distribution~\cite{Fleming:2007qr,Fleming:2007xt} that is also known from massless quark invariant mass-based event-shape distributions. As we show in Sec.~\ref{sec:renormalizationtopdecay}, $F$ also remains the only source of leading power non-perturbative effects for the decay sensitive differential cross section~(\ref{eq:diffcrossmulti}) and also provides the non-perturbative corrections to the $b$-jet, because we define the $b$-jet system as all hadronic final states in the top quark hemisphere.

We conclude this section by noting that the form of the factorization in Eq.~(\ref{eq:bHQET_cross_section_new-5}) involves the complete top hemisphere invariant mass variable $s_t^*$ of Eq.~(\ref{eq:hemisphere_offshellness_bHQET}) (and likewise $s_{\bar t}^*$ for the antitop). This is adequate for the case where the top decay is treated inclusively. However, when considering in addition the kinematics of the semileptonic top quark decay, the invariant mass of the top hemisphere and the one of the hadronic b-jet, which is radiated with an angle w.r.\ to the $\vec{n}$-direction, are affected by the ultra-collinear plus ($k_t^{*+}$) and the minus ($k_t^{*-}$) momenta entering the jet function in a different (angle-dependent) way. It then becomes mandatory to distinguish between the plus and minus momentum contributions of the top hemisphere off-shellness defined as\footnote{\label{foot:offshells} We remind the reader that $s_t^{*+}+s_t^{*-}=s_t^*$ gives the total top hemisphere off-shellness in Eq.~(\ref{eq:hemisphere_offshellness_bHQET}) and that the choice $s_t^{*+}=s_t^{*-}=s_t^*/2$ ensures that the top hemisphere rest frame 
is the frame were $v_n^\mu=(1,\vec{0})$.}
\begin{align} 
\label{eq:tophemisphere_offshellness_bHQET}
s_t^{*+} \equiv  v_{n}^- k_t^{*+}\,, &&  
s_t^{*-} \equiv v_{n}^+ k_t^{*-}\,.
\end{align}
For the derivation of the factorization for the top quark decay in Sec.~\ref{sec:factorizationtopdecay}, which involves a differential treatment and further factorization of  $J_{B_t}$, we will therefore consider $J_{B_t}$ as a function of  
the plus and the minus off-shellness variables,
\begin{align}
\label{eq:Jbtarguments}
 J_{B_t} \Big( s_t^* \minus v_n^-\ell^+   \Big) =  J_{B_t} \Big( s_t^{*+} \plus s_t^{*-} \minus v_n^-\ell^+   \Big) \to J_{B_t} \Big( s_t^{*+}  \minus v_n^-\ell^+, s_t^{*-} \Big)\,,
\end{align}
This makes manifest that only the plus contribution is affected by the convolution with the (plus momentum distribution of the) hemisphere soft function.

\subsection{Inclusive Semileptonic B decay in the Endpoint Region}
\label{sec:Bdecayfactorization_new}

In this section we briefly review the derivation of the SCET factorization theorem for inclusive semileptonic $\bar{B} \to X_u \ell \bar{\nu}_\ell$ decays in the kinematic endpoint region, where the entire hadronic final state forms a jet and which was first derived in Refs.~\cite{Bauer:2001yt,Bosch:2004th}\footnote{
We remind the reader that the term inclusive semileptonic decay refers to this simple jet clustering. We also note that the authors of Ref.~\cite{Bauer:2001yt} considered the process $\bar{B} \to X_s \gamma$, but the QCD manipulations for the derivation of the factorization theorem are identical to the ones for $\bar{B} \to X_u \ell \bar{\nu}_\ell$.}. In the derivation of the factorization theorem for the differential top quark decay in Sec.~\ref{sec:factorizationtopdecay} the description of top-antitop production and the $b$-jet top decay dynamics requires the introduction of two different light-like directions, one for top-antitop dijet axis and one for the $b$-jet direction. Both directions are indistinguishable for the large-angle soft radiation as the top decay products are boosted in the top direction, but the difference is relevant for the ultra-collinear radiation, which is soft in the top rest frame. Anticipating this notational complication we use in the following the light-like reference vectors $n'^\mu=(1,\vec{n}')$ and $\bar{n}'^\mu = (1, - \vec{n}')$ for the $X_u$-jet direction (which is the analogue of the $b$-jet for the top quark decay). The light-cone components with respect to $n'$ and $\bar n'$ are written in the form
$p^\mu=(\hat{p}^+, \hat{p}^-, \hat{p}^\perp) = (n' \cdott p , \bar{n}' \cdott p, \hat{p}^\perp )$ with hats to make the difference manifest. We also note that in the following we assume that the $\bar B$ velocity is $v^\mu$, rather than adopting the $\bar B$ rest frame, as a preparation for the treatment of the boosted top quark decay in Sec.~\ref{sec:factorizationtopdecay} where it is useful not to adopt a particular frame. To be consistent with the terminology we use for the top production we therefore refer to the radiation, that is typically called 'soft' in the context of heavy meson decays also as ultra-collinear.

The differential decay rate in full QCD, which integrates to the $\bar B$ meson's width, reads  
\begin{align}
\mathrm{d} \Gamma (\bar{B} \to X_u \ell \bar{\nu}_{\ell}) =& \, \frac{1}{2 m_B} \, \frac{\mathrm{d}^3 p_\ell}{(2\pi)^3} \, \frac{1}{2 E_\ell} \, \frac{\mathrm{d}^3 \pn}{(2\pi)^3} \, \frac{1}{2 E_{\nu_\ell}} \sum_{X_u} (2\pi)^4 \, \delta^{(4)} (p_B \minus p_\ell \minus \pn \minus P_{X_u}) \\
&\times  \bra{\bar{B}} \mathcal{H}_W^\dagger (0) \ket{X_u \ell \bar{\nu}_\ell} \bra{X_u \ell \bar{\nu}_\ell} \mathcal{H}_W (0) \ket{\bar{B}} \ , \notag
\end{align}
including the  $\bar{B} \to X_u \ell \bar{\nu}_\ell$ matrix element with the 
weak effective Hamiltonian operator
\begin{align} 
\label{eq:weak_Hamiltonian_new_2}
\mathcal{H}_W = -\frac{4 G_F}{\sqrt{2}} \, V_{ub} \, (\bar{u} \gamma^\mu P_L b) (\bar{\ell} \gamma_\mu P_L \nu_\ell) \,,
\end{align}
where the quark current operator $\mathcal{J}^\mu_{bu} = (\bar{u} \gamma^\mu P_L b)$ describes the hadronic $b\to u$ part of the decay. 
 For later convenience we also introduce the total leptonic momentum 
\begin{align} 
q = p_\ell + \pn\,,
\end{align}
so that $q^2$ is the virtual $W$ invariant mass and we write the $\bar B$ momentum as 
\begin{align}
p_B=m_B v
\end{align}
with the 4-velocity $v^\mu$.
The sum over $X_u$ includes the phase space integration over all hadronic final state particles containing the $u$ quark produced through the QCD $b\to u$ current. As we treat the hadronic final state inclusively, we also introduce the total hadronic momentum variable 
\begin{align} 
H = P_{X_u}=p_B-  p_\ell - \pn = p_B-q \,,
\end{align}
so that we can factorize the total phase space in terms of the three-body phase space $\Pi_3(p_B; p_\ell, \pn, H) $ for the leptonic momenta and the hadronic momentum $H^\mu$, the sum over all hadronic final state particles $X_u$ and an additional integration over the total hadronic invariant mass $H^2$.
Given that the final state can be written as a tensor product of hadronic and leptonic states, $\ket{X_u \ell \bar{\nu}_\ell}=\ket{X_u}\ket{\ell \bar{\nu}_\ell}$,   and   the form of the weak Hamiltonian in Eq.~(\ref{eq:weak_Hamiltonian_new_2}), the decay rate can be decomposed into the product of a leptonic and a hadronic tensor
\begin{align} 
\label{eq:decomposition_leptonic_hadronic_tensor_new_2}
\mathrm{d} \Gamma (\bar{B} \to X_u \ell \bar{\nu}_{\ell}) = \mathrm{d} H^2 \, \mathrm{d} \Pi_3(p_B ; p_\ell, \pn, H) \, \Big( \frac{4 G_F}{\sqrt{2}} \Big)^2 \, |V_{ub}|^2 \, \tilde{L}_{\mu \nu} (p_\ell, \pn) \, W^{\mu \nu} (p_B, q) \,.
\end{align}
The momentum space expression $\tilde{L}^{\mu \nu} (p_\ell, p_{\nu_\ell})$ of the leptonic tensor is obtained from $L^{\mu \nu} = \sum_{\mathrm{spins}} \bra{0}  \, \big( \bar{\ell} \gamma^\mu P_L \nu_\ell \big)^\dagger (0) \ket{\ell \bar{\nu}_\ell} \bra{\ell \bar{\nu}_\ell}  \big( \bar{\ell} \gamma^\nu P_L \nu_\ell \big) (0) \ket{0}$. The hadronic tensor is given by
\begin{align}
\label{eq:BdecayWmunu_new}
W^{\mu \nu} = \frac{1}{2 m_B} \, \frac{1}{2\pi} \sum_{X_u} (2\pi)^4 \, \delta^{(4)} (m_B v \minus q \minus P_{X_u})  \, \bra{\bar{B}} \mathcal{J}^{\dagger \mu}_{bu} (0) \ket{X_u} \bra{X_u} \mathcal{J}^\nu_{bu} (0) \ket{\bar{B}}\,,
\end{align}
and can be further factorized in the endpoint region. In analogy to the factorization theorem for boosted top pair production, we first match the QCD current $\mathcal{J}^\mu_{bu}$ onto corresponding HQET-SCET heavy-to-light effective currents, which integrates out local effects at the bottom mass scale. The matching relation reads
\begin{align} 
\label{eq:EFT_heavy_to_light_currents_B_decays_new}
\mathcal{J}^\mu_{bu} (0) = \sum_{j=1}^3 \, C_j(m_b,\hat{v}^+ \hat{H}^-) \, \big[ \bar{\chi}_{n'} Y_{n',+}^{{\rm uc},\dagger} \, \Gamma^\mu_j \, h^{(b)}_v \big] (0) \, + \, {\cal O}(\Lambda_{\rm QCD}/m_b) \ ,
\end{align}
where $\bar{\chi}_{n'} = \bar{\xi}_{n'}  W_{n'}$ is the SCET ($u$-quark) jet field defined in analogy to the SCET jet fields in Eq.~(\ref{eq:SCET_currents-3}) (but defined w.r.\ to the light-like reference vector $n'$ pointing in the $X_u$-jet direction and with the proper adaptations of the flavor number scheme appropriate for 		the $
\bar B$ decay). $h^{(b)}_v$ is the HQET bottom quark field\footnote{Here we use the superscript (b) for the HQET bottom quark field. For the HQET top quark field we suppress this superscript to keep notation simple.} defined in analogy to bHQET heavy quark fields in Eq.~(\ref{eq:SCET-bHQET-current-match-2})  with velocity label $v^\mu$ which we identify with the 4-velocity of the $\bar{B}$ meson. The 
dependence of the Wilson coefficient on $\hat{v}^+ \hat{H}^-\approx 2v \cdot H\approx m_b$ arises from the lepton momentum transfer~\cite{Bauer:2000yr}. 
It is convenient but not mandatory to work in the $\bar B$ rest frame, where $v = (1, \vec{0})$ (with $\hat{v}^+=\hat{v}^-=1$, $\hat{v}^\perp=0$) and the $b$-quark is very slowly moving (with velocity $\sim\Lambda_{\rm QCD}/m_b$). 
For the direction of the collinear reference vector 
$n'^\mu=(1,\vec{n}')$ we adopt the direction of the total hadronic final state momentum $\vec{n}' =\vec{H}/|\vec{H}|$. Adopting the $\bar B$ rest frame, the hadronic final state jet is back-to-back to the leptonic final state in the endpoint region so that $\vec{n}' = - \vec{q}/|\vec{q}|$ and we also have 
$\hat{v}^+ \hat{H}^-=\hat{H}^-$, which is the notation commonly used in the flavor physics literature. However, we note that in a general frame the association of $\vec{n}'$ with the hadronic momentum, and the definitions of $n'^\mu$ and $\bar n'^\mu$ given above, represent the correct choices.
This choice for $n'$ also fixes the large label momentum of the SCET jet field as
\begin{align} 
\label{eq:SCET-B-largelabel}
\tilde{P}_{X_{n'}}^\mu =  (\hat{\tilde{P}}_{X_{n'}}^+,\hat{\tilde{P}}_{X_{n'}}^-,\hat{\tilde{P}}_{X_{n'}}^\perp)=(0,\hat{H}^-,0) = (0,m_b \hat{v}^- - \hat{q}^-,0) 
\,.
\end{align}
For the heavy-to-light currents 
we employ the Dirac basis
\begin{align}
\label{eq:diracbasis}
\Gamma_j^\mu \in P_R \Big \{ \gamma^\mu, v^\mu, \frac{n'^\mu}{\hat{v}^+} \Big \} \,,
\end{align}
of Refs.~\cite{Pirjol:2002km, Chay:2002vy} where also the expressions of the Wilson coefficients at the one-loop order can be found.  
The soft Wilson line $Y_{n',+}^{{\rm uc},\dagger}$ appears since
we again employ the soft decoupled jet fields $\bar{\chi}_{n'}$. This decoupling is not performed for the heavy bottom quark field $h^{(b)}_v$.
Using the $\bar{B}$ meson states defined in HQET, $\ket{\bar{B}_v} = m_B^{-1/2} \ket{\bar{B}}$, and the factorization $\ket{X_u}=\ket{X_{n'}X_{\rm uc}}=\ket{X_{n'}}\ket{X_{\rm uc}}$, this yields the following form of the hadronic tensor 
\begin{align}
\label{eq:BdecayWmunufactorization}
W^{\mu \nu}(v,p_B,q) &= \frac{1}{2(2\pi)}   \sum_{j,j'=1}^3  \, C_j(m_b,\hat{v}^+ \hat{H}^-) C_{j'}(m_b,\hat{v}^+ \hat{H}^-) \\ 
&\times 
 \sum_{X_{n'},X_{\rm uc}} (2\pi)^4 \, \delta^{(4)} (m_b v \plus \overline{\Lambda} v \minus q \minus P_{X_{n'}} \minus P_{X_{\rm uc}}) \notag \\
&\times \bra{\bar{B}_v}  \big( \bar{h}^{(b)}_v Y_{n',-}^{\rm uc} \bar{\Gamma}^\mu_{j'} \chi_{n'} \big) (0) \ket{X_{n'}X_{\rm uc}} \bra{X_{n'}X_{\rm uc}} \big( \bar{\chi}_{n'} \Gamma^\nu_j Y_{n',+}^{{\rm uc},\dagger} h^{(b)}_v \big) (0) \ket{\bar{B}_v} \,, \notag
\end{align}
where we wrote the $\bar{B}$ meson mass as
\begin{align}
\label{eq:Bmass}
m_B = m_b+\overline{\Lambda} \,.
\end{align}
Analogous to the derivation in the previous section, we now disentangle the fields in the hadronic tensor into the distinct collinear and soft sectors. 
For this purpose we use the SCET Fierz relation
\begin{align} 
\label{eq:Fierz_relation_new2}
(\xi_{n'})_\alpha (\bar{\chi}_{n'})_\beta= \frac{1}{2} \bigg[\,  \mathrm{tr} \Big[ \xi_{n'} \bar{\chi}_{n'} \frac{\slashed{\bar{n}}'}{2} \Big] &  \Big( \frac{\slashed{n}'}{2} \Big)_{\alpha \beta} 
 - \mathrm{tr} \Big[ \xi_{n'} \bar{\chi}_{n'} \frac{\slashed{\bar{n}}'}{2} \gamma_5 \Big]\Big( \frac{\slashed{n}'}{2} \gamma_5 \Big)_{\alpha \beta} 
\\ &
- \mathrm{tr} \Big[ \xi_{n'} \bar{\chi}_{n'} \frac{\slashed{\bar{n}}'}{2} \gamma_\perp^\mu \Big] \,  \Big( \frac{\slashed{n}'}{2} \gamma_{\perp, \mu} \Big)_{\alpha \beta} \bigg] \,, \notag
\end{align}
for the two collinear jet fields and the HQET relation in Eq.~(\ref{eq:HQET_Fierz_relation_new}) for the two heavy quark fields, where in both cases only the respective first terms leads to non-vanishing $\bar{B}$ meson matrix elements at leading order in the power counting. In addition we decompose all momenta into large label ($\tilde{P}$) and small residual dynamical ($k$) parts,
\begin{align}\label{eq:variabledecompo-B}
	&P_{X_{n'}} = \tilde{P}_{X_{n'}} + K_{X_{n'}}\,,  &   &P_{X_{\rm uc}} = K_{X_{\rm uc}}\,, 
\end{align}
and introduce residual total momentum variables $k_{n'}^\mu$ and $k_{\rm uc}^\mu$ through the identity
\begin{align} 
\label{eq:introduction_of_integration_variables-B}
1 =  \int \mathrm{d}^4 k_{n'} \, \delta^{(4)} (k_{n'} \minus K_{X_{n'}}) 
	\int \mathrm{d}^4 k_{\rm uc} \, \delta^{(4)} (k_{\rm uc} \minus K_{X_{\rm uc}}) \,.
\end{align} 
Using that our definitions of the light-like reference vector $n'^\mu$ and the large label momenta imply that the total perp hadronic momentum vanishes, the hadronic tensor adopts the form
\begin{align}
\label{eq:BdecayWmunufactorization-3}
W^{\mu \nu}&(v,p_B,q)   = \frac{1}{2(2\pi)}  H_{d}^{\mu\nu}(q,v,m_b)  \\
&\times \sum_{X_{n'}}\int \mathrm{d}^4 k_{n'} \, \delta^{(4)} (k_{n'} \minus K_{X_{n'}}) \,   \frac{1}{4 N_c \hat{H}^-} \, \mathrm{Tr} \Big[ \bra{0}   \slashed{\bar{n}}' \chi_{n'} (0) \ket{X_{n'}} \bra{X_{n'}} \bar{\chi}_{n'} (0) \ket{0} \Big] \notag \\
&\times \sum_{X_{\rm uc}} \int \mathrm{d}^4 k_{\rm uc} \, \delta^{(4)} (k_{\rm uc} \minus K_{X_{\rm uc}})  \,  \mathrm{Tr} \Big[ \bra{\bar{B}_v}  \big( \bar{h}^{(b)}_v Y_{n',-}^{\rm uc} \big) (0) \ket{X_{\rm uc}} \bra{X_{\rm uc}} \big(  Y_{n',+}^{{\rm uc},\dagger} h^{(b)}_v \big) (0) \ket{\bar{B}_v} \Big]\ . \notag \\
&\times 2 \, (2\pi)^4 \, \delta ( m_b \hat{v}^+ \plus \overline{\Lambda} \hat{v}^+ \minus \hat{q}^+ \minus \hat{k}_{n'}^+ \minus \hat{k}_{\rm uc}^+) \, \delta (\overline{\Lambda}\hat{v}^- \minus \hat{k}_{n'}^- \minus \hat{k}_{\rm uc}^-) \, \delta^{(2)} (m_b \hat{v}^\perp \plus \overline{\Lambda} \hat{v}^\perp \minus \hat{k}_{n'}^\perp \minus \hat{k}_{\rm uc}^\perp) \,, \notag
\end{align}
where we defined 
\begin{align} 
\label{eq:Hddef}
H_{d}^{\mu\nu}(q,v,m_b) & \equiv \hat{H}^- \sum_{j,j'=1}^3  \mathrm{tr}\bigg[ \! \frac{P_v}{2} \, \bar{\Gamma}^\mu_{j'}\,  \frac{\slashed{n}'}{2}\, \Gamma^\nu_j  \! \bigg]  H_{jj'}(m_b,\hat{v}^+ \hat{H}^-)   \,, \notag \\
H_{j j'}(m_b,\hat{v}^+ \hat{H}^-) & \equiv C_j(m_b,\hat{v}^+ \hat{H}^-) \, C_{j'} (m_b,\hat{v}^+ \hat{H}^-)  
\end{align}
At leading order in the power counting in the endpoint region we can also replace $\hat{H}^-\frac{n'^\mu}{2}$ by $H^\mu$ and $\hat{v}^+ \hat{H}^-$ by $2 v\cdot H$, since the additional contributions of the residual momenta are power-suppressed. This replacement simplifies the practical treatment of the 3-body phase space in Eq.~(\ref{eq:decomposition_leptonic_hadronic_tensor_new_2}).
In the next step we introduce the final state inclusive SCET jet function
\begin{align}
\label{eq:SCETjetfctsdef}
 J_{n'} (\hat{H}^- k_{n'}^+) = \sum_{X_n'}  \delta^{(4)} (k_{n'} \minus K_{X_{n'}})  \frac{(2\pi)^3}{4 N_c \hat{H}^-} \, \mathrm{Tr} \big[ \bra{0}  \slashed{\bar{n}}' \chi_{n'} (0) \ket{X_{n'}} \bra{X_{n'}} \bar{\chi}_{n'} (0) \ket{0} \big] \,,
\end{align}
for the $u$-quark jet field matrix elements and employ the $\delta$-functions in the last line of Eq.~(\ref{eq:BdecayWmunufactorization-3}) to perform the integrations over $\hat{k}_{n'}^-$ and $\hat{k}_{n'}^\perp$ to obtain
\begin{align}
W^{\mu \nu}(v,&p_B,q)  =    H_{d}^{\mu\nu}(q,v,m_b) \,  
\int \mathrm{d} \hat{k}_{n'}^+ \,  J_{n'} (\hat{H}^- \hat{k}_{n'}^+) \, \delta ( m_b \hat{v}^+\plus \overline{\Lambda}\hat{v}^+ \minus \hat{q}^+ \minus \hat{k}_{n'}^+ \minus \hat{k}_{\rm uc}^+)  \\
&\times \frac{1}{2} \,  \sum_{X_{\rm uc}} \int \mathrm{d} \hat{k}_{\rm uc}^+ \, \mathrm{d} \hat{k}_{\rm uc}^- \, \mathrm{d}^2 \hat{k}_{\rm uc}^\perp \, \delta (\hat{k}_{\rm uc}^+ \minus \hat{K}_{X_{\rm uc}}^+) \, \delta (\hat{k}_{\rm uc}^- \minus \hat{K}_{X_{\rm uc}}^-) \,  \delta^{(2)} (\hat{k}_{\rm uc}^\perp \minus \hat{K}_{X_{\rm uc}}^\perp) \notag  \\ 
&\times \mathrm{Tr} \Big[ \bra{\bar{B}_v}  \big( \bar{h}^{(b)}_v Y_{n',-}^{\rm uc} \big) (0) \ket{X_{\rm uc}} \bra{X_{\rm uc}} \big(  Y_{n',+}^{{\rm uc},\dagger} h^{(b)}_v \big) (0) \ket{\bar{B}_v} \Big] \ . \notag 
\end{align}
Finally, we also integrate over $\hat{k}_{\rm uc}^-$ and $\hat{k}_{\rm uc}^\perp$, which eventually yields the factorized form of the hadronic tensor
\begin{align}
\label{eq:BdecayWfinal}
W^{\mu \nu}(v,\,p_B,q)  =  & H_{d}^{\mu\nu}(q,v,m_b) \int\! \mathrm{d} \hat{k}_{n'}^+ \,\mathrm{d} \hat{k}_{\rm uc}^+ \,  \delta ( m_b  \hat{v}^+\plus \overline{\Lambda} \hat{v}^+ \minus \hat{q}^+ \minus \hat{k}_{n'}^+ \minus \hat{k}_{\rm uc}^+)  \\
& \times J_{n'} (\hat{H}^- \hat{k}_{n'}^+) \, S_{\mathrm{shape}} (\hat{v}^+,\hat{k}_{\rm uc}^+)  \notag \\
= &  
 H_{d}^{\mu\nu}(q,v,m_b) \int\! \mathrm{d} \hat{\ell}^+  \, J_{n'} (\hat{H}^- ( m_b \hat{v}^+ \plus \overline{\Lambda} \hat{v}^+ \minus \hat{q}^+ \minus \hat{\ell}^+) ) \, S_{\mathrm{shape}} (\hat{v}^+,\hat{\ell}^+)  \notag
\end{align}	
where the shape function $S_{\mathrm{shape}}$, which is defined as~\cite{Ligeti:2008ac}
\begin{align}
\label{eq:fullshapefct}
S_{\mathrm{shape}} (\hat{v}^+,\hat{\ell}^+) &= \frac{1}{2} \sum_{X_{\rm uc}} \delta (\hat{\ell}^+ \minus \hat{K}_{X_{\rm uc}}^+)  \,  \mathrm{Tr} \Big[ \bra{\bar{B}_v}  \big( \bar{h}^{(b)}_v Y_{n',-}^{\rm uc} \big) (0) \ket{X_{\rm uc}} \bra{X_{\rm uc}} \big(  Y_{n',+}^{{\rm uc},\dagger} h^{(b)}_v \big) (0) \ket{\bar{B}_v} \Big]  \notag  \\
&= \frac{1}{2} \, \mathrm{Tr} \Big[ \bra{\bar{B}_v}  \bar{h}^{(b)}_v  (0) \, \delta (i n' \cdott D^{\rm uc}  \minus \bar{\Lambda}\hat{v}^+ \plus\hat{\ell}^+ ) \,    h^{(b)}_v  (0) \ket{\bar{B}_v} \Big]  \,,
\end{align}
describes the light-cone (plus w.r.\ to the jet) momentum distribution of the ultra-collinear (or soft for a $\bar B$ meson at rest) radiation arising from the $\bar{B}$ meson decay. 
The factorization formula of Eqs.~(\ref{eq:decomposition_leptonic_hadronic_tensor_new_2}) together with (\ref{eq:BdecayWfinal}) is frame-independent and valid for any choice of the $\bar B$ meson velocity $v^\mu$. While the SCET jet function $J_{n'}$ is manifest boost-invariant, the shape function is not, as it describes the distribution of the soft (and in general ultra-collinear) light-cone momentum $\hat{K}_{X_{\rm uc}}\cdot n'$, which is frame-dependent. Assuming that $\Lambda^\mu_{\,\,\nu}$ transforms the boosted $\bar B$ meson into its rest frame (i.e. $\Lambda^\mu_{\,\,\nu}\,v^\nu=v^\mu_{\rm rest}=(1,\vec{0})$), it is straightforward to show that the light-light $X_u$-jet reference vector transforms as $\Lambda^\mu_{\,\,\nu}\,n'^\nu=  \hat{v}^+ n'^\mu_{\rm rest}$ so that $(\hat{\ell}^+)_{\rm rest}=\hat{\ell}^+/\hat{v}^+ $. Since the  $\bar B$ meson matrix elements in the shape function definition (other than the $\delta$ term) are boost-invariant, this implies the relation 
\begin{align}
\label{eq:shapefctrel}
S_{\mathrm{shape}} (\hat{v}^+,\hat{\ell}^+)=\frac{1}{\hat{v}^+}\, S_{\mathrm{shape}} \bigg(1,\frac{\hat{\ell}^+}{\hat{v}^+}\bigg) \,,
\end{align}
for the boosted and rest frame shape functions, which we have also cross checked by an explicit computation at ${\cal O}(\alpha_s)$. 
The analogous relation also holds for the shape function's renormalization $Z$-factor (and its inverse) and the anomalous dimension. 

Let us also comment on the role of non-perturbative effects. In analogy to the double hemisphere invariant mass factorization theorem in the double resonant region of Eq.~(\ref{eq:bHQET_cross_section_new-5}), the sensitivities of the matching coefficient $H_{d}^{\mu\nu}$ and the inclusive massless quark SCET jet function $J_{n'}$ to infrared momenta are at most quadratic and thus power-suppressed. In contrast, 
the shape function $S_{\rm shape}$ is differential and defined from vacuum matrix elements which involve $\bar{B}$ meson matrix elements.
It thus contains non-perturbative effects associated to powers of $
\Lambda_{\rm QCD} \hat{v}^+/\hat{\ell}^+\sim m_b\Lambda_{\rm QCD}/H^2$ that must be resummed to all orders in the endpoint region where $H^2\sim  m_b \Lambda_{\rm QCD}$.
Similar to the double hemisphere soft function in Eq.~(\ref{eq:nonpertF}) the partonic and non-perturbative contributions of the shape function can also be factorized~\cite{Ligeti:2008ac}
\begin{align}
\label{eq:nonpertFshape}
S_{\mathrm{shape}} (\hat{v}^+,\hat{\ell}^+) = 
\int\!\! \mathrm{d} \hat{k}^+\, 
\hat S_{\text{shape}} (\hat{v}^+,\hat{\ell}^+ - \hat{k}^+) F_{\text{shape}}(\hat{v}^+,\hat{k}^+) \,,
\end{align}
where $\hat S_{\text{shape}}$ is the partonic shape function for an on-shell external massive quark state, see Eq.~(\ref{eq:renormalozed_shape_function_app}), and $F_{\text{shape}}$ is a non-perturbative light-cone distribution function. 
For the treatment of the semileptonic top quark decay discussed in Sec.~\ref{sec:factorizationtopdecay} the analogue of the shape function is the momentum distribution of the (top) ultra-collinear radiation arising from the entire top production and decay process, where the production stage radiation (which in addition includes the large-angle soft radiation) is controlled by the factorization theorem in Eq.~(\ref{eq:bHQET_cross_section_new-5}). Thus for the semileptonic top decay it is 
the entire top quark hemisphere (defined by a measurement of its invariant mass $M_t$) which takes the role of $\bar B$ meson state. This top quark hemisphere state is perturbatively computable up to the non-perturbative contributions that arise from the double hemisphere soft function in Eq.~(\ref{eq:nonpertF}).

Note that the factorization formulae in Eqs.~(\ref{eq:decomposition_leptonic_hadronic_tensor_new_2}) and (\ref{eq:BdecayWfinal}), which are valid in the small $H^2$ region, can be easily adapted to describe the inclusive semileptonic decay of an {\it on-shell} top quark through two modifications: one replaces the non-perturbative $\bar B$ meson shape function $S_{\mathrm{shape}}$ in Eq.~(\ref{eq:fullshapefct}) by the partonic shape function $\hat S_{\text{shape}}$ for the on-shell top quark
and the lepton tensor $\tilde{L}_{\mu \nu} (p_\ell, \pn)$ for $\ell^-\bar{\nu}_\ell$ final state in Eq.~(\ref{eq:decomposition_leptonic_hadronic_tensor_new_2}) by the one for $\ell^+\nu_{\ell}$ final states given in Eq.~(\ref{eq:leptonME}). The resulting factorization formula has been the basis of a fixed-order (non-resummed) calculation of the differential on-shell top quark semileptonic decay in Ref.~\cite{Gao:2012ja} using the phase space slicing method with respect to inclusive $b$-jet invariant mass $H^2$, which comprises the entire final state hadronic system. Going beyond the treatment of the top quark as an on-shell particle is the subject of the subsequent main part of this article.

Furthermore, we point out that the factorization formula, when evaluated at tree-level in the fixed-order expansion, reduces to the exact QCD tree-level result for the top quark on-shell decay. This is because at tree-level the leading term in the small $H^2$ expansion, on which the power counting of the factorization is based, fully captures the exact tree-level result which is proportional to $\delta(H^2)$ (for a massless final state quark). This implies that for distributions that arise already at tree-level,\footnote{This includes essentially all phenomenologically relevant differential semileptonic top decay observables that have been considered so far in the literature, including classic observables such as the charged lepton energy or the $b$-jet lepton invariant mass.
A counter example is the $H^2$ distribution itself, which is proportional to $\delta(H^2)$ at tree-level. It is highly sensitive to the value of $\alpha_s$ as it is generated entirely by the QCD radiation.} the non-singular corrections that are not captured by the factorization (which are power-suppressed in the small $H^2$ region, but can be sizeable in other parts of the phase space) can only arise in terms of $\alpha_s$-suppressed matching corrections. Since the tree-level contribution (which generates the LL and NLL resummation results) is valid in the entire top decay phase space, this has the useful phenomenological consequence that the non-singular corrections that need to be added to the factorization formula remain relatively small  corrections for the entire spectrum of such distributions. 
In other words, the small $H^2$ region dominates all distributions that are tree-level generated. This remains to be true also for the factorization we derive for the decaying off-shell and resonant top quark in Sec.~\ref{sec:jetfctfactorization}, where the tree-level and on-shell kinematics still governs the major quantitative aspects of the entire process. 
We come back to this aspect in our phenomenological discussion in Sec.~\ref{sec:pheno}.

\section{Off-shell Top Quark Semileptonic Decay in the Endpoint Region}
\label{sec:jetfctfactorization}

\begin{figure}[!t]%!htbp
	\centering
	\includegraphics[scale=0.30]{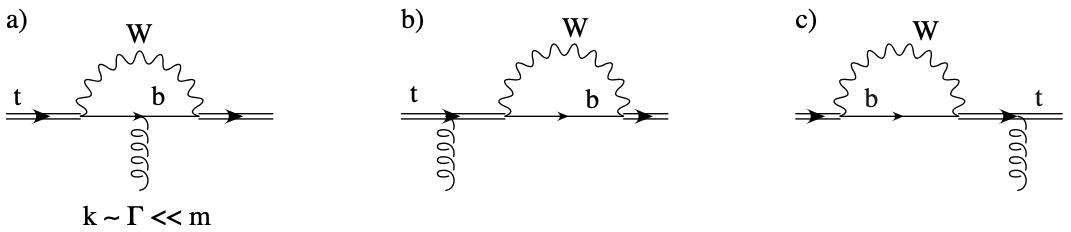}
	\caption{Example of the cancellation of soft gluon attachments to the decay products.}
	\label{fig:Self-energy}
\end{figure}

The form of the bHQET factorization theorem for the double hemisphere invariant mass distribution in the resonance region in
Sec.~\ref{sec:hemispherefactorization_new} applies for stable as well as unstable top quarks when considering only the leading order electroweak effects when treating the top decay in an inclusive manner~\cite{Fleming:2007qr,Fleming:2007xt}. It was shown in these references that in this case
the only leading order  effect of the top quark decay is that the $i\epsilon$ term in the bHQET top quark propagators is replaced by the top quark width term $i\Gamma_t/2$,
\begin{align}
\label{eq:HQETunstable}
\frac{P_v}{v.k+i\epsilon} \to \frac{P_v}{v.k+i\frac{\Gamma_t}{2}}\,.
\end{align} 
So, when using the optical theorem for the inclusive bHQET jet function, see Eq.~(\ref{eq:bHQET_jet_function_app_2}) and also Refs.~\cite{Jain:2008gb,Clavero:2024yav}, the top quark width can be implemented simply by the replacement $s\to s+i\Gamma_t$ in the jet function correlator prior to taking the imaginary part via the optical theorem~\cite{Fleming:2007xt,Hoang:2019fze}. Alternatively, using that the bHQET jet function correlator is analytic, except for its cut along the positive $s$ axis,  we can also obtain the unstable top inclusive bHQET function from the stable top quark case through the relation
\begin{align}
\label{eq:unstablebHQETjetfunction_convolution}
J_B^{\Gamma_t}(s) = \int \mathrm{d} s' \, J_B^{\Gamma_t=0} ( s') \, \frac{\Gamma_t}{\pi ((s - s')^2 \plus \Gamma^2_t)} \,.
\end{align}
Thus the leading power effects of the top quark decay are confined exclusively to the bHQET jet functions. This treatment is correct to all orders in QCD at leading order in the bHQET power counting and at leading order in the inclusive top decay, see Tab.~\ref{tab:modetable}. The basis of this simple outcome is that ultra-collinear gluon attachments to the inclusive top quark decay (i.e.\ soft gluons with momenta $k^\mu\ll m_t$ in the top rest frame) cancel to all orders in the strong coupling. In the matching of full QCD plus the electroweak interactions to bHQET, which describes the ultra-collinear QCD dynamics, this cancellation arises in the Wilson coefficients of the leading power $i \bar h_{v_n}(v_{n}\cdot A_{n}^{\rm uc})h_{v_n}$ and $i \bar h_{v_{\bar n}}(v_{\bar n}\cdot A_{\bar n}^{\rm uc})h_{v_{\bar n}}$ operators. The relevant diagrams for this matching calculation at ${\cal O}(\alpha_s)$ are shown in Fig~\ref{fig:Self-energy}. Since these operators are part of the heavy quark covariant derivative, this cancellation is ensured by ultra-collinear gauge invariance. 
The only net leading order effects in the matching arises in the imaginary part of the on-shell self-energy which yields the propagator replacement in Eq.~(\ref{eq:HQETunstable}).\footnote{The same mechanism is also known for ultra-soft gluon interactions for the fully inclusive production cross section of a color-singlet toponium state in $e^+e^-$ collisions.~\cite{Melnikov:1993np}.}   

This gauge cancellation also applies to the large-angle soft radiation which is still soft in the top rest frame, see the momentum counting in the third column of Tab.~\ref{tab:modetable}. Physically, however, for the large-angle soft radiation the cancellation has an even stronger reason, since it cannot distinguish between a boosted top or the $b$-jet (or likewise between a boosted antitop and its decay products). This is because at leading power, the interaction to either the top or the bottom quark yield the same eikonal propagator, so that both are represented by the same light-like color line. This mechanism is not affected by the additional radiation of the leptons within the top quark jet, which leads to an angle between the $b$-jet direction and the $t\bar t$ (thrust) axis in the top rest frame, since from the perspective of the soft radiation in the $e^+e^-$ rest frame the top decay products are boosted in the top directions. 
This insensitivity to the intrinsic structure of the top quark jet has already been accounted for by the soft-collinear decoupling relations in Eqs.~(\ref{eq:SCET_currents-3}) and (\ref{eq:SCET-bHQET-current-match-2}) and does not rely on whether the top quark decays semileptonically or hadronically, as long as the hadronic decay products arising from the top decay are treated inclusively. As a consequence, the description of the large-angle soft radiation through the hemisphere soft function $S_{\rm hemi}$ based on the matrix elements involving $n$ and $\bar n$ Wilson lines in Eq.(\ref{eq:nf5hemisphere_soft_function_new}) remains valid even when we consider differential observables for the (hadron inclusive) semileptonic top decay.
Furthermore, since from the perspective of the decaying top quark, the large-angle soft radiation is strongly boosted in the antitop direction (see again the large-angle soft momentum scaling in the top rest frame in third column of Tab.~\ref{tab:modetable}), the $\ell^+$ momentum distribution already encoded in the hemisphere soft function is the only momentum component of large-angle soft radiation relevant for the total momentum of the $b$-jet.

As a consequence, for the discussions on the differential semileptonic top quark decay in the following subsections, we only have to
consider in detail how the decay affects the dynamics of the top-ultra-collinear modes within the bHQET jet function $J_{B_t}$. As far as the large-angle soft radiation in the top quark hemisphere is concerned, it is sufficient to just keep track of how its total momentum component $\ell^+$ affects the kinematics of the total top-ultra-collinear momentum and the momentum of the $b$-jet. 

To proceed, the first essential step is to add the top quark decay into the formulation of the bHQET jet function. 
For this purpose we first rederive in Sec.~\ref{sec:inclusivefactorization} the convolution formula of Eq.~(\ref{eq:unstablebHQETjetfunction_convolution}), not using analytic properties of the jet function and the optical theorem, but from the definition of the jet function with explicit sums over the final states given in Eq.~(\ref{eq:bHQETjetfct2}). To the best of our knowledge this derivation has not been provided in the literature, and is thus interesting by itself. 
Together with the manipulations used in the factorization for the inclusive semileptonic $\bar B$ decay in Sec.~\ref{sec:Bdecayfactorization_new}, we can then continue in Sec.~\ref{sec:factorizationtopdecay}
with the derivation of the factorization for the top decay in the small $b$-jet mass region.

\subsection{Factorization of the Top Decay for the Inclusive Case}
\label{sec:inclusivefactorization}

To prove the convolution relation of Eq.~(\ref{eq:unstablebHQETjetfunction_convolution}), using an explicit description of the top quark decay in the resonance region, it is convenient to first rewrite the stable top inclusive bHQET jet function defined from heavy quark field matrix elements with an explicit sum over all final states,
\begin{align}
\label{eq:bHQETjetfct2}
J_{B_t}^{\Gamma_t=0}(2 v\cdot k)
= &\frac{1}{8\pi N_c \,m_t} \sum_{X_n} (2\pi)^4\,  \delta^{(4)} (p \minus P_{X_n}) \\
&\times \mathrm{Tr} \Big[ \bra{0} \bar{T} [W_{n,+}^{\mathrm{uc},\dagger} h_v] (0) \ket{X_n} \bra{X_n} T [\bar{h}_v W_{n,-}^{\mathrm{uc}}] (0) \ket{0} \Big] \,, \notag
\end{align}
in terms of matrix elements only involving ultra-collinear Wilson lines. Here, the total 4-momentum entering the jet function is $p = m_t v+k$, and $P_{X_n}$ is the momentum of all final state particles. Note that the definition in Eq.~(\ref{eq:bHQETjetfct2}) differs slightly from the one given in Eq.~(\ref{eq:bHQETjetfcts-2}), where the measurement $\delta$-function only containes the momenta residual with respect to the large label momentum $m_t v$. When considering details concerning the top quark decay, where the light decay products can become very energetic because they acquire parts of the top quark's large label momentum, the definition in Eq.~(\ref{eq:bHQETjetfct2}) is more suitable.
For simplicity we take a frame where $v^\mu=v_n^\mu=(1,\vec 0)$ and $k^{\mu}=(k^{0},\vec 0)=( s/2,\vec 0)$ and $2 v\cdot k=2k_0=s$. The desired expression is obtained by rewriting the heavy top quark fields in terms of decoupled top fields using static Wilson line of Eq.~(\ref{eq:ucWilsonlines}), $h_v = Y_{v,-}^{\rm uc}\,  h_v^{(0)}$, $\bar h_v =   \bar h_v^{(0)}\,Y^{{\rm uc},\dagger}_{v,+}$, and by factorizing the entire $n$-collinear final state in terms of on-shell decoupled top and gluonic (as well as light quark) final states $\ket{X_n}=\ket{(t_{v,s})^{(0)}}\ket{X_{\rm uc}}$, where $
s$ is the top spin index. Since the top quark is on-shell, its possible momenta have the form $p_t^\mu = m_t v^\mu + (0,\vec{k}_t)$ at leading order in the bHQET power counting. 
We can now use $\sum_{X_n}=\sum_{X_{\rm uc}} \sum_{s}\frac{1}{2}\int d^3\vec{k}_t/(2\pi)^3$ for the final state phase space and can rewrite
the global energy-momentum conserving $\delta$-function in the form 
$\delta^{(4)} (p-P_{X})=\delta(k^{0}-K^0_{X_{\rm uc}})\delta^{(3)} (\vec{k}_t \plus \vec{K}_{X_{\rm uc}})$. 
We can also compute the 'sterile' top operator matrix elements $\sum_\alpha \sum_s \bra{0}(h_{v}^{(0)}(0))^a_\alpha \ket{(t_{v,s})^{(0)}}\bra{(t_{v,s})^{(0)}} (\bar{h}_{v}^{(0)}(0))^b_\alpha\ket{0}=\mbox{tr}[2P_v]\delta^{ab}=4\delta^{ab}$, where $\alpha$ is a Dirac and $a$, $b$ are color indices.\footnote{The factor $4$ involves the HQET norm $\langle t_{v,s}(\vec k)|t_{v,s}(\vec p)\rangle = 2v^0\delta^{(3)}(\vec k-\vec p)$ for heavy quark states, which also yields the factor $1/2$ in the top quark phase space integral contained in $\sum_{X_n}$.}  Carrying out the $\vec{k}_t$ integration, we arrive at  
\begin{align}
\label{eq:stable_bHQET_JF}
J_{B_t}^{\Gamma_t=0}(2 k^{0})
 =& \frac{1}{2 N_c \,m_t} \sum_{X_{\rm uc}} \, \delta(k^{0}-K^0_{X_{\rm uc}})\\
&\times \mathrm{tr}_c \Big[ \bra{0}  \bar{T} [W_{n,+}^{\mathrm{uc}, \dagger}  Y_{v,-}^{\rm uc}] (0) \ket{X_{\rm uc}} \bra{X_{\rm uc}}  T [Y^{{\rm uc},\dagger}_{v,+} W_{n,-}^{\mathrm{uc}}] (0) \ket{0} \Big] \,, \notag
\end{align}
where $\mathrm{tr}_c$ stands for a pure color trace.
This is the alternative expression for the stable quark inclusive bHQET jet function we are interested in and which has already been introduced in Ref.~\cite{Jain:2008gb}.
While $k^{0}=s/2$ is the jet off-shellness,  it is the total energy $K^0_{X_{\rm uc}}$ of the non-top final state particles that is measured from the matrix elements.  

We now derive the convolution formula~(\ref{eq:unstablebHQETjetfunction_convolution}) starting again from the formula in Eq.~(\ref{eq:bHQETjetfct2}), extending it such as to account for the top quark decay at the operator level. There are two essential modifications. First, we need to separate the final state $b$-quark and $W$-boson arising from the top quark decay from the (ultra-collinear) gluon and light quark final state particles,
\begin{align}
\ket{X_n} & \to \ket{bW}\ket{X_{\rm uc}} \,.
\end{align}
This is in close analogy to the decomposition for the stable top case above, but we note that the $b$-$W$ invariant mass is in general not equal to the top mass $m_t$. The second modification extends the form of the (anti)time-ordered QCD operator matrix elements by explicitly including the electroweak $t\to bW$ decay operator
\begin{align} 
T [\bar{h}_v W_{n,-}^{\mathrm{uc}}] (0) & \to  \int \!\mathrm{d}^4 z_1 \, T \, \mathcal{O}_\Gamma(z_1)  \, [\bar{h}_v W_{n,-}^{\mathrm{uc}}] (0) \,,   \\
\bar{T} [W_{n,+}^{\mathrm{uc}, \dagger} h_v] (0)  & \to  \int \!\mathrm{d}^4 z_2 \,\bar{T}  \,[W_{n,+}^{\mathrm{uc}, \dagger} h_v] (0) \, \mathcal{O}_\Gamma^\dagger(z_2)  \,,  \notag
\end{align}
with
\begin{align} 
\label{eq:decay_op}
\mathcal{O}_\Gamma(z) &= \bigg( \frac{i e}{\sqrt{2} \,s_w} \bigg) \, V_{tb}\, \big( \bar{b}^{(0)} \, \gamma^\mu P_L \, W^-_\mu\, t^{(0)} \big)(z) \\
&= e^{-i m_t v\cdot z}\,\bigg( \frac{i e}{\sqrt{2} \,s_w} \bigg)\, V_{tb} \, \big( \bar{b}^{(0)} \, \gamma^\mu P_L \, W^-_\mu\, h_v^{(0)} \big)(z) \,. \notag
\end{align}
Since we know that all ultra-collinear gluon (soft) interactions with the inclusive (hard) top decay cancel due to gauge invariance, the decay operator is formulated in terms of 'sterile' bottom and top quark fields. Likewise the final state $b$ quark state is treated as sterile. This approximation is adequate for the inclusive bHQET jet function, but we will of course have to consider regular and fully interacting bottom and top quark fields for the description of the differential top quark decay. This yields the following expression of the bHQET jet function:
\begin{align}
J_{B_t}^{\Gamma_t}(2 k^{0}) =&\,  \frac{|V_{tb}|^2}{8\pi \, N_c \, m_t }  \, \Big( \frac{e}{\sqrt{2} \,s_w} \Big)^2 \, \int \mathrm{d}^4 z_1 \int \mathrm{d}^4 z_2    \\
 &\times 
 \mathrm{e}^{-im_t v \cdot (z_1 - z_2)} \sum_{bW}\sum_{X_{\rm uc}} (2\pi)^4 \, \delta^{(4)} (p\minus p_b \minus q_W\minus K_{X_{\rm uc}})  \notag \\
& \times  \mathrm{Tr} \Big[ \bra{0} \bar{T}   [W_{n,+}^{\mathrm{uc}, \dagger} h_v] (0) \, [\bar{h}_v^{(0)} \gamma^\mu P_L W^+_\mu b^{(0)}] (z_2) \ket{bW}\ket{X_{\rm uc}} \notag \\
&\times \bra{bW} \bra{X_{\rm uc}} T  [\bar{b}^{(0)} \gamma^\nu P_L W^-_\nu h_v^{(0)}](z_1) \, [\bar{h}_v W_{n,-}^{\mathrm{uc}}] (0)   \ket{0} \Big] \notag \,.
\end{align}
Now, we again rewrite the jet function bHQET top quark fields in terms of decoupled fields, $h_v = Y_{v,-}^{\rm uc}\,  h_v^{(0)}$, $\bar h_v =   \bar h_v^{(0)}\,Y^{{\rm uc},\dagger}_{v,+}$, which allows us to compute the contraction of the decoupled (free) bHQET quark fields
within each (anti-)time-ordered matrix elements in terms of the unstable bHQET top quark propagators
\begin{align} \label{eq:bHQET_quark_propagators}
& \bra{0} T \big\{ ( h_v^{(0)})^a_\delta (z_1)  ( \bar{h}_v^{(0)})^b_\rho (0)          \big\} \ket{0} = \delta^{ab} \, (P_v)_{\delta \rho} \, \int \frac{\mathrm{d}^4 k_1}{(2\pi)^4} \, \frac{i}{v \cdott  k_1 \plus  i \frac{\Gamma_t}{2}} \, \mathrm{e}^{-i  k_1 \cdot z_1} \,, \\
&\bra{0} \bar{T} \big\{ ( h_v^{(0)})^a_\delta (0)  ( \bar{h}_v^{(0)})^b_\rho (z_2)          \big\} \ket{0} = \delta^{ab} \, (P_v)_{\delta \rho} \, \int \frac{\mathrm{d}^4  k_2}{(2\pi)^4} \, \frac{(-i)}{v \cdott  k_2 \minus i \frac{\Gamma_t}{2}} \, \mathrm{e}^{+i  k_2 \cdot z_2} \notag \,,
\end{align}
In addition, the top decay requires the color indices on the $b$-quark fields in the matrix elements to be the same, which allows us to write
\begin{align}
(b^{(0)})^a_\alpha (z_2) \, (\bar{b}^{(0)})^b_\beta (z_1) = \frac{\delta^{a b}}{N_c} \, (b^{(0)})^d_\alpha (z_2) \, (\bar{b}^{(0)})^d_\beta (z_1) \,.
\end{align}
These rearrangements completely separate the top decay part from the remaining contributions related to the residual gluon radiation and we can thus write the jet function in the form
\begin{align}
J_{B_t}^{\Gamma_t}(2 k^{0}) =&\,  \frac{|V_{tb}|^2}{8\pi \, N_c \, m_t }  \, \Big( \frac{e}{\sqrt{2} \,s_w} \Big)^2  \int \mathrm{d}^4 z_1\int \mathrm{d}^4 z_2 \,   \mathrm{e}^{-im_t v  \cdot (z_1 - z_2)}\,  \sum_{bW}\sum_{X_{\rm uc}}\,|\mathcal{M}_{td}|^2 \\
&\times (2\pi)^4 \, \delta^{(4)} (m_t v + k\minus p_b \minus q_W\minus K_{X_{\rm uc}}) \notag \\
&\times \int \frac{\mathrm{d}^4 k_1}{(2\pi)^4} \, \frac{1}{v \cdott k_1 \plus  i \frac{\Gamma_t}{2}} \, \mathrm{e}^{-i k_1 \cdot z_1} \int \frac{\mathrm{d}^4 k_2}{(2\pi)^4} \, \frac{1}{v \cdott k_2 \minus  i \frac{\Gamma_t}{2}} \, \mathrm{e}^{i k_2 \cdot z_2} \notag \\
& \times   \mathrm{tr}_c \Big[ \bra{0} \bar{T}   [W_{n,+}^{\mathrm{uc},\dagger} Y_{v,-}^{\rm uc}] (0) \ket{X_{\rm uc}} \bra{X_{\rm uc}} T [Y_{v,+}^{{\rm uc},\dagger} W_{n,-}^{\mathrm{uc}}] (0) \ket{0} \Big]  \notag \,,
\end{align}
where $|\mathcal{M}_{td}|^2$ denotes the reduced squared matrix element for the top decay
\begin{align}
|\mathcal{M}_{td}|^2 &= \frac{1}{N_c} \,  \mathrm{Tr} \Big[ \bra{0} \bar{T} [ P_v \gamma^\mu P_L W^+_\mu b^{(0)}] (z_2) \ket{b\, W} \bra{b \, W} T  [\bar{b}^{(0)} \gamma^\nu P_L W^-_\nu P_v ](z_1) \ket{0} \Big]  \\
& =\mathrm{e}^{i (p_b + q_W)\cdot (z_1 - z_2)} \, \mathrm{tr} \big[ P_v \gamma^\mu P_L u(p_b) \bar{u} (p_b) \gamma^\nu P_L \big] \, \epsilon_\nu^*(q_W) \,  \epsilon_\mu (q_W) \notag \\
& = \mathrm{e}^{i (p_b + q_W)\cdot (z_1 - z_2)} |\tilde{\mathcal{M}}_{td}|^2  \,. \notag 
\end{align}
We can then carry out the integration over $z_1$ and $z_2$, which yields two energy-momentum conserving $\delta$-functions for the top decay kinematics that also set $k_2=k_1$, and do the $k_2$ integral. Now, with $k_t\equiv k_1$, which parametrizes the off-shellness 4-momentum of the decaying top quark, we arrive at 
\begin{align}
J_{B_t}^{\Gamma_t}(2 k^{0}) =&\,  \frac{|V_{tb}|^2}{8\pi \, N_c \, m_t }  \, \Big( \frac{e}{\sqrt{2} \,s_w} \Big)^2  \int \mathrm{d}^4 k_t \, \frac{1}{(k_t^0)^2 \plus  (\frac{\Gamma_t}{2})^2}   \\
&\times \sum_{X_{\rm uc}}   \delta^{(4)} (k\minus k_t\minus K_{X_{\rm uc}}) \, \sum_{bW} \, (2\pi)^4\, \delta^{(4)} (m_t v + k_t\minus p_b \minus q_W) \,|\tilde{\mathcal{M}}_{td}|^2 \notag \\
& \times   \mathrm{tr}_c \Big[ \bra{0} \bar{T}   [W_{n,+}^{\mathrm{uc}, \dagger} Y_{v,-}^{\rm uc}] (0) \ket{X_{\rm uc}} \bra{X_{\rm uc}} T [Y_{v,+}^{{\rm uc},\dagger} W_{n,-}^{\mathrm{uc}}] (0) \ket{0} \Big]  \notag \,.
\end{align}
By carrying out the $b$-$W$ phase space sum over the reduced squared matrix element, we obtain the total top quark decay rate
\begin{align}
& \sum_{bW} \,\Big( \frac{e}{\sqrt{2} \,s_w} \Big)^2 |V_{tb}|^2  |\tilde{\mathcal{M}}_{td}|^2  \,(2\pi)^4\,\delta^{(4)} (m_t v + k_t\minus p_b \minus q_W)  \\
=& \sum_{\rm b, W \,spins}
\int \mathrm{d} \Pi_2(m_t v+k_t; p_b,q_W)  \,\Big( \frac{e}{\sqrt{2} \,s_w} \Big)^2 |V_{tb}|^2   \, |\tilde{\mathcal{M}}_{td}|^2 
 = 2\,  \Gamma_t \,, \notag
\end{align}
which we approximate to be the equal to on-shell top quark width $\Gamma_t$. As we are in the resonance region, where the $b$-$W$ invariant mass is very close to $m_t$ the invariant mass dependence of the decay rate is very small and we can set it to the on-shell width at leading order in the bHQET power counting. Furthermore we can
rewrite the global energy-momentum conserving $\delta$-function as
$\delta^{(4)} (k - k_t - K_{X_{\rm uc}})=\delta(k^{0} - k_t^0 - K^0_{X_{\rm uc}}) \delta^{(3)}(\vec{k}_t + \vec{K}_{X_{\rm uc}})$. This allows us to integrate over the top $3$-momentum $\vec{k}_t$ (which is simply set to  $\vec{k}_t=-\vec{K}_{X_{\rm uc}}$).
Next, we introduce the variable $k_{\rm uc}^0$ for the energy of the non-top final state particles via the identity
$ 1 = \int \mathrm{d}k_{\rm uc}^0 \delta(k_{\rm uc}^0-K^0_{X_{\rm uc}})$. We then arrive at
\begin{align}
J_{B_t}^{\Gamma_t}(2 k^{0}) =&\,  \frac{1}{8\pi \, N_c \, m_t }   
\int \mathrm{d} k_t^0 \, \mathrm{d} k_{\rm uc}^0  \, \delta (k^{0}\minus k_t^0 \minus k^0_{\rm uc}) \,\frac{2\Gamma_t}{(k_t^0)^2 \plus  (\frac{\Gamma_t}{2})^2}   \notag \\
& \times  \sum_{X_{\rm uc}} \delta(k_{\rm uc}^0-K^0_{X_{\rm uc}})  \mathrm{tr}_c \Big[ \bra{0} \bar{T}   [W_{n,+}^{\mathrm{uc}, \dagger} Y_{v,-}^{\rm uc}] (0) \ket{X_{\rm uc}} \bra{X_{\rm uc}} T [Y_{v,+}^{{\rm uc},\dagger} W_{n,-}^{\mathrm{uc}}] (0) \ket{0} \Big]  \notag \,,
\end{align} 
where in the last line we can now spot the stable top quark bHQET jet function of Eq.~(\ref{eq:stable_bHQET_JF}). The final result thus reads
\begin{align}
J_{B_t}^{\Gamma_t}(2 k^{0}) =&\,  \frac{1}{4\pi}   
\int \mathrm{d} k_t^0 \, \mathrm{d} k_{\rm uc}^0  \, \delta (k^{0}\minus k_t^0 \minus k^0_{\rm uc}) \,\frac{2\Gamma_t}{(k_t^0)^2 \plus  (\frac{\Gamma_t}{2})^2} \,  J_{B_t}^{\Gamma_t=0}(2 k^{0}_{\rm uc})  \notag \\  
=&\,  2   
\int  \mathrm{d} k_{\rm uc}^0 \,  J_{B_t}^{\Gamma_t=0}(2 k^{0}_{\rm uc})  \, \frac{\Gamma_t}{\pi ( 4(k^{0}-k_{\rm uc}^0)^2 \plus  \Gamma_t^2)} 
\,,
\end{align} 
which is equivalent to Eq.~(\ref{eq:unstablebHQETjetfunction_convolution}).

\subsection{Derivation of the Factorization Formula for the Resonant Top Quark Decay}
\label{sec:factorizationtopdecay}

We are now ready to discuss the derivation of the novel endpoint factorization formula for the  semileptonic, boosted and off-shell top quark decay, where we will merge and combine aspects of the factorization derivations for the unstable top inclusive bHQET jet function from Sec.~\ref{sec:inclusivefactorization} and of the semileptonic $\bar B$ decay in the endpoint region from Sec.~\ref{sec:Bdecayfactorization_new}. We remind the reader that the $b$-jet emerging from the top decay is defined from clustering all hadrons in the top hemisphere into the jet and that we treat the antitop hemisphere inclusively.

The starting point is again the formula for the inclusive bHQET top quark jet function 
\begin{align}
\label{eq:bHQETjettop1}
&J_{B_t}^{\Gamma_t} (v^- k^+,v^+ k^-)=\\
&= \frac{1}{8\pi \, N_c \, m_t} \sum_X (2\pi)^4 \, \delta^{(4)} (m_t\, v \plus k \minus P_X)  \mathrm{Tr} \Big[\! \bra{0} \overline{T}\, \big( W_{n,+}^{\mathrm{uc}, \dagger} \,    h_v \big) (0) \ket{X} \bra{X} T \big( \bar{h}_v \, W_{n,-}^{\mathrm{uc}} \big) (0) \ket{0} \!\Big], \notag
\end{align}
where the total 4-momentum entering the jet function is $p = m_t v+k$. In contrast to Sec.~\ref{sec:inclusivefactorization}, we do not adopt any particular frame, and the following discussion and results apply in the $e^+e^-$ c.m.\ as well as the hemisphere rest frame. We remind the reader that within the double hemisphere mass distribution~(\ref{eq:bHQET_cross_section_new-5}) we have 
\begin{align}
\label{eq:koffshelldef}
& v^- k^+=s_t^{*+} - v^-\ell^+\,, 
&v^+ k^-= s_t^{*-}\,,
\end{align}
and we write $J_{B_t}^{\Gamma_t}$ as a function of the plus and minus off-shellness variables
defined in Eq.~(\ref{eq:Jbtarguments}). We remind the reader that $v$ stands for the top velocity $v_n$ in Eq.~(\ref{eq:vlabeldef-2}).
We also recall that for the large-angle soft momentum $\ell^\mu$ only the plus momentum $\ell^+$ needs to be accounted for as the contributions of $\ell^-$ and $\ell^\perp$ are $m_t/Q$ power-suppressed and can be set to zero. As already mentioned in the discussion of the double hemisphere factorization,  this also implies that $k^\perp$ does not contribute at leading power and can be set to zero, since the total perp momentum of the top hemisphere is zero for our choice of the light-cone reference vector $n$, (see footnote~\ref{foot:kperp}).
But we have to keep in mind that even for $k^\perp=0$, the perp momentum with respect to the $b$-jet direction, 
$\hat{k}^\perp$ is in general non-zero.

To account for the differential top quark decay we follow the approach for the unstable top inclusive jet function discussed in Sec.~\ref{sec:inclusivefactorization}
and separate the hadronic particles (containing the light $b$-quark) and the leptons in the final state,
\begin{align}
\ket{X} & \to \ket{\ell^+ \nu_\ell}\ket{X_{b}}\,.
\end{align}
The hadronic state $\ket{X_{b}}$ contains all hadronic particles in the top hemisphere except for those coming from large-angle soft radiation.
We also render the top decay operator explicit in the (anti)time-ordered matrix elements
\begin{align} 
\label{eq:Tproducttopdecaysemileptonic}
T [\bar{h}_v W_{n,-}^{\mathrm{uc}}] (0) & \to  \int \!\mathrm{d}^4 z_1 \, T \, \mathcal{\tilde O}_\Gamma(z_1)   \, [\bar{h}_v W_{n,-}^{\mathrm{uc}}] (0) \,,   \\
\bar{T} [W_{n,+}^{\mathrm{uc}, \dagger} h_v] (0)  & \to  \int \!\mathrm{d}^4 z_2 \,\bar{T}  \,  [W_{n,+}^{\mathrm{uc}, \dagger} h_v] (0) \, \mathcal{\tilde O}_\Gamma^\dagger(z_2)  \,, \notag
\end{align}
where ($\tilde{g}_{\mu \nu} = g_{\mu \nu} - q_\mu q_\nu / M_W^2$)
\begin{align} 
\label{eq:decay_operatortilde}
\mathcal{\tilde O}_\Gamma(z) &= \bigg( \frac{i e}{\sqrt{2} \,s_w} \bigg)^2 V_{tb} \, \big[ \big( \bar{b} \, \gamma^\mu P_L \, t \big) \, \big( \bar{\nu}_\ell \, \gamma^\sigma  P_L \, \ell    \big) \big] (z) \, \frac{(-i)}{q^2 \minus M_W^2 \plus i M_W \Gamma_W} \bigg( g_{\mu \sigma} \minus \frac{q_\mu q_\sigma}{M_W^2} \bigg) \notag \\
&= -i \, \frac{4G_F}{\sqrt{2}} V_{tb} \, \frac{M_W^2 \, \tilde{g}_{\mu \sigma}}{M_W^2 \minus q^2 \minus  i M_W \Gamma_W} \, \big[ \big( \bar{b} \, \gamma^\mu P_L \, t \big) \,  \big( \bar{\nu}_\ell \, \gamma^\sigma  P_L \, \ell    \big) \big] (z) \,.
\end{align}
The top and bottom quark fields appearing in $\mathcal{\tilde O}_\Gamma$ are at this point still full QCD fields. The expressions in Eq.~(\ref{eq:Tproducttopdecaysemileptonic}) involving the local 4-fermion operator $[(\bar{b}\gamma^\mu P_L t)(\bar{\nu}_\ell \gamma^\sigma  P_L \ell) ] (z)$ are obtained from the time-ordered product of the electroweak $t\to bW$ and $W\to \ell\nu_{\ell}$ operators at space-time points $z$ and $z'$, respectively, integrated over $z$ and $z'$. Since the $W$ and the leptons are sterile w.r.\ to QCD and we only consider electroweak effects at leading order, we can contract the $W$ fields yielding a propagator and carry out the $z'$ integration which fixes the $W$ propagator momentum as 
\begin{align} 
q = p_\ell + \pn\,.
\end{align}
This yields the effective operator $\mathcal{\tilde O}_\Gamma$ which is still local with respect to QCD. 
At this point we can proceed in analogy to the $\bar B$ meson decay factorization by introducing the momentum variable $H$ for the total hadronic part of the top hemisphere momentum $M_t=m_t v+k^*$,
\begin{align} 
\label{eq:Hmomdef}
H = P_{X_b}+\ell=M_t -  q= m_t v+k+\ell-q \,,
\end{align}
and we factorize the total physical phase space in terms of the three-body phase space $\mathrm{d}\Pi_3(M_t; p_\ell, \pn, H)$, the sum over all hadronic final state particles $X_b$ and the integration over $H^2$. Note, that the additional total large-angle soft momentum $\ell^\mu$ included in the definition of $H^\mu$ is mandatory, so that the phase space integration correctly involves the physical hadronic top hemisphere momentum. Within the phase space factorization for the momentum $m_t v+k$ entering the jet function, adding $\ell^\mu$ just corresponds to a shift of an integration variable, since for a given $k$ also $\ell=k^*-k$ is fixed, so that we have $\mathrm{d}\Pi_3(M_t=m_t v+k^*; p_\ell, \pn, H)=\mathrm{d}\Pi_3(m_t v+k; p_\ell, \pn, H-\ell=P_{X_b})$. In addition we match the QCD $t\to b$ current onto corresponding HQET-SCET heavy-to-light effective currents, which integrates out local decay effects at the top mass scale. The matching relation is in close analogy to Eq.~(\ref{eq:EFT_heavy_to_light_currents_B_decays_new}) and reads
\begin{align} 
\label{eq:EFT_heavy_to_light_currents_t_decay}
(\bar{b}\gamma^\mu P_L t) (z) = \mathrm{e}^{i (\hat{H}^- \frac{n'}{2}  - m_t v   )\cdot z}\, \sum_{j=1}^3 \, C_j(m_t,\hat{v}^+ \hat{H}^-) \, \big[ \bar{\chi}_{n'} Y_{n',+}^{{\rm uc},\dagger} \, \Gamma^\mu_j \, h_v \big] (z)  +  {\cal O}(\Gamma_t/m_t),
\end{align}
where $h_v$ is the bHQET top quark field already appearing in the jet function definition of Eq.~(\ref{eq:bHQETjettop1}). Likewise, $\bar{\chi}_{n'}$ is the SCET ($b$-quark) jet field, where we treat the bottom quark as massless.
For the direction of the collinear reference vector 
$n'^\mu=(1,\vec{n}')$ we can adopt the direction of the total hadronic final state momentum $\vec{n}' =\vec{H}/|\vec{H}|$, which fixes the large label momentum of the SCET jet field as 
\begin{align} 
\label{eq:SCET-top-largelabel}
\tilde{P}_{X_{n'}}^\mu =  (\hat{\tilde{P}}_{X_{n'}}^+,\hat{\tilde{P}}_{X_{n'}}^-,\hat{\tilde{P}}_{X_{n'}}^\perp)=(0,\hat{H}^-,0) = (0,m_t \hat{v}^- - \hat{q}^-,0)  \,,
\end{align}
in analogy to the treatment of the $\bar B$ decay. We also recall that $2v\cdot H\approx \hat{v}^+ \hat{H}^-\approx m_t$ in the endpoint region. 
The $b$-quark jet field is once more treated as soft decoupled, which implies the appearance of the ultra-collinear Wilson line $Y_{n',+}^{{\rm uc},\dagger}$. The choice of the Dirac basis and the form of the Wilson coefficients are also analogous to our treatment of the $\bar B$ endpoint decay, see Eq.~(\ref{eq:diracbasis}) and Refs.~\cite{Pirjol:2002km, Chay:2002vy}. Computing the leptonic matrix elements,
\begin{align}
\label{eq:leptonME}
L^{\sigma \nu} (z_1, z_2) &= \sum_{\mathrm{spins}} \bra{0}  \, \big( \bar{\nu}_\ell \gamma^\sigma P_L \ell \big)^\dagger (z_2) \ket{\ell^+ \nu_\ell} \bra{\ell^+ \nu_\ell}  \big( \bar{\nu}_\ell \gamma^\nu P_L \ell \big) (z_1) \ket{0}  \\
&= \mathrm{e}^{iq \cdot (z_1 - z_2)} \, \mathrm{tr} \big[ \slashed{p}_{\nu_\ell}  \gamma^\nu  P_L  \slashed{p}_\ell  \gamma^\sigma  P_L \big]  \notag \\
&= \mathrm{e}^{iq \cdot (z_1 - z_2)} \, 2 \, \big[ p_{\nu_\ell}^\sigma p_\ell^\nu + p_{\nu_\ell}^\nu p_\ell^\sigma  - (p_{\nu_\ell} \! \cdot \! p_\ell) g^{\sigma \nu} - i \varepsilon^{\sigma \nu \delta \lambda} \, (p_\ell)_\delta \, (p_{\nu_\ell})_\lambda \big] \notag \\
&\equiv \mathrm{e}^{iq \cdot (z_1 - z_2)} \tilde{L}^{\sigma \nu} (p_\ell, p_{\nu_\ell}) \ , \notag
\end{align}
and writing $\ket{X_b}=\ket{X_{n'}X_{uc}}=\ket{X_{n'}}\ket{X_{uc}}$ for the hadronic state, we can now decompose the (differential) jet function into a hadronic and a leptonic part,
\begin{align} 
\label{eq:decomposition_bHQETjetfct_new}
\mathrm{d} J_{B_t}^{\Gamma_t} (v^- k^+,v^+ k^-) =&\,   \bigg( \frac{e}{\sqrt{2} \,s_w} \bigg)^4 |V_{tb}|^2\, \mathrm{d} H^2 \, \mathrm{d} \Pi_3( M_t ; p_\ell , \pn, H)\\
&\times  \tilde{L}^{\sigma \nu} (p_\ell, \pn) \, \frac{\tilde{g}_{\rho \sigma} \,  \tilde{g}_{\mu \nu}}{|q^2 \minus M_W^2 + i M_W \Gamma_W|^2} \, W^{\rho \mu} (n, v, k ,q) \ , \notag
\end{align}	
where the hadronic tensor $W^{\rho \mu}(n, v, k ,q)$ is given by
\begin{align}
W^{\rho \mu}&(n,v,k,q) =  \frac{1}{8\pi \, N_c \, m_t }  \, \sum_{j,j'=1}^3  \, C_j(m_t,\hat{v}^+ \hat{H}^-) C_{j'}(m_t,\hat{v}^+ \hat{H}^-) \\
&\times
 \sum_{X_{n'},X_{uc}} \! (2\pi)^3 \, \delta^{(4)} (m_t v \plus k \minus q \minus P_{X_{n'}} \minus P_{X_{uc}})  \,
%\\
%&\times 
\int \! \mathrm{d}^4 z_1 \int \! \mathrm{d}^4z_2 \, \mathrm{e}^{i (\hat{H}^- \frac{n'}{2} + q  - m_t v   )\cdot (z_1 - z_2)} \notag \\
&\times \mathrm{Tr} \Big[ \! \bra{0} \overline{T} \big[     W_{n,+}^{\mathrm{uc}, \dagger} \, h_v \big] (0)  \, \big[ \bar{h}_v Y_{n',-}^{\rm uc} \bar{\Gamma}^\rho_{j'} \chi_{n'}  \big] (z_2)     \ket{X_{n'} X_{uc}} \notag  \\
&\times \!\bra{X_{n'} X_{uc}} T  \big[\bar{\chi}_{n'} \Gamma^\mu_j Y_{n',+}^{{\rm uc},\dagger} h_v \big] (z_1) \,  \big[ \bar{h}_v \, W_{n,-}^{\mathrm{uc}} \big] (0)  \ket{0}  \Big]  \,. \notag 
\end{align}	
In the next step we disentangle the $n'$- and ultra-collinear fields in the hadronic tensor into the distinct collinear and soft sectors using the SCET and HQET Fierz relations in Eqs.~(\ref{eq:Fierz_relation_new2}) and~(\ref{eq:HQET_Fierz_relation_new}), respectively, where we remind the reader that in both cases only the respective first terms contribute at leading order in the power counting. We also decompose the $n'$-collinear and soft momenta into large label ($\tilde{P}$) and residual dynamical ($k$) parts and introduce the residual total collinear and soft momentum variables according to Eqs.~(\ref{eq:variabledecompo-B}) and (\ref{eq:introduction_of_integration_variables-B}). This gives the follwing form for the hadronic tensor
\begin{align} \label{eq:Wtopfact1}
W^{\rho \mu}&(n, v, k ,q)  = \frac{H_{d}^{\rho\mu}(q,v,m_t) }{8\pi \, N_c \, m_t }  
\,  \int \! \mathrm{d}^4 z_1 \! \int \! \mathrm{d}^4 z_2 \,  \mathrm{e}^{-i (\hat{H}^- \frac{n'}{2} + k_{n'} +  q  -m_t v )\cdot (z_2 - z_1)}   \\ 
&\times \sum_{X_{n'}} \int \mathrm{d}^4 k_{n'} \, \delta^{(4)} \big( k_{n'} \! - \! K_{X_{n'}} \big)  \frac{1}{4N_c \hat{H}^-} \, \mathrm{Tr} \big[ \bra{0}  \slashed{\bar{n}}' \chi_{n'} (0) \ket{X_{n'}}  \bra{X_{n'}} \bar{\chi}_{n'} (0) \ket{0} \big]  \notag \\
&\times \sum_{X_{uc}}  \int \mathrm{d}^4k_{uc} \, \delta^{(4)} \big( k_{uc} \! - \! K_{X_{uc}} \big) \bra{0} \overline{T} \, \big[    (W_{n,+}^{\mathrm{uc}, \dagger} )^{kl} \, ( h_v)^l_\lambda \big] (0) \, \big[(\bar{h}_v)_\alpha^a (Y_{n',-}^{\rm uc})^{ab} \big] (z_2) \ket{ X_{uc}} \notag  \\
&\times \!\bra{ X_{uc}} T \, \big[ (Y_{n',+}^{{\rm uc},\dagger})^{bt} (h_v)^t_\alpha \big] (z_1)\, \big [ (\bar{h}_v)^m_\lambda \,( W_{n,-}^{\mathrm{uc}})^{mk} \big] (0)  \ket{0}  \notag \\
&\times  (2\pi)^3 \, \delta^{(4)} \Big( m_t v \plus  k \minus q \minus  \hat{H}^- \frac{n'}{2} \minus k_{n'} \minus k_{uc}    \Big)  \,, \notag
\end{align}
with the hard decay Wilson coefficient $H_{d}^{\rho\mu}(q,v,m_t)$ as defined in Eq.~(\ref{eq:Hddef}) and where the second line can be written in terms of the inclusive $b$-quark jet function using Eq.~(\ref{eq:SCETjetfctsdef}).\footnote{We remind the reader that in this article we treat the $b$-quark as massless. Finite $b$ quark mass effects will also affect the phase space boundaries, see Ref.~\cite{Regner}.
} In this expression for the hadronic tensor the top quark decay takes place at the space-time position $z_{1,2}$. Since this makes the physical momentum accounting somewhat intransparent, we now shift the positions of the operators describing the top quark decay to $0$ and then change variables $z_{1,2} \to -z_{1,2}$, so that the bHQET top jet fields are at positions $z_{1,2}$. This yields an additional factor $\mathrm{e}^{-ik_{uc}\cdot(z_2-z_1) }$ which can be combined with the exponential in the first line of Eq.~(\ref{eq:Wtopfact1}) to yield $\mathrm{e}^{i k\cdot (z_2 - z_1)}$ and represents the residual momentum insertion into the jet.
We thus arrive at  
\begin{align}
\label{eq:Wtopfactalmost}
&W^{\rho \mu}(n, v, k ,q)  =  \frac{H_{d}^{\rho\mu}(q,v,m_t) }{8\pi \, N_c \, m_t } 
\,  \int \! \mathrm{d}^4 z_1 \mathrm{d}^4 z_2 \,  \mathrm{e}^{i k\cdot (z_2 - z_1)}\, \int \frac{\mathrm{d}^4 k_{n'}}{(2\pi)^3} \, J_{n'} (\hat{H}^- \hat{k}_{n'}^+)   \\ 
&\times \sum_{X_{uc}}  \int \mathrm{d}^4k_{uc} \, \delta^{(4)} \big( k_{uc} \! - \! K_{X_{uc}} \big) \bra{0} \overline{T}  \, \big[    (W_{n,+}^{\mathrm{uc}, \dagger} )^{kl} \, ( h_v)^l_\lambda \big] (z_2) \,  \big[ (\bar{h}_v)_\alpha^a (Y_{n',-}^{\rm uc})^{ab} \big] (0) \ket{ X_{uc}} \notag  \\
&\times \!\bra{ X_{uc}} T \big[ (Y_{n',+}^{{\rm uc}\dagger})^{bt} (h_v)^t_\alpha \big] (0) \, \big[ (\bar{h}_v)^m_\lambda \,( W_{n,-}^{\mathrm{uc}})^{mk} \big] (z_1)  \ket{0}  \notag \\[2mm]
&\times (2\pi)^3 \,2 \,  \delta ( m_t \hat{v}^+ \plus  \hat{k}^+ \minus \hat{q}^+ \minus  \hat{k}_{n'}^+ \minus \hat{k}_{uc}^+  )   \,  \delta (  \hat{k}^- \minus  \hat{k}_{n'}^- \minus \hat{k}_{uc}^-  )  \,  \delta^{(2)} ( m_t \hat{v}_\perp \plus  \hat{k}_\perp \minus \hat{q}_\perp \minus  \hat{k}_{n',\perp} \minus \hat{k}_{uc,\perp}  ). \notag
\end{align}	
In the final step we perform the integrations over $\hat{k}_{n'}^-$, $\hat{k}_{uc}^-$,  $\hat{k}_{n'}^\perp$ and $\hat{k}_{uc}^\perp$, which simply constrains the total minus and perp momenta (with respect to $n'$) through the last two $\delta$-functions, to arrive at 
\begin{align}
\label{eq:Wtopfactfin}
W^{\rho \mu}(n, v, k ,q)  = &  H_{d}^{\rho\mu}(q,v,m_t) 
\, \int \mathrm{d} \hat{k}^+_{n'} \, \mathrm{d} \hat{k}_{uc}^+ \, \delta ( m_t \hat{v}^+ \plus  \hat{k}^+ \minus \hat{q}^+ \minus  \hat{k}_{n'}^+ \minus \hat{k}_{uc}^+  )  \\
& \times   J_{n'} (\hat{H}^- \hat{k}_{n'}^+) \, \tilde{S}_{ucs} (\hat{v}^+,v^-, \gamma^2, v^- k^+ + v^+ k^- , \hat{k}_{uc}^+) \ , \notag 
\end{align}	
with the ultra-collinear-soft (ucs) function being defined as
\begin{align}
\label{eq:Sucsdef}
\tilde{S}_{ucs} &(\hat{v}^+,v^-, \gamma^2, v^- k^+ + v^+ k^-, \hat{a}^+) = \,  \frac{1}{8\pi \, N_c \, m_t}   \sum_{X_{uc}} \delta \big( \hat{a}^+ \! - \! \hat{K}^+_{X_{uc}} \big) \\
&\times \int \! \mathrm{d}^4 z_1 \! \int \! \mathrm{d}^4 z_2  \, \mathrm{e}^{i k\cdot (z_2 - z_1)} \, \bra{0} \overline{T} \big[    (W_{n,+}^{\mathrm{uc}, \dagger} )^{kl} \, ( h_v)^l_\lambda \big] (z_2) \,  \big[ (\bar{h}_v)_\alpha^a (Y_{n',-}^{\rm uc})^{ab} \big] (0)   \ket{ X_{uc}}     \notag     \\   
&\times           \bra{ X_{uc}} T \big[ (Y_{n',+}^{{\rm uc},\dagger})^{bt} (h_v)^t_\alpha \big](0) \,  \big[ (\bar{h}_v)^m_\lambda \,( W_{n,-}^{\mathrm{uc}})^{mk} \big] (z_1)  \ket{0} \notag \\[1mm]
=&\,  \frac{1}{8\pi \, N_c \, m_t} \int \! \mathrm{d}^4 z_1 \! \int \! \mathrm{d}^4 z_2  \, \mathrm{e}^{i k\cdot (z_2 - z_1)} \,\bra{0} \overline{T} \big(   [(W_{n,+}^{\mathrm{uc}, \dagger} )^{kl} \, ( h_v)^l_\lambda ] (z_2)\, [(\bar{h}_v)_\alpha^a (Y_{n',-}^{\rm uc})^{ab}] (0) \big) \notag\\[1mm]
& \qquad \times 
\delta (\hat{a}^+ \minus n' \cdott \hat{P}) T\big( [ (Y_{n',+}^{{\rm uc},\dagger})^{bt} (h_v)^t_\alpha] (0)\,
[ (\bar{h}_v)^m_\lambda \,( W_{n,-}^{\mathrm{uc}})^{mk}] (z_1) \big) \ket{0} \,. \notag 
\end{align}
The factorized form of the hadronic tensor in Eq.~(\ref{eq:Wtopfactfin}) and the ucs function represent two of the main results of this work.

\subsection{The Ultra-Collinear-Soft Function}
\label{sec:Sucsproperties}

The ultra-collinear-soft (ucs) function  $\tilde{S}_{ucs}$ 
describes the light-cone momentum $\hat{a}^+=a\cdot n'$ distribution (w.r.\ to the direction of the emitted $b$-jet) of the gluons that are soft 
in the heavy quark rest frame and thus arises from the analogous modes contained in the shape function for the $\bar B$ meson decay. However, there are important differences.
The ucs function is generated by ultra-collinear Wilson lines describing coherent radiation coming from the hard processes of boosted top-antitop production and the top quark decay as well as the top quark propagation with velocity $v$. It thus describes QCD effects which are commonly associated to non-factorizable contributions from the perspective of the factorized narrow width approach, where the cross section is approximated by on-shell top production followed by an on-shell top quark decay. The ucs function can thus be used to study non-factorizable QCD effects from the operator level. Furthermore,
the state that contains the top quark decay is not a non-perturbative hadronic bound state but a top hemisphere jet controlled by an invariant mass measurement which is described by the factorization theorem for boosted $t\bar t$ production derived in Sec.~\ref{sec:hemispherefactorization_new}. As a consequence, the matrix elements appearing in the second and third lines of Eq.~(\ref{eq:Sucsdef}) are non-local time-ordered products of decay and production operators where the total momentum of the decaying (off-shell and resonant) top plus (ultra-collinear) gluon system is $p=m_t v+k$, inserted at positions $z_{1,2}$, such that also a top quark propagator arises.\footnote{The large label momentum contribution $m_t v$ is accounted for already in the matching relation~(\ref{eq:EFT_heavy_to_light_currents_t_decay}).} This implies another major difference, namely that the large top quark width $\Gamma_t$ renders the ucs function computable in perturbation theory, in contrast to the $\bar B$ meson shape function. 
In addition, the dependence on the light-like vector $\bar n^\mu=(1,-\vec{n})$ through the ultra-collinear Wilson lines $W_{n,-}^{\rm uc}$ and $W_{n,+}^{{\rm uc},\dagger}$ associated to the hard production process provides an additional complication that does not arise in the description of the $\bar{B}$ meson decay where only a dependence on the $n'$ direction appears. 

The dependence on the two light-like directions  $\bar n^\mu=(1,-\vec{n})$ and $n'^\mu=(1,\vec{n}')$ entails that the ucs function depends on $\hat{v}^+=v\cdot n'$ and $v^-=v\cdot\bar n$ as well as on the  relative angle between $\vec{n}$ and $\vec{n}'$, which we parameterize by the variable
\begin{equation}
	\gamma^2 = \frac{2}{\bar{n} \cdot n'}\,.
\end{equation}
It can take values between $1$ (when the $b$-jet $3$-momentum is parallel to the top hemisphere $3$-momentum) and infinity (when the $b$-jet $3$-momentum is anti-parallel to the top hemisphere $3$-momentum), i.e. $\gamma^2\in[1,\infty)$.
Furthermore, the ucs function has soft and collinear singularities with respect to both directions, which also depend on $\gamma^2$. The factorization formula of Eqs.~(\ref{eq:decomposition_bHQETjetfct_new}) together with (\ref{eq:Wtopfactfin}) is frame-independent and valid for any choice of the reference velocity $v^\mu$. As for the $\bar B$ shape function the matrix elements in Eq.~(\ref{eq:Sucsdef}) are boost-invariant, and it is thus straightfoward to derive the relation between the ucs function for a general frame and the one where $v^\mu_{\rm rest}=(1,\vec{0})$ (which we call the rest frame for simplicity). 
Assuming that $\Lambda^\mu_{\,\,\nu}$ boosts from a general frame into the rest frame along $\vec{n}$ (i.e. $\Lambda^\mu_{\,\,\nu}\,v^\nu=v^\mu_{\rm rest}=(1,\vec{0})$) and recalling that $\Lambda^\mu_{\,\,\nu}\,n'^\nu=  \hat{v}^+ n'^\mu_{\rm rest}$, $\Lambda^\mu_{\,\,\nu}\,n^\nu=  (1/v^-) n^\mu_{\rm rest}$, $\Lambda^\mu_{\,\,\nu}\,\bar n^\nu=  v^- \bar n^\mu_{\rm rest}$, we find that 
$(\hat{\ell}^+)_{\rm rest}= \hat{\ell}^+/\hat{v}^+$, $(k^+)_{\rm rest}=v^-k^+$,  $(k^-)_{\rm rest}=k^-/v^-$ and $(\gamma^2)_{\rm rest}=v^-\hat{v}^+\gamma^2$. This implies the relation
\begin{align}
\label{eq:ucsfctrel}
	\tilde{S}_{ucs} (\hat{v}^+,v^-, \gamma^2, v^- k^+ + v^+ k^-, \hat{a}^+)=
\frac{1}{\hat{v}^+}\, \tilde{S}_{ucs} \bigg(1,1,  \gamma^2 v^-\hat{v}^+,v^- k^+ + v^+ k^-, \frac{\hat{a}^+}{\hat{v}^+} \bigg)\,,
\end{align}
where we also recall that $v^+=1/v^-$.

To acquire an intuitive feeling for the content of the ucs function let us briefly discuss its explicit form at tree-level, where the collinear and soft Wilson lines in the matrix elements of the ucs function reduce to unit operators and no ultra-collinear gluons are radiated. The calculation reduces 
to the evaluation of the (anti-)time-ordered vacuum matrix elements of bHQET top quark fields given in Eq.~(\ref{eq:bHQET_quark_propagators}) which gives a bHQET top quark propagator and its complex conjugate.
Combining all exponential factors and carrying out the  $z_{1,2}$ integrations, we can use the resulting delta functions to do the $k_{1,2}$ integrals in the configuration space propagators, which fixes $k_1 = k_2 =k$. The tree-level result for the ucs function then reads
\begin{align}
\label{eq:Sucstree}
\tilde{S}^{\text{tree}}_{\rm ucs} &(\hat{v}^+,v^-, \gamma^2, v^+ k^-+v^- k^+, \hat{a}^+) 
=\frac{\mathrm{Tr} [P_v]}{8\pi \,N_c  \, m_t} \, \frac{\delta (\hat{a}^+) }{|v\cdott k \plus i \frac{\Gamma_t}{2}|^2}    \\
	&= \frac{1}{\pi\, m_t} \, \frac{\delta (\hat{a}^+)}{|v^+ k^-+v^- k^+ \plus i \Gamma_t|^2} 
=  \frac{1}{4\pi\, m_t \, \hat{v}^+ } \, \frac{\delta (\bar{a}^+)}{|\Delta|^2}        \notag \,,
\end{align}
which exhibits the tree-level bHQET jet function with the unstable top quark's Breit-Wigner mass distribution times the tree-level (perturbative) meson decay shape function. At tree-level an explicit dependence on $\hat{v}^+$, $v^-$ and $\gamma^2$ does not yet arise in the ucs function $\tilde{S}_{ucs}$. In the last equality we introduced the notations
\begin{align}
\label{eq:Deltaelldefs}
\Delta \equiv v\cdott k \plus i \frac{\Gamma_t}{2}\,,
\qquad 
\bar{a}^+ \equiv \frac{\hat{a}^+}{\hat{v}^+}\,,
\end{align} 
%the off-shellness variable $\Delta = (s + i\Gamma_t)/2$ as well as $\bar{\ell}^+ = \hat{\ell}^+/\hat{v}^+$. This notation 
which will be convenient for the discussion of the ${\cal O}(\alpha_s)$ corrections and the renormalization of the ucs function below. The simple factorized structure exhibited in the tree-level result of Eq.~(\ref{eq:Sucstree}) is not maintained at higher orders due to the non-factorizable character of the ucs function.

The dependence of the ucs function on two light-like directions $\bar{n}$ and $n'$ causes a significant technical complication in the evaluation of the convolutions with the $b$-jet function in the factorization formula~(\ref{eq:Wtopfactfin}) and the hemisphere soft function in the double hemisphere factorization of Eq.~(\ref{eq:bHQET_cross_section_new-5}). Physically this complication arises from the fact that the large-angle soft $\ell^+$ momentum (in the $e^+e^-$ frame) contributes to the invariant masses of the entire hemisphere and the $b$-jet in a different (and $\gamma^2$-dependent) way. 
This complexity also affects the renormalization procedure (and the subsequent summation of the large logarithms). To see this more explicitly, let us have a look on the full $k^\pm$-dependence of the factorization formula for the hadronic tensor in Eq.~(\ref{eq:Wtopfactfin}). (We remind the reader that $k$ parametrizes the total residual momentum of the ultra-collinear sector, see Eq.~(\ref{eq:koffshelldef}).) The $\delta$-function appearing in the first line controls the total plus momentum (w.r.\ to the $n'$ direction) of the $b$-jet and  depends on the light cone momentum component $\hat{k}^+=k\cdot n'$. It is related to $k^\pm$ by
\begin{equation} 
\label{eq:tildekplusdef}
\hat{k}^+= \hat{k}^+ (\gamma^2, k^+, k^-)= \frac{1}{\gamma^2} k^+ \plus \Big( 1 \minus \frac{1}{\gamma^2} \Big) k^-\,,
\end{equation}
and thus depends on the angle between the $b$-jet and top directions. We recall that we have $k^\perp=0$ for our choice of the light-cone reference vector $n$, (see footnote~\ref{foot:kperp}).
For a decay event where, in the $e^+e^-$ frame, the $b$-jet and the leptons are both well collimated within the top hemisphere in the $\vec{n}$ direction, we have $1/\gamma^2\sim (m_t/Q)v^-$ and $1-1/\gamma^2\sim (m_t/Q) v^+$, so that the $k^+$ and the $k^-$ terms in $\hat{k}^+$ both contribute at leading order in the power counting.
Due to the convolution over $v^-k^+$ with respect to the hemisphere soft function, see Eqs.~(\ref{eq:bHQET_cross_section_new-5}) and (\ref{eq:Jbtarguments}), and because we want to maintain the canonical notation for the renormalization for the SCET jet function $J_{n'}$, 
it is thus mandatory to introduce a modified shifted definition of the ucs function, 
where the entire dependence on $v^-k^+$ is shifted into ucs function:
\begin{align}
\label{eq:Sucsdefnotilde}
S_{ucs}(\hat{v}^+,v^-, \gamma^2, s^+, s^-, \hat{\ell}^+) & \equiv \tilde{S}_{ucs} \Big(\hat{v}^+,v^-, \gamma^2, s^+ + s^-, \hat{\ell}^+ \plus \hat{k}^+ \Big(\gamma^2, \frac{s^+}{v^-}, v^- s^-\Big)\Big)  \\
& = \tilde{S}_{ucs} \Big(\hat{v}^+,v^-, \gamma^2, s^+ + s^-, \hat{\ell}^+ \plus \frac{1}{\gamma^2} \frac{s^+}{v^-} \plus \Big( 1 \minus \frac{1}{\gamma^2} \Big)v^- s^-\Big)\,. \notag
\end{align}
Here, we employ the boost-invariant off-shellness variables $s^+=v^- k^+$ and $s^- = v^+ k^-$ following Eq.~(\ref{eq:tophemisphere_offshellness_bHQET}), and we note that the momentum variable $\hat{\ell}^+$ introduced here is unrelated to the momentum $\ell^\mu$ of the large-angle soft radiation. While the original definition of $\tilde{S}_{ucs}$ in Eq.~(\ref{eq:Sucsdef}) is convenient for the computation of the higher order corrections, it is mandatory to use $S_{ucs}$ in Eq.~(\ref{eq:Sucsdefnotilde}) for the purpose of renormalization and for the computation of the cross section. We note that for transparency in the following discussions we frequently use $\hat{a}^+$ as the last argument for $\tilde{S}_{ucs}$, when the distinction to $S_{ucs}$ is relevant. 

In terms of this modified ucs function, the hadronic tensor can be written as 
\begin{align} \label{eq:hadronic_tensor_semileptonic_top_decay}
W^{\rho \mu}&(n, v, k ,q)  =\,  H_{d}^{\rho\mu}(q,v,m_t) \\
&\times \int \mathrm{d} \hat{k}^+_{n'} \, \mathrm{d} \hat{\ell}^+ \, \delta ( m_t \hat{v}^+ \minus \hat{q}^+ \minus  \hat{k}_{n'}^+ \minus \hat{\ell}^+  )   \, J_{n'} (\hat{H}^- \hat{k}_{n'}^+) \, S_{ucs}(\hat{v}^+,v^-, \gamma^2, v^- k^+, v^+ k^-, \hat{\ell}^+)  \notag \\
=&\,  H_{d}^{\rho\mu}(q,v,m_t) \,
\int \mathrm{d} \hat{\ell}^+  \,  J_{n'} (\hat{H}^- (m_t \hat{v}^+ \minus \hat{q}^+\minus  \hat{\ell}^+) ) \, S_{ucs} (\hat{v}^+,v^-, \gamma^2, v^- k^+, v^+ k^-, \hat{\ell}^+) \ . \notag
\end{align}
which exhibits a more canonical form, and shifts the complexity arising from having the two light-like directions $n$ and $n'$ entirely into ucs function $S_{ucs}$.

\subsection{Final Cross Section Result and Renormalization Consistency}
\label{sec:renormalizationtopdecay}

The factorized form of the hadronic tensor in Eq.~(\ref{eq:hadronic_tensor_semileptonic_top_decay}) represents the main result of this work.  Combined with Eqs.~(\ref{eq:bHQET_cross_section_new-5}) and (\ref{eq:decomposition_bHQETjetfct_new}), we can now write down the complete factorization formula for the 
double differential hemisphere mass distribution with an additional measurement of an observable $X={\cal X}(M_t,p_\ell , \pn, H)$ determined 
from the leptons and the total momentum $H^\mu$ of all hadrons arising from the top quark decay within the top hemisphere:  
\begin{align}
\label{eq:dhmdecayfinal}
\frac{\mathrm{d}^3\sigma}{\mathrm{d} M_t^2 \mathrm{d} M_{\bar{t}}^2 \mathrm{d} X} =
&\,  \sigma_0 \, \int \mathrm{d} H^2 \, \mathrm{d} \Pi_3(M_t= m_t v_n+k^* ; p_\ell , \pn, H) \, \delta(X-{\cal X}(M_t, p_\ell , \pn, H))
 \notag \\
&\times \, \Big( \frac{e}{\sqrt{2} \,s_w} \Big)^4 |V_{tb}|^2\,
 \tilde{L}^{\sigma \nu} (p_\ell, \pn) \, \frac{\tilde{g}_{\rho \sigma} \,  \tilde{g}_{\mu \nu}}{|q^2 \minus M_W^2 + i M_W \Gamma_W|^2} \notag \\ 
& \times  H_Q(Q) \, H_{m} \Big(m_t,\frac{Q}{m_t}\Big) \, H_{d}^{\rho\mu}(q=p_\ell + \pn,v,m_t)    \\
& \times \int \mathrm{d} \ell^+ \,  \mathrm{d} \ell^-  \mathrm{d} \hat{\ell}^+ \,   
	J_{B_{\bar{t}}}^{\Gamma_t} \Big(s_{\bar t}^* \minus \frac{Q}{m_t} \ell^-\Big) \,  S_{\mathrm{hemi}}(  \ell^+,  \ell^- )  \notag \\
&\mbox{\hspace{3mm}} \times J_{n'} \Big(\hat{H}^- (  m_t \hat{v}_n^+ \minus \hat{q}^+ \minus \hat{\ell}^+  )  \Big) \, 
	S_{ucs}\Big(\hat{v}_n^+,v_n^-, \gamma^2, \frac{s_t^{*}}{2} \minus  \frac{Q}{m_t} \ell^+, \frac{s_t^{*}}{2}, \hat{\ell}^+\Big)\,.  \notag
\end{align}
It is a combination of the factorization formulae for the double hemisphere invariant mass distribution and the inclusive semileptonic heavy meson decay in the endpoint region, with $S_{ucs}$ being the novel factorization function where both these factorizations are connected.  
We remind the reader concerning the definitions of the physical off-shellness variables given in Eqs.~(\ref{eq:hemisphere_offshellness_bHQET}) and (\ref{eq:tophemisphere_offshellness_bHQET}) and footnote~\ref{foot:offshells}, the reference velocities $v_{n/\bar n}$ in Eq.~(\ref{eq:vlabeldef-2}) and the total top hemisphere hadronic momentum $H=M_t- p_\ell - \pn$. The formulation (and the 3-body phase space $\Pi_3$) above is in terms of momentum (light-cone) variables defined in the $e^+e^-$ frame. Since we have formulated the top quark decay using manifest Lorentz-covariant factorization functions $H_{d}^{\rho\mu}$, $J_{n'}$ and the ucs function $S_{ucs}$, which satisfies the scaling relation~(\ref{eq:ucsfctrel}), one can switch for the top decay kinematics for $p_\ell$, $\pn$ and $H$ to the (simpler) hemisphere rest frame where $M_t^\mu=(M_t,\vec{0})$ (which implies $s_t^{*+}=s_t^{*-}=s_t^{*}/2$) by simply setting $v_n^-=\hat{v}_n^+=1$ above wherever they appear explicitly.  The explicit factors $Q/m_t$ must remain untouched.
We also note that the generalization to account for a multi-differential top decay observable $\vec X=\vec{\cal X}(M_t,p_\ell,\pn, H)$ or to also consider in addition  
a semileptonically decaying antitop quark in the $\bar n$ sector, is in principle straightforward. 

The various functions appearing in Eq.~(\ref{eq:dhmdecayfinal}) and their renormlization are well-known from Refs.~\cite{Bauer:2001yt,Bosch:2004th,Fleming:2007qr,Fleming:2007xt} (and collected at ${\cal O}(\alpha_s)$ in Apps.~\ref{app:apprenor}  and \ref{app:NLO_corrections}) 
with the exception of the novel ucs function $S_{ucs}$. Fortunately, the renormalization of the ucs function is already determined by the renormalization properties of the other factorization functions due to consistency conditions~\cite{Fleming:2007xt} that arise from the fact that e.g.\ current renormalization can also be achieved through the renormalization of the soft, jet and ucs functions. Thus (within the $\overline{\rm MS}$ scheme) all higher order contributions in the respective $Z$-factors are related, such that the $Z$-factor of one of the factorization functions is determined by the those of the others. This provides the renormalization condition for the ucs function and represents a powerful cross check for the results of its ${\cal O}(\alpha_s)$ corrections which we determine in Sec.~\ref{sec:ucsoftcomputations}.

As the ucs function shares renormalization properties of the inclusive bHQET top quark jet function $J_{B_t}$, let us first discuss the consistency condition for $J_{B_t}$ emerging from the double hemisphere mass factorization formula of Eq.~(\ref{eq:bHQET_cross_section_new-5}).
The explicit expressions for the bHQET (5-flavor theory) renormalization relations for all bare and UV-finite functions $J_{B_t}$, $S_{\text{hemi}}$ and the bHQET $t\bar t$ current, expressed in terms of Wilson coefficient $H_m=|C_m|^2$, are given in Eqs.~(\ref{eq:renormalization_H}) and (\ref{eq:renormalization_functions_appendix}), where for the double hemisphere soft function $S_{\rm hemi}$ the $Z$-factor factorizes symmetrically into two single hemisphere factors $Z_s$.
Anticipating that the plus ($s_t^+=v_n^-k^+$) and minus ($s_t^-=v_n^+k^-$) off-shellness dependencies in the ucs function differ due to the additional angle-dependence and that only the plus contribution is involved in the convolution with the soft hemisphere function, we now write the renormalization of $J_{B_t}$ using the notation of Eq.(\ref{eq:Jbtarguments}),
\begin{align} 
\label{eq:renJBtmod}
J_{B_{t}}^{\mathrm{bare}} (s_t^+,s_t^-) &= \int \mathrm{d} s'^+_t \, Z_{J_{B_{t}}} (s_t^+ \minus s'^+_t, \mu) \, J_{B_{t}} (s'^+_t,s_t^- ,\mu)\,. 
\end{align}
Renormalization consistency within the factorization theorem 
implies that (within the $\overline{\rm MS}$ scheme) the higher order corrections of all the $Z$-factors cancel, which yields
\begin{align}
\int \mathrm{d} s'^+ \, \frac{m_t}{Q} \Big|Z_{C_m}\Big(m_t,\frac{Q}{m_t},\mu\Big)\Big| \, Z_s \Big(  \frac{m_t}{Q} (s^+- s'^+ ),\mu \Big)   \, Z_{J_{B_t}} (s'^+ \minus s''^{+},\mu) = \delta (s^{+}  \minus s''^{+}) \,,
\end{align} 
where we used $v_n^-=Q/m_t$.
This is equivalent to
\begin{align}
Z_{J_{B_t}} (s^+,\mu) =  \frac{m_t}{Q} \, \Big|Z_{C_m}\Big(m_t,\frac{Q}{m_t},\mu\Big)\Big|^{-1} \, Z_s^{-1} \Big(  \frac{m_t}{Q} s^+, \mu \Big) \,,
\end{align}
which gives the bHQET jet function renormalization factor shown in Eq.~(\ref{eq:Z_factor_functions_app}) in terms of the bHQET current and hemisphere soft function $Z$-factors. 

As the ucs function also shares renormalization properties of the shape function of the $\bar B$ meson endpoint factorization formula (up to the trivial change from 4-flavor to the 5-flavor QCD) let us also have a closer look at Eq.~(\ref{eq:BdecayWfinal}). The explicit expressions for the HQET renormalization relations for the functions $J_{n'}$, $S_{\text{shape}}$ and the heavy-to-light current, expressed in terms of Wilson coefficients $C_j$ are again given in Eqs.~(\ref{eq:renormalization_H}) and (\ref{eq:renormalization_functions_appendix}). The emerging consistency condition reads
\begin{align}
\label{eq:Bdecayconsist1}
\hat{H}^- \int \mathrm{d} \hat{\ell}^{'+}  \, Z_{C_j}^2(\hat{v}^+ \hat{H}^-,\mu) \,  Z_{J_{n'}} ( \hat{H}^- (\hat{\ell}^+ \minus \hat{\ell}^{'+}),\mu) \, Z_{\mathrm{shape}} (\hat{v}^+,\hat{\ell}^{'+} \minus \hat{\ell}^{''+},\mu) = \delta (\hat{\ell}^+ \minus  \hat{\ell}^{''+} ) \,.
\end{align}
This is equivalent to
\begin{align}
\label{eq:Bdecayconsist2}
Z_{\mathrm{shape}} (\hat{v}^+,\hat{\ell}^{+},\mu) = \hat{H}^- \, Z_{C_j}^{-2}(\hat{v}^+ \hat{H}^-,\mu) \,  Z_{J_{n'}}^{-1} (\hat{H}^- \hat{\ell}^{+},\mu)   \,,
\end{align}
and gives the shape function renormalization factor shown in Eq.~(\ref{eq:Z_factor_functions_app}) in terms of the current and SCET jet function $Z$-factors. We note that Eqs.~(\ref{eq:Bdecayconsist1}) and (\ref{eq:Bdecayconsist2}) are formulated in a general frame, which yields the additional dependence on $\hat{v}^+$. The form of Eq.~(\ref{eq:Bdecayconsist2}) also reconfirms (as $\hat{v}^+ \hat{H}^-$ is boost-invariant) that the relation between the shape function $Z$ factors in a general and the rest frame (with $\hat{v}^+=1$) reads
$Z_{\mathrm{shape}} (\hat{v}^+,\hat{\ell}^{+},\mu)=\frac{1}{\hat{v}^+} Z_{\mathrm{shape}} \Big(1,\frac{\hat{\ell}^{+}}{\hat{v}^+},\mu\Big)$
in analogy to Eq.~(\ref{eq:shapefctrel}).

Because the ucs function $S_{ucs}(\hat{v}^+,v^-, \gamma^2, s^+, s^-, \hat{\ell}^+)$ in the factorization theorem in Eq.~(\ref{eq:dhmdecayfinal}) shares the renormalization properties of the shape function concerning the variable $\hat{\ell}^+$ as well as of the bHQET jet function concerning the variable $s^+$, the renormalization relation for bare and renormalized ucs functions can now be easily written down as
\begin{align}  
\label{eq:renormalization_Sucs}
S_{ucs}^{\mathrm{bare}}(\hat{v}^+,v^-, \gamma^2, s^+, s^-, \hat{\ell}^+) = \int & \mathrm{d} s'^+   \int \mathrm{d} \hat{\ell}'^+ \, Z_{J_{B_{t}}} (s^+ \minus s'^+, \mu ) \, Z_{	{\mathrm{shape}}} (\hat{v}^+, \hat{\ell}^+ \minus \hat{\ell}'^+, \mu) \notag \\
&\times  S_{ucs} (\hat{v}^+,v^-, \gamma^2, s'^+, s^-, \hat{\ell}'^+, \mu) \,. 
\end{align}
This implies that the renormalization group equation has the form
\begin{align}
\mu \frac{\mathrm{d}}{\mathrm{d} \mu}  S_{ucs} (\hat{v}^+,v^-, \gamma^2,& s^+, s^-, \hat{\ell}^+, \mu)  =  \int \mathrm{d} s'^+ \gamma_{J_{B}} ( s^+ \minus s'^+, \mu) \, S_{ucs} (\hat{v}^+,v^-, \gamma^2, s'^+, s^-, \hat{\ell}^+, \mu) \notag \\
& \quad  \plus \int \mathrm{d} \hat{\ell}'^+ \, \gamma_{S_{\mathrm{shape}}} (\hat{\ell}^+ \minus \hat{\ell}'^+, \mu) \, S_{ucs} (\hat{v}^+,v^-, \gamma^2, s^+, s^-, \hat{\ell}'^+, \mu) \,,
\end{align}
where $\gamma_{J_{B}}$ and $\gamma_{S_{\mathrm{shape}}}$ are the anomalous dimensions of the bHQET jet function and the shape function, respectively, see App.~\ref{app:apprenor} for their definition. These anomalous dimensions have been computed up to three-loops~\cite{Jain:2008gb,Clavero:2024yav,Becher:2005pd,Bruser:2019yjk}, and we have collected them in App.~\ref{app:NLO_corrections} for the convenience of the reader. 
We emphasize that the $S_{ucs}$ which appears here, is related to the original ucs function matrix element $\tilde{S}_{ucs}$ of Eq.~(\ref{eq:Sucsdef}) through relation~(\ref{eq:Sucsdefnotilde}). The non-trivial two-dimensional character of this relation, which induces an angle ($\gamma^2$) dependent combination of the UV divergences in $Z_{J_{B_t}}$ and $Z_{\mathrm{shape}}$, prohibits the recombination of $\gamma_{J_{B}}$ and $\gamma_{S_{\mathrm{shape}}}$ into a single simple anomalous dimension.
It is also a reflection of the soft rest-frame dynamics sensitive to details of the production and decay process which entails the  non-factorizable properties of the off-shell effects encoded in the ucs function. However, the factorized form of Eq.~(\ref{eq:renormalization_Sucs}) implies that the RG evolution associated to the production stage radiation (contained in the bHQET jet function evolution) and the one associated to the top quark decay and the $b$-jet production (contained in the shape function evolution) remain factorized. This is consistent with the factorized cross-section level implementation of production and decay stage parton showering in multipurpose Monte-Carlo event generators.  

The relation of Eq.~(\ref{eq:renormalization_Sucs}) allows us to determine the UV divergences of the ucs function. 
At ${\cal O}(\alpha_s)$ the UV divergences coming from $Z$-factors of the shape function (class I) and the bHQET jet function (class II) can be written as
\begin{align}
	\tilde{S}_{\mathrm{ucs,div}}^{(\alpha_s)} = 
	\tilde{S}_{\mathrm{ucs,div}}^{(\alpha_s),\mathrm{I}} + \tilde{S}_{\mathrm{ucs,div}}^{(\alpha_s),\mathrm{II}}\,.
\end{align}
Both terms can be obtained quite easily since the tree-level ucs function (see Eq.~(\ref{eq:Sucstree})) to be inserted in Eq.~(\ref{eq:renormalization_Sucs}) contains a $\delta$-function involving both convolution variables. The results read
{
\allowdisplaybreaks
\begin{align}
\tilde{S}_{\mathrm{ucs,div}}^{(\alpha_s),\mathrm{I}} &(\hat{v}^+,v^-, \gamma^2, s^++s^-, \hat{a}^+ )  = S_{\mathrm{ucs,div}}^{(\alpha_s),\mathrm{I}} (\hat{v}^+,v^-, \gamma^2, s^+, s^-, \hat{\ell}^+) \\
&=  \int \mathrm{d} \hat{\ell}'^+  \, Z_{S_{\mathrm{shape}}}^{(\alpha_s)} (\hat{v}^+, \hat{\ell}^+ \minus \hat{\ell}'^+,\mu) \, \bigg[
\frac{1}{4\pi m_t} \,  \frac{1}{|\Delta|^2}\,
\delta\Big( \hat{\ell}'^+ \plus \hat{k}^+ \Big(\gamma^2, \frac{s^+}{v^-}, v^- s^-\Big)\Big) \bigg]     \notag \\
&=  \frac{1}{4\pi \,m_t \,|\Delta|^2}\, Z_{S_{\mathrm{shape}}}^{(\alpha_s)} (\hat{v}^+, \hata,\mu) \,,\notag \\
\tilde{S}_{\mathrm{ucs,div}}^{(\alpha_s),\mathrm{II}} &(\hat{v}^+,v^-, \gamma^2, s^++s^-, \hat{a}^+ )  = S_{\mathrm{ucs,div}}^{(\alpha_s),\mathrm{II}} (\hat{v}^+,v^-, \gamma^2, s^+, s^-, \hat{\ell}^+) \notag \\
&=  \int \mathrm{d} s'^+   \, Z_{J_{B_{t}}}^{(\alpha_s)} (s^+ \minus s'^+, \mu ) \, \bigg[ \frac{1}{\pi m_t } \,  \frac{1}{|s'^+ + s^-+i\Gamma_t|^2}\, \delta\Big( \hat{\ell}^+ \plus  
\frac{1}{\gamma^2} \frac{s'^+}{v^-} \plus \Big( 1 \minus \frac{1}{\gamma^2} \Big)v^- s^-
\Big) \bigg]     \notag \\
&=  \frac{1}{\pi m_t}\,  \frac{\gamma^2 v^-}{|2\Delta- \bara \gamma^2 \hat{v}^+ v^- |^2} \,  Z_{J_{B_{t}}}^{(\alpha_s)} (\bara \gamma^2 \hat{v}^+ v^-, \mu) \,, \notag
\end{align}
}%
where the superscript $(\alpha_s)$ stands for the  ${\cal O}(\alpha_s)$ corrections and
\begin{align}
	\label{eq:ahatdef}
	\hata &\equiv \hat{\ell}^+ \plus \hat{k}^+ \Big(\gamma^2, \frac{s^+}{v^-}, v^- s^-\Big) \,.
\end{align}
Moreover, we employ the variables $\Delta=(s^+ + s^-+i\Gamma_t)/2=(s+i\Gamma_t)/2$ and 
$\bara = \hata/\hat{v}^+$ already introduced in Eq.~(\ref{eq:Deltaelldefs}). Using the explicit form of the one-loop $Z$-factor corrections for the shape and the bHQET jet functions given in Eqs.~(\ref{eq:Z_factor_functions_app}), we obtain
\begin{align} 
\label{eq:ucs_divergentpart}
\tilde{S}_{\mathrm{ucs,div}}^{(\alpha_s),\mathrm{I}} (\hat{v}^+, v^-, \gamma^2, s, \bara) = 
& \frac{1}{4\pi m_t \hat{v}^+ |\Delta|^2 }\,  \frac{\alpha_s C_F}{4\pi}  \bigg\{ \delta (\bara) \Big[ -\! \frac{2}{\epsilon^2} \! + \! \frac{2}{\epsilon} \Big]  \plus \frac{4}{\epsilon} \,  \frac{1}{\mu} \mathcal{L}_0 \Big(\frac{\bara}{\mu} \Big) \bigg \} \,, \\
\tilde{S}_{\mathrm{ucs,div}}^{(\alpha_s),\mathrm{II}} (\hat{v}^+, v^-, \gamma^2, s, \bara)   = 
& \frac{1}{4\pi m_t \hat{v}^+ }\,  \frac{\alpha_s C_F}{4\pi}  \bigg\{ \frac{\delta (\bara)}{|\Delta|^2}  \Big[  \frac{2}{\epsilon^2} \! + \! \frac{2}{\epsilon} \minus \frac{4}{\epsilon} \log (\gamma^2 \hat{v}^+ v^-) \Big] \notag \\
& \quad \minus\, \frac{16}{\epsilon} \, \frac{1}{|\bara \gamma^2 \hat{v}^+ v^- \minus 2 \Delta|^2}  \, \frac{1}{\mu} \mathcal{L}_0 \Big(\frac{\bara}{\mu} \Big) \bigg \} \,,
\end{align}
where we use the conventional notation
\begin{align} \label{eq:plus_distributions}
	\mathcal{L}_n(x) = \bigg[ \frac{\theta (x) \, \log^n (x)}{x} \bigg]_+ \,,
\end{align}
with $n \geq 0$ for the plus distributions.
This result for the divergenc terms is indeed reproduced by the explicit diagrammatic computation of the ucs function discussed in the next subsection, which reconfirms the consistency of the factorization theorem in Eq.~(\ref{eq:dhmdecayfinal}). The interesting conceptual aspect of this consistency is that the UV divergences of the ucs function are fully aware of the angle dependent difference of how the large-angle soft radiation enters the hemisphere and $b$-jet invariant masses. 

\begin{figure}[!t]%!htbp
	\centering
	\includegraphics[scale=0.23]{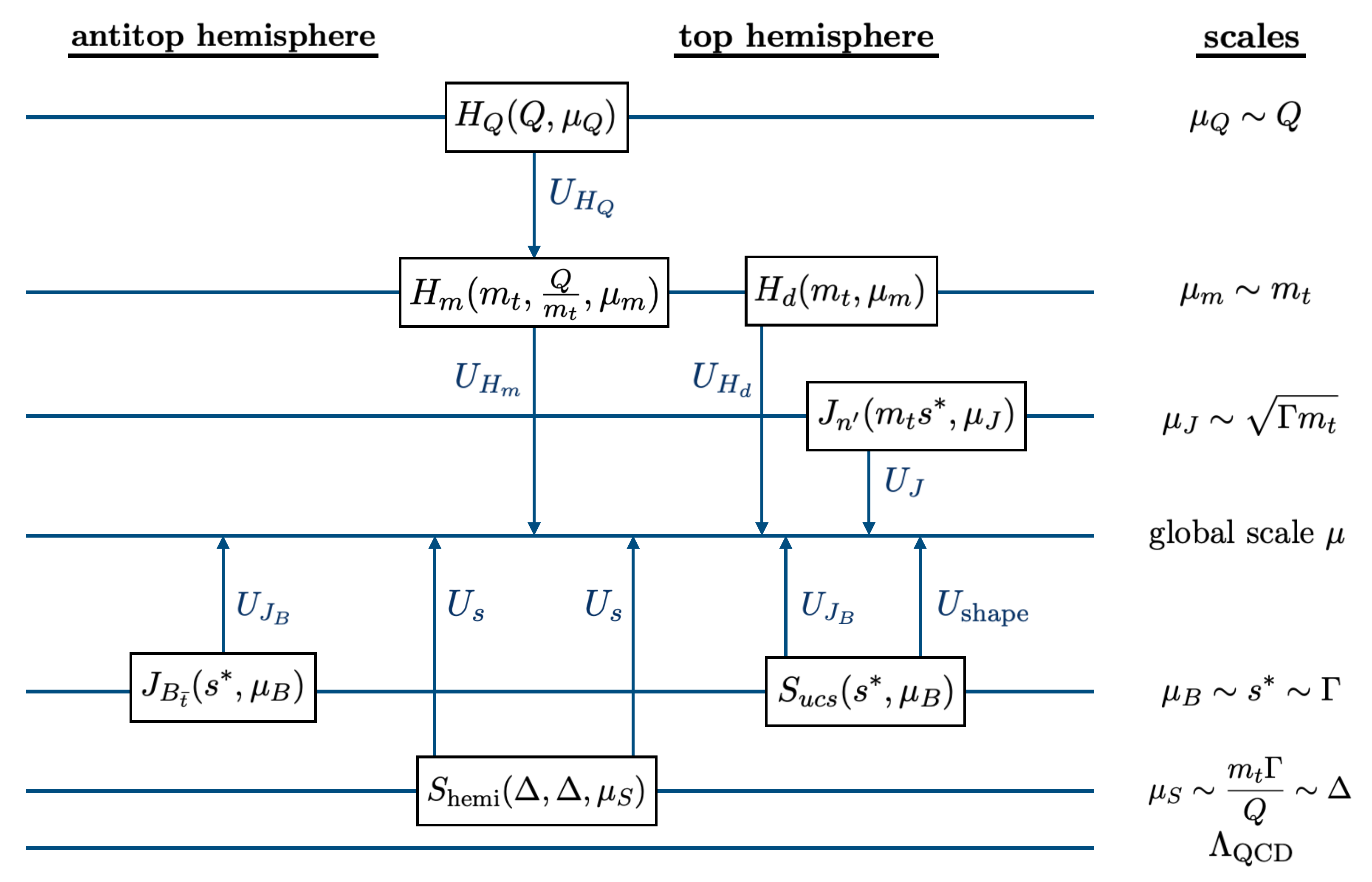}
	\caption{Matching factors, factorization functions and RG evolution as shown in the factorization formula (\ref{eq:dhmdecayfinalrge}) for boosted top-antitop pair production with a semileptonic top quark decay in the top mass sensitive endpoint region, and where the hemisphere masses $M_t$ and $M_{\bar t}$ are in the resonance region. The $b$-jet is defined as the top hemisphere without the lepton momenta. The evolution of $U_{H_Q}$, $U_{H_m}$ and $U_{H_d}$ are local, while those in $U_J$, $U_{J_B}$, $U_S$ and $U_\mathrm{shape}$ involve 
	convolutions. The global renormalization scale $\mu$, to which all matching factors and functions appearing below the hard scattering scale $Q$ evolve, 
	may be set to any other scale below $m_t$ due to renormalization consistency.  }
	\label{fig:scales}
\end{figure}

The final expression for the factorization theorem with all RG evolution factors and renormalization scales displayed explicitly then reads
{
\allowdisplaybreaks
\begin{align}
\label{eq:dhmdecayfinalrge}
&\frac{\mathrm{d}^3\sigma }{\mathrm{d} M_t^2 \mathrm{d} M_{\bar{t}}^2 \mathrm{d} X}  =
\,  \sigma_0 \, \int \mathrm{d} H^2 \, \mathrm{d} \Pi_3(M_t= m_t v_n+k^* ; p_\ell , \pn, H) \, \delta(X-{\cal X}(M_t, p_\ell , \pn, H))
\notag \\
&\quad \times \, \Big( \frac{e}{\sqrt{2} \,s_w} \Big)^4 |V_{tb}|^2\,
\tilde{L}^{\sigma \nu} (p_\ell, \pn) \, \frac{\tilde{g}_{\rho \sigma} \,  \tilde{g}_{\mu \nu}}{|q^2 \minus M_W^2 + i M_W \Gamma_W|^2}  \\ 
&\quad \times  H_Q(Q,\mu_Q) \, U_{H_Q}(Q,\mu_Q,\mu_m)\, \notag  \\
& \quad \times H_{m} \Big(m_t,\frac{Q}{m_t},\mu_m\Big)\, U_{H_m}\Big(\frac{Q}{m},\mu_m,\mu\Big) \, 
H_{d}^{\rho\mu}(q=p_\ell + \pn,v,m_t,\mu_m) \, U_{H_d}(\hat{v}_n^+\hat{H}^-,\mu_m,\mu) \notag  \\
& \quad \times \int \mathrm{d} \ell^+  \mathrm{d} \ell'^+ \,  \mathrm{d} \ell^- \mathrm{d} \ell'^- \mathrm{d} \hat{\ell}^+ \mathrm{d} \hat{\ell}'^+ \mathrm{d} s_{\bar t} \, \mathrm{d} s_{t}\, \mathrm{d} p^2 \, \,
U_{J_B}\Big(s_{\bar t}^* \minus \frac{Q}{m_t} \ell^--s_{\bar t},\mu,\mu_B\Big) J_{B_{\bar{t}}}^{\Gamma_t}(s_{\bar t},\mu_B)  \notag \\
& \qquad \times U_s(\ell^+-\ell'^+,\mu,\mu_S) U_s(\ell^-- \ell'^-,\mu,\mu_S) S_{\mathrm{hemi}}(  \ell'^+,  \ell'^-,\mu_S )  \notag \\
& \qquad  \times   U_{J}\Big(\hat{H}^- (  m_t \hat{v}_n^+ \minus \hat{q}^+ \minus \hat{\ell}^+ )-p^2,\mu,\mu_J  \Big) J_{n'}(p^2,\mu_J)  \notag \\
& \qquad  \times 
U_{J_B}\Big(\frac{s_t^{*}}{2} \minus  \frac{Q}{m_t} \ell^+ - s_t,\mu,\mu_B\Big)
U_{\rm shape}(\hat{\ell}^+-\hat{\ell}'^+,\mu,\mu_B)
S_{ucs}\Big(\hat{v}_n^+,v_n^-, \gamma^2, s_t , \frac{s_t^{*}}{2}, \hat{\ell}'^+,\mu_B\Big)\,,  \notag
\end{align}
}
where $\mu_Q$, $\mu_m$, $\mu_J$, $\mu_B$ and $\mu_S$ are the natural renormalization scales with a canonical scaling indicated in the right part of Fig.~\ref{fig:scales}. There are a few comments in order.
First, the choice of the global renormalization scale $\mu$, which is located  between the $b$-jet invariant mass and the hemisphere offshellness in Fig.~\ref{fig:scales}, can be chosen at any scale, as the factorization theorem is strictly $\mu$-invariant. Second, we can in principle adopt different natural and even global renormalization scales for the factorization functions of the top and the antitop hemispheres since the offshellness of the hemisphere invariant masses, $M_t-m_t$ and $M_{\bar t}-m_t$, may differ. We note, however, that large hierarchies between $M_t-m_t$ and $M_{\bar t}-m_t$ should be avoided as this may cause non-global logarithms in the hemisphere soft function $S_{\rm hemi}$. 

We emphasize that the factorization formula~(\ref{eq:dhmdecayfinalrge}) is valid at the hadron level and thus accounts for non-perturbative effects. The only factorization function sensitive to non-perturbative effects at leading power in the resonance region is the double hemisphere soft function $S_{\rm hemi}$. As already mentioned in Sec.~\ref{sec:hemispherefactorization_new},
these non-perturbative effects arise from the restriction on the radiation that is soft in the $e^+e^-$ rest frame through its assignment to either the top or the antitop hemisphere, and they also govern the non-perturbative effects associated to the $b$-jet 
definition we employ. They can be factored from the partonic soft function 
$\hat{S}_{\mathrm{hemi}}$ in the form of the non-perturbative soft distribution function $F$ shown in Eq.~(\ref{eq:nonpertF}) and can be determined from invariant mass based $e^+e^-$ dijet eventshape distributions such as thrust~\cite{Abbate:2010xh,Benitez:2024nav}, $C$ parameter~\cite{Hoang:2014wka} or heavy jet mass~\cite{Hoang:2025uaa}. This will also allow for conceptual tests of hadronization models in Monte-Carlo event generators for observables differential in the top quark decay in the line of the studies in Refs.~\cite{Hoang:2018zrp,Hoang:2024zwl}.

Finally, we also point out that in the double resonant region the factorization theorem is tree-level exact for the top quark decay in the same way as the factorization of Eqs.~(\ref{eq:decomposition_leptonic_hadronic_tensor_new_2}) and (\ref{eq:BdecayWfinal})
for the on-shell top quark decay. As already pointed out at the end of Sec.~\ref{sec:Bdecayfactorization_new},  this implies that non-singular corrections not captured by the factorization formula are always $\alpha_s$-suppressed for decay sensitive distributions that arise already at tree-level, so that the factorization (including the summation of logarithms and the non-singular corrections) can be applied for the entire spectra of such distributions. We come back to this in our discussion in Sec.~\ref{sec:pheno}.

\subsection{Computation of the Ultra-Collinear-Soft Function at NLO}
\label{sec:ucsoftcomputations}

\begin{figure}[!t]%!htbp
\centering
\includegraphics[scale=0.23]{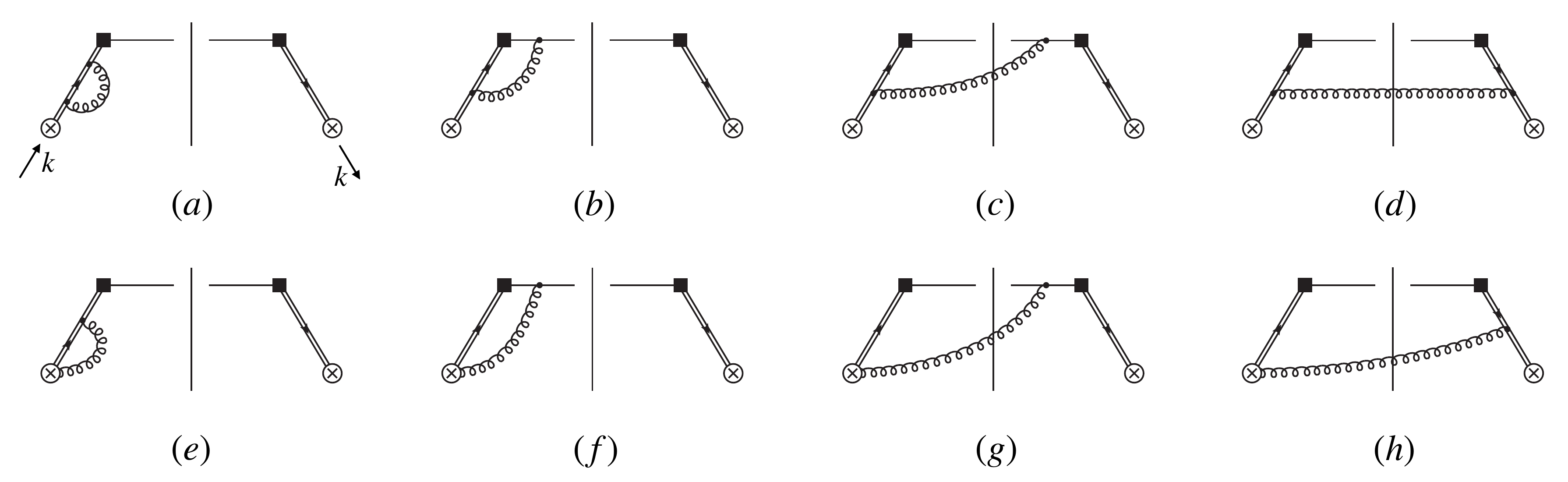}
\caption{One-loop diagrams for the computation of the NLO QCD corrections to the ultra-collinear-soft function.}
\label{fig:one_loop_diagrams_Sucs}
\end{figure}

We now discuss the computation of the $\mathcal{O}(\alpha_s)$ corrections to the ultra-collinear-soft function $\tilde{S}_{ucs}$. The corresponding one-loop diagrams are shown in Fig.~\ref{fig:one_loop_diagrams_Sucs}, where double lines denote heavy top quarks and single lines represent soft Wilson lines, $Y_{n',-}^{\mathrm{uc}}$ and $Y_{n',+}^{\mathrm{uc},\dagger}$, respectively. We stress that all diagrams except for diagram (d) have a complex conjugated mirror diagram, which are not shown in Fig.~\ref{fig:one_loop_diagrams_Sucs}.

\indent We can distinguish between two different classes of diagrams. The class I diagrams are depicted in the first row of Fig.~\ref{fig:one_loop_diagrams_Sucs} ((a) to (d)) and incorporate diagrams of the type encountered in the computation of the partonic shape function for the semileptonic $\bar{B}$ decays at $\mathcal{O}(\alpha_s)$ \cite{Bauer:2003pi, Bosch:2004th}. The partonic shape function involves matrix elements of on-shell external $b$ quark states, so that the virtual corrections in diagrams (a) and (b) yield scaleless integrals that vanish in dimensional regularization\footnote{The proper way to include the heavy quark self-energy in Fig.~\ref{fig:one_loop_diagrams_Sucs} (a) is to add the wavefunction $Z$-factor to the tree-level on-shell diagram. For heavy quark fields these wavefunction contributions have the form $1/\epsilon_{UV} - 1/\epsilon_{IR}$ and effectively exchange an infrared divergence with an ultraviolet one \cite{Bauer:2003pi}.}. The computation of the ultra-collinear soft function, on the other hand, involves an off-shellness (including a finite top quark width) for the heavy quark propagators which gives rise to non-vanishing virtual corrections. Moreover, we emphasize that we do not need to account for self-energy diagrams for the Wilson lines since, in Feynman gauge, they are proportional to $n'^2 =0$. The computation of diagrams (a) to (d) is rather straightforward and we refer to Ref.~\cite{Regner} for the details. The sum of the first class of diagrams in Feynman gauge is given in the appendix, see Eq.~(\ref{eq:NLO_corrections_Sucs_class1}).

Note that the sum of diagrams (a) to (d) is not separately gauge-invariant and that also the sum of their $1/\epsilon$ divergences does not agree with those related to the shape function shown in Eq.~(\ref{eq:ucs_divergentpart}). In particular,
the last term in Eq.~(\ref{eq:NLO_corrections_Sucs_class1}), which involves the factor $1/(\Delta - \Delta^*) = 1/(i \Gamma_t)$, arises from the real radiation diagram (d). In the context of the shape function computation for an on-shell external heavy quark this diagram yields an IR $1/\epsilon$ divergence and distributions in $\bar{a}^+$. In the ucs function the top width acts as an IR regulator so that this divergence does not arise and, furthermore, the result is a regular function and not a distribution.

The class II diagrams are shown in the second row of Fig.~\ref{fig:one_loop_diagrams_Sucs} ((e) to (h)) and involve gluons arising from the collinear Wilson lines of the top quark jet function. The loop integral of diagram (e) was already encountered in the computation of the bHQET jet function at the one-loop level and can therefore be readily evaluated using the results of Ref.~\cite{Fleming:2007xt}. This leaves the computation of the remaining three diagrams, which are more involved. The loop integral of diagram (f) is of the general form 
\begin{align} \label{eq:loop_integral_diagram_f}
I^{(f)} \sim (\bar{n} \cdott n') \,  \tilde{\mu}^{2\epsilon} \! \!  \int \! \mathrm{d}^d k  \,  \frac{1}{[v \cdott k \minus \Delta]_-} \, \frac{1}{[-\bar{n} \cdott k]_-} \, \frac{1}{[n' \! \cdot \! k]_-} \, \frac{1}{[-k^2]_-} \,,
\end{align}
where we use the shorthand notation $[x]_\pm \equiv [x \pm i0^+]$. To solve this integral, it is convenient to introduce Schwinger parameters such that, after doing the Wick rotation\footnote{Note that we needed to use a different sign convention for the propagators in the loop integral (\ref{eq:loop_integral_diagram_f}) as compared to the phase space integral in Eq.~(\ref{eq:PS_integral_diagram_g}) to ensure that after the Wick rotation we can safely introduce the Schwinger parameters $\Gamma(\nu)/A^\nu = \int_0^\infty \mathrm{d}a\,  a^{\nu-1} \mathrm{e}^{-a A}$ with $A> 0$ for Euclidean kinematics.}, the $d$-dimensional Euclidean loop momentum integral has the form of a Gaussian integral that can be easily solved and gives rise to a Schwinger parameter representation in terms of Symanzik polynomials (see e.g.~\cite{Weinzierl:2022eaz}). Alternatively, the loop integral in Eq.~(\ref{eq:loop_integral_diagram_f}) can also be solved using the Feynman parameter representation. This integral representation has the advantage that the Cheng-Wu theorem \cite{Cheng:1987ga} can be applied which facilitates the computation.
The real-emission contributions (diagrams (g) and (h)), however, turn out to be more difficult to compute. To illustrate the issue, it is instructive to take a closer look at the generic form of diagram (g), which is given by a phase space integral of the form 
\begin{align} \label{eq:PS_integral_diagram_g}
I^{(g)} \sim (\bar{n} \cdott n') \, \tilde{\mu}^{2\epsilon} \! \!  \int \! \mathrm{d}^d k \, \delta (k^2) \, \theta (k^0) \, \delta (\hat{k}^+ \! - \!  \hat{a}^+) \,  \frac{1}{[\Delta \! - \! v \! \cdot \! k]_+} \, \frac{1}{[\bar{n} \! \cdot \! k]_+} \, \frac{1}{[n' \! \cdot \! k]_-} \,.
\end{align}
It is natural to use a light cone decomposition for the momenta to solve the integral. However, the propagators and the measurement function depend on the light-like vectors $\bar{n}^\mu=(1,-\vec{n})\eqqcolon (1, \vec{\bar{n}})$ and $n'^\mu=(1,\vec{n}')$ in addition to the dependence on the off-shellness and the top quark width, so that an immediate light cone decomposition does not lead to a straightforward computation. Therefore, in a first step, we boost into a new frame\footnote{We note that the spatial directions of $n^\mu=(1,\vec n)$ and $\bar n^\mu=(1,-\vec n)$ are in general not any longer back-to-back after a boost into another frame.} where $\vec{n}'=-\vec{\bar{n}}$ so that $n'\cdot\bar{n}=2$ and the transverse space is purely space-like such that is can be parameterized employing spherical coordinates. The most convenient solution is a symmetric boost that yields the same boost factor for  $n'^\mu$ and $\bar{n}^\mu$. It turns out that this factor is just $1/\gamma$ such that $\tilde v^+=\gamma \hat{v}^+$ and $\tilde{v}^- = \gamma v^-$, where $\tilde v^\pm$ are defined in the new frame. The same method can also be applied to diagram (h).   
For details we again refer to Ref.~\cite{Regner}. The sum of all class~II diagrams is shown in Eq.~(\ref{eq:NLO_corrections_Sucs_class2}).
As for the first class of diagrams, also class~II exhibits new interesting analytic structures. The most prominent is the shifted Breit-Wigner function 
$\Gamma_t/|\bar{a}^+\gamma^2 \hat{v}^+ v^-  \minus 2\Delta|^2$ and the additional dependence on $\gamma^2 \hat{v}^+ v^-$ that parametrizes the angle between $\vec{\bar{n}}$ and $\vec{n}'$.

To conclude the calculation of the $\mathcal{O}(\alpha_s)$  corrections to the ucs function, we need to discuss the possibility of 0-bin subtractions related to the overlap of the collinear and soft modes. Such a soft-collinear overlap arises with the $n'$-collinear modes constituting the $b$-jet function $J_{n'}$ and with the large-angle soft modes in the hemisphere soft function $S_{\rm hemi}$.
From the perspective of the $b$-jet collinear modes, the dynamics of the ucs function is soft, and the associated soft-collinear overlap is already accounted for in the 0-bin subtraction of $J_{n'}$~\cite{Bauer:2011uc,Procura:2014cba,Hoang:2019fze}. 
From the perspective of the large-angle hemisphere soft modes the ucs modes are collinear, so we need to consider the associated 0-bin subtractions for the ucs function. They are obtained by 
using large-angle soft scaling for the momenta in the loop and phase space integrals of the ucs function.
The resulting 0-bin prescription can be most easily seen considering the top hemisphere rest frame, where the ultra-collinear momenta are soft and the large-angle soft radiation is collimated in the antitop direction, $k^\mu\approx k^+ \frac{\bar n^\mu}{2}$. This implies the replacements $v\cdot k \to 1/2 v^- k^+$ and $n'\cdot k\to (\bar n\cdot n')/2 \,k^+= k^+/\gamma^2$ appearing in the loop and phase space integrands.
For the class~I diagrams (a) to (d), the 0-bin contributions are subleading in the power counting and can be neglected. For the class~II diagrams (e) to (h), they yield leading power contributions which, however, lead to scaleless integrals, so that the 0-bin contributions do not yield any additional modification.
To see this for the virtual diagrams, consider the 0-bin of diagram (f) in Eq.~(\ref{eq:loop_integral_diagram_f}),
\begin{align} 
I^{(f)}_{\mathrm{0-bin}} \sim (\bar{n} \cdott n') \,  \tilde{\mu}^{2\epsilon} \! \!  \int \! \mathrm{d}^d k  \,  \frac{1}{[\frac{1}{2} v^- k^+ \minus \Delta]_-} \, \frac{1}{[-k^-]_-} \, \frac{1}{[\frac{1}{\gamma^2} k^+]_-} \, \frac{1}{[-k^2]_-} \,.
\end{align}
Introducing Feynman parameters to combine the propagators and performing the integration over the loop momentum leads to an integral representation in which one of the Feynman parameter integrals is scaleless. The 0-bin of diagram (e) appears in the computation of the jet function and is known to vanish.
For the real-emission diagrams the 0-bin replacements also modify the measurement $\delta$-function. For diagram (g) in Eq.~(\ref{eq:PS_integral_diagram_g}) this gives
\begin{align} 
I^{(g)}_{\mathrm{0-bin}} \sim \gamma^2\,  (\bar{n} \cdott n')  \tilde{\mu}^{2\epsilon} \! \!  \int \! \mathrm{d}^d k \, \delta (k^2) \, \theta (k^0) \, \delta (k^+ \minus  a^+) \,  \frac{1}{[\Delta \! - \! \frac{1}{2} v^- k^+]_+} \, \frac{1}{[ k^-]_+} \, \frac{1}{[\frac{1}{\gamma^2} \,  k^+]_-} \,.
\end{align}
Performing a light cone decomposition of the phase space integral with respect to the top and antitop jet directions, it is straightforward to show that the 0-bin contribution again leads to a scaleless integral and vanishes. In a similar fashion, one can also show that the 0-bin contribution of diagram (h) vanishes.

The sum of the UV-divergences of the class~I and II diagrams agrees 
with the UV-divergences of $\tilde{S}_{\mathrm{ucs,div}}^{(\alpha_s)}$ in the previous subsection from renormalization consistency. We note once more that the UV-divergences obtained from the explicit calculation of the class~I and II Feynman diagrams does (in general) not agree with the terms obtained in  
$\tilde{S}_{\mathrm{ucs,div}}^{(\alpha_s),\mathrm{I}}$ and $\tilde{S}_{\mathrm{ucs,div}}^{(\alpha_s),\mathrm{II}}$, respectively, which are not individually gauge-invariant. 
The final result for the renormalized ucs function up to $\mathcal{O}(\alpha_s)$ reads
{\allowdisplaybreaks
\begin{align} \label{eq:renormalized_NLO_corrections_Sucs_complete}
&\tilde{S}_{ucs} (\hat{v}^+,v^-, \gamma^2, s, \bara)= \frac{1}{4\pi m_t \hat{v}^+} \, \frac{1}{|\Delta|^2} \bigg \{ \delta (\bara) \\
&\plus \frac{\alpha_s C_F}{4\pi}  \bigg\{     \delta (\bara) \bigg[ \minus \frac{4\pi^2}{6} \plus 8   \minus  4 \log \bigg( \frac{-2\Delta}{\mu} \bigg) \minus 4 \log \bigg( \frac{-2\Delta^*}{\mu} \bigg) \plus 4\,  \mathrm{Li}_2 \bigg( \frac{1}{\tilde{\gamma}^2} \bigg) \notag \\
& \quad +\! 4 \log \bigg( \frac{-2\Delta}{\mu} \bigg) \log \big( \tilde{\gamma}^2 \big)  \plus  4 \log \bigg( \frac{-2\Delta^*}{\mu} \bigg) \log \big( \tilde{\gamma}^2 \big) 
 - 2 \log^2 \big( \tilde{\gamma}^2 \minus 1 \big) \plus 2 \log^2 \bigg( 1-\frac{1}{\tilde{\gamma}^2} \bigg)  \bigg] \notag \\
& \quad -\!  \frac{2}{\mu} \mathcal{L}_0 \Big( \frac{\bara}{\mu} \Big)  \bigg[  \log \bigg( \frac{\bara \minus 2\Delta}{\mu} \bigg) \plus \log \bigg( \frac{\bara \minus 2\Delta^*}{\mu} \bigg) \bigg] \notag \\
& \quad +\! \frac{|\Delta|^2}{|\bara \tilde{\gamma}^2  \minus 2\Delta|^2} \, \frac{8}{\mu} \mathcal{L}_0 \Big( \frac{\bara}{\mu} \Big) \bigg[  2 \log \bigg( \frac{\bara \tilde{\gamma}^2 \minus 2\Delta}{\mu} \bigg)  \notag \\
& \quad\quad +\! 2 \log \bigg( \frac{\bara \tilde{\gamma}^2 \minus 2\Delta^*}{\mu} \bigg) \minus \log \bigg( \frac{\bara \minus 2\Delta}{\mu} \bigg) \minus \log \bigg( \frac{\bara \minus 2\Delta^*}{\mu} \bigg)  \bigg] \notag  \\
& \quad +\! \frac{1}{\mu} \mathcal{L}_1 \Big( \frac{\bara}{\mu} \Big) \bigg[ \frac{16 |\Delta|^2}{|\bara \tilde{\gamma}^2 \minus 2\Delta |^2} \minus 4 \bigg] 
 +  \frac{2}{\Delta \minus \Delta^*} \bigg[ \log \bigg( \frac{\bara \minus 2\Delta}{\mu} \bigg) \minus  \log \bigg( \frac{\bara \minus 2\Delta^*}{\mu} \bigg) \bigg] \notag \\
& \quad +\! \frac{2\tilde{\gamma}^2}{|\bara \tilde{\gamma}^2  \minus 2\Delta|^2} \, \frac{\Delta \plus \Delta^*}{\Delta \minus \Delta^*} 
\times \bigg[ (\bara \tilde{\gamma}^2 \minus 2\Delta^*)  \bigg( 2 \log \bigg( \frac{\bara \tilde{\gamma}^2 \minus 2\Delta}{\mu} \bigg) \minus \log \bigg( \frac{\bara \minus 2\Delta}{\mu} \bigg) \bigg) \notag \\
& \quad  ~~~ -\! (\bara \tilde{\gamma}^2 \minus 2\Delta) \bigg( 2 \log \bigg( \frac{\bara \tilde{\gamma}^2 \minus 2\Delta^*}{\mu} \bigg) \minus \log \bigg( \frac{\bara \minus 2\Delta^*}{\mu} \bigg) \bigg) \bigg] \notag \\
&  \quad +\!  \frac{4 \tilde{\gamma}^2}{|\bara \tilde{\gamma}^2  \minus 2\Delta|^2} \,
\bigg[ \Delta \bigg( \log \bigg( \frac{\bara \minus 2\Delta}{\mu} \bigg) \minus 2 \log \bigg( \frac{\bara \tilde{\gamma}^2 \minus 2\Delta}{\mu} \bigg) \bigg) \notag \\
& \quad  +\!   \Delta^* \bigg( \log \bigg( \frac{\bara \minus 2\Delta^*}{\mu} \bigg) \minus 2 \log \Big( \frac{\bara \tilde{\gamma}^2 \minus 2\Delta^*}{\mu} \Big) \bigg) \bigg] \bigg\}   + \Big[ \delta (\bar{a}^+) \,  \frac{\Delta \plus \Delta^*}{|\Delta|^2  } +\delta'(\bara) \Big] \, \delta m_t(R)   \bigg \}.  \notag 
\end{align}
}
where $\delta m_t(R) = m_t^{\mathrm{pole}} - m_t(R)$ is the residual mass term arising in a top mass renormalization scheme $m_t(R)$ other than the pole mass. We also used the short-hand notation 
\begin{align} \label{eq:gammatilde}
\tilde{\gamma}^2 \equiv \gamma^2 \hat{v}^+ v^-\,.
\end{align}
The complicated dependence on the off-shellness momentum variable $s$, which is related to the external momentum entering the ucs function matrix element, the light-cone momentum $\bar{a}^+$, coming from the coherent momentum sum of the top-ultra-collinear radiation arising from the top production, propagation and decay stages, and the angular variable $\gamma^2$, is a reflection of the non-factorizable character of the ucs function. It poses a substantial technical challenge for the practical evaluation of the fully RG resummed cross section in Eq.~(\ref{eq:dhmdecayfinalrge}), accounting also for the convolutions involving the non-perturbative hemisphere soft distribution function $F$ shown in Eq.~(\ref{eq:nonpertF}).

\section{Phenomenlogical Evaluation at NLO}
\label{sec:pheno}

We conclude this article with a brief phenomenological ${\cal O}(\alpha_s)$  fixed-order evaluation of the factorization formula~(\ref{eq:dhmdecayfinal}) in the top quark pole mass $m_t^{\rm pole}$ scheme and without imposing a gap subtraction for the large-angle hemisphere soft function $S_{\rm hemi}$ (i.e.\ $\delta m_t(R)=\delta_{\rm gap}(R)=0$). 
We postpone the computation and analysis of the consistently resummed result at N${}^2$LL$+{\cal O}(\alpha_s)$ and the impact of infrared renormalon subtractions as well as the non-perturbative hemisphere soft function contributions to future publications. 
Using only the singular fixed-order corrections and leaving the dominant ${\cal O}(\Lambda_{\rm QCD})$ renormalons unsubtracted, we carry out an at least qualitative analysis of the NLO QCD corrections. Concretely, we discuss the impact of the QCD correction on the $b$-jet lepton invariant mass $M_{j_b\ell}
\equiv \sqrt{(H+p_\ell)^2}$ for events in the top and antitop (hemisphere) double resonant region. The latter is ensured by imposing cuts on the top and antitop hemisphere invariant masses, $|M_{t,\bar t}-m_t|\le \Delta M_t$. Similar invariant mass cuts were employed in Ref.~\cite{Liebler:2015ipp} in an analysis of fiducial (total) cross sections for $e^+e^-\to W^+W^- b\bar b$ in the double and single resonance regions with the aim to explore the possibilities of a top quark width measurement at a future $e^+e^-$ collider from the double and single resonant fiducial cross sections. In their study fat $b$-jets were also studied, but observables differential in the top quark decay were not considered.

For the ${\cal O}(\alpha_s)$  fixed-order evaluation of the factorization formula in Eq.~(\ref{eq:dhmdecayfinal}) the convolutions over the light-cone momenta 
$\ell^+$, $\ell^-$ and $\hat\ell^+$ can all be carried out trivially because there is always at least one $\delta$-function that constrains each of them to a particular value. The result, treating the top decay 3-body-phase space $\mathrm{d} \Pi_3(M_t; p_\ell, \pn, H)$ in the hemisphere rest frame, can be written down in compact form and reads 
{
\allowdisplaybreaks
\begin{eqnarray}
\label{eq:NLOfactorization}
\lefteqn{ 
\left.\frac{\mathrm{d}^3\sigma}{\mathrm{d} M_t^2 \mathrm{d} M_{\bar{t}}^2 \mathrm{d} X}\right|_{\rm NLO, sing} = }  \\
& = &  \sigma_0 \, \int \mathrm{d} H^2 \, \mathrm{d} \Pi_3(M_t= m_t v_n+k^* ; p_\ell , \pn, H) \, \delta(X-{\cal X}(M_t, p_\ell , \pn, H))
\notag \\
&&\times \, \Big( \frac{e}{\sqrt{2} \,s_w} \Big)^4 |V_{tb}|^2 \, \frac{1}{|(p_\ell+p_{\nu_{\ell}})^2 \minus M_W^2 + i M_W \Gamma_W|^2} \notag \\
&&\times  \bigg \{  2\, \hat{H}^- \,  \hat{p}_{\nu_\ell}^+  (p_\ell \cdott  v) \,  \frac{1}{4 \pi m_t} \, \frac{1}{|\Delta_t|^2} \,  \delta ( H^2) \, J_{B_{\bar{t}}}^{\mathrm{tree}} (s_{\bar{t}}^*)       \notag \\ 
&& \quad + 2\, \hat{H}^- \,  \hat{p}_{\nu_\ell}^+  (p_\ell \cdott  v)  \, \frac{1}{4 \pi m_t} \, \frac{1}{|\Delta_t|^2} \,  \delta ( H^2) \bigg[  J_{B_{\bar{t}}}^{(\alpha_s)} (s_{\bar t}^*) \plus \frac{1}{Q} \, \mathrm{Im} \Big[ \hat{\mathcal{S}}_{\mathrm{hemi}}^{(\alpha_s)} \Big( \frac{m_t}{Q} (s_{\bar{t}}^* \plus i \Gamma_t) \Big) \Big] \bigg]  \notag \\
&&  \quad+  2\, \hat{H}^- \, \hat{p}_{\nu_\ell}^+  (p_\ell \cdott  v)   \, J_{B_{\bar{t}}}^{\mathrm{tree}} (s_{\bar{t}}^*)  \bigg[  \frac{1}{4 \pi m_t} \, \frac{1}{|\Delta_t|^2}    \, \delta(H^2) \Big[ H_Q^{(\alpha_s)} (Q)  \plus H_{m}^{(\alpha_s)} \Big(m_t,\frac{Q}{m_t}\Big) \Big] \notag \\
&& \hspace{4.7cm}+ \frac{1}{\hat{H}^-} \, \tilde{S}_{ucs}^{(\alpha_s)}\Big(1,1, \gamma^2, s_t^{*} , \frac{H^2}{\hat{H}^-} \Big) \plus \frac{1}{4 \pi m_t} \, \frac{1}{|\Delta_t|^2} \, J_{n'}^{(\alpha_s)} (H^2)  \notag\\
&& \hspace{4.7cm} + \frac{1}{4 \pi m_t} \, \frac{4}{|s_t^* \minus \gamma^2\frac{H^2}{\hat{H}^-} \plus i \Gamma_t|^2} \, \frac{\gamma^2 m_t}{Q \hat{H}^-} \,  S_{\mathrm{hemi}}^{(\alpha_s)} \Big( \frac{\gamma^2 m_t}{Q} \, \frac{H^2}{\hat{H}^-} \Big)   \bigg] \notag \\
&& \quad +   \frac{1}{4 \pi m_t} \, \frac{1}{|\Delta_t|^2} \,  \delta ( H^2) \, J_{B_{\bar{t}}}^{\mathrm{tree}} (s_{\bar{t}}^*)   \,  \hat{H}^-     \bigg[  2 \hat{p}_{\nu_\ell}^+  (p_\ell \cdott  v)\,  H_{11}^{(\alpha_s)}(m_t,\hat{H}^-) \notag \\
&& \hspace{6cm}  + \Big( \hat{p}_{\nu_\ell}^+ (p_\ell \cdott v) \plus \hat{p}_\ell^+ (\pn \cdott v) \minus \frac{q^2}{2}   \Big) \, H_{12}^{(\alpha_s)}(m_t,\hat{H}^-) \notag \\
&& \hspace{6cm} + 2 \hat{p}_\ell^+ \hat{p}_{\nu_\ell}^+ \, H_{13}^{(\alpha_s)}(m_t,\hat{H}^-) \bigg] \bigg \}\,, \notag
\end{eqnarray}
}%
where the tree-level usc function (see Eq.~(\ref{eq:Sucstree})) has been written out explicitly, 
$J_{B_{\bar{t}}}^{\mathrm{tree}}(s_{\bar{t}}^*)$ is the tree-level bHQET antitop quark jet function and $\Delta_t\equiv (s_t^*+i\Gamma_t)/2$, and we recall the definitions for the hemisphere off-shell variables in Eq.~(\ref{eq:hemisphere_offshellness_bHQET}). The terms $F^{(\alpha_s)}$ refer to the ${\cal O}(\alpha_s)$ corrections of the other factorization functions and hard factors. The corresponding expressions are all collected in App.~\ref{app:NLO_corrections} (see
Eq.~(\ref{eq:renormalized_NLO_corrections_Sucs_complete}) for the ucs function). 
The analogue ${\cal O}(\alpha_s)$ fixed-order result for the on-shell top quark decay obtained from the factorization formula in Eqs.~(\ref{eq:decomposition_leptonic_hadronic_tensor_new_2}) and (\ref{eq:BdecayWfinal})  is shown in Eq.~(\ref{eq:ontopNLOfactorization}).
We have suppressed the renormalization scale $\mu$ in all their arguments. We recall that
for $e^+e^-\to t\bar t$ the tree-level result is $\alpha_s$-independent and that all explicit logarithmic scale-dependence cancels in the sum of all ${\cal O}(\alpha_s)$ corrections so that the only scale-dependence arises through the value of the strong coupling.
Details concerning the evaluation of the  3-body phase space $\mathrm{d} H^2\mathrm{d} \Pi_3(M_t; p_\ell, \pn, H)$ integrals in the top hemisphere rest frame are presented in App.~\ref{app:phasespace}. For simplicity we treat the $W$-boson in the narrow width limit ($\Gamma_W=0$) and recall that we set the bottom quark mass to zero. 

\begin{figure}[!t]%!htbp
	\centering
	\includegraphics[width=0.48\linewidth]{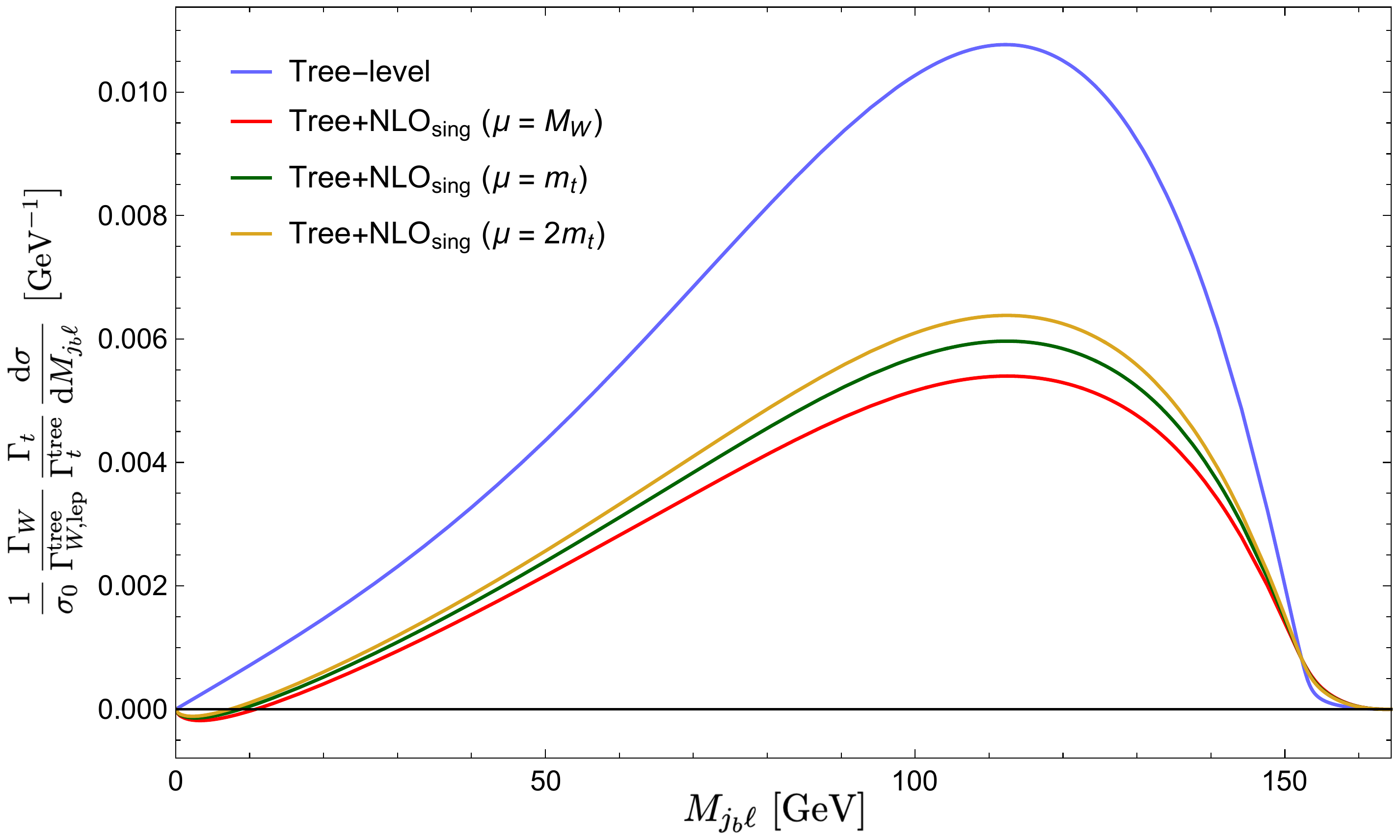}
	\hspace{5pt}
	\includegraphics[width=0.48\linewidth]{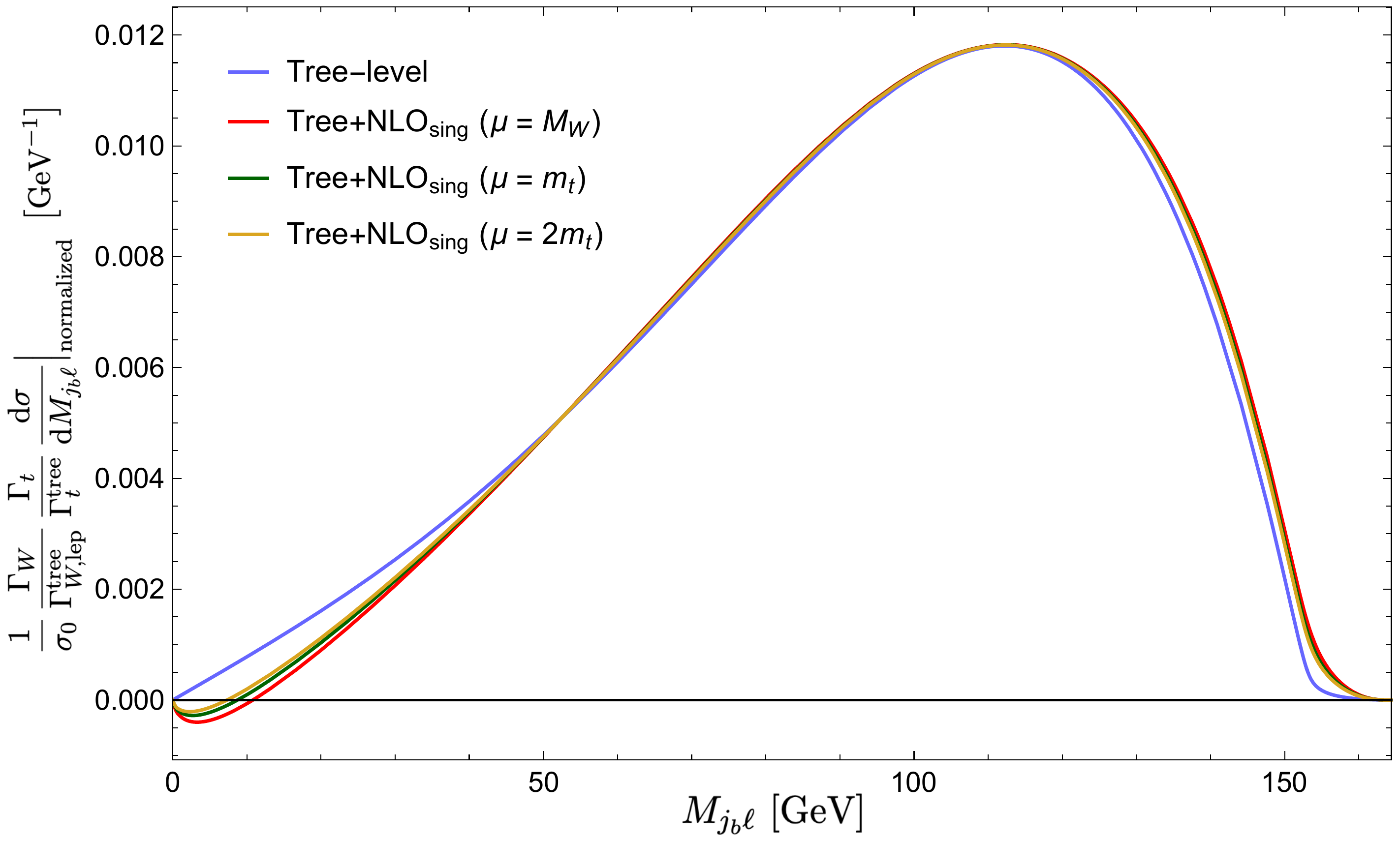}
	\caption{Left panel: Tree-level (blue) and ${\cal O}(\alpha_s)$ fixed order evaluation of the factorization theorem for the $b$-jet lepton invariant mass $M_{j_b\ell}$ distribution of Eq.~(\ref{eq:NLOfactorization}) for $Q=700$~GeV, $m_t=173$~GeV and $\Gamma_t=1.42$~GeV and cuts on the top and antitop hemisphere invariant masses $|M_{t,\bar t}-m_t|<10$~GeV in the pole mass scheme and without gap subtraction. The $W$-boson is treated in the narrow width limit $\Gamma_W\to 0$, and we used the tree-level (partial) widths $\Gamma_{W,{\rm lep}}^{\rm tree}=G_F M_W^3/(6\pi\sqrt{2})$ and $\Gamma_t^{\rm tree}=G_F m_t^3/(8\pi\sqrt{2})(1-M_W^2/m_t^2)^2(1+2M_W^2/m_t^2)$ for normalization. At ${\cal O}(\alpha_s)$ we used the renormalization scales $\mu=M_W$ (red), $m_t$ (green) and $2m_t$ (yellow). The right panel shows all distributions normalized to unity to visualize the impact of the ${\cal O}(\alpha_s)$ corrections on the shape of the distribution.
}
	\label{fig:Mbldist}
\end{figure}

\begin{figure}[!t]%!htbp
	\centering
	\includegraphics[width=0.48\linewidth]{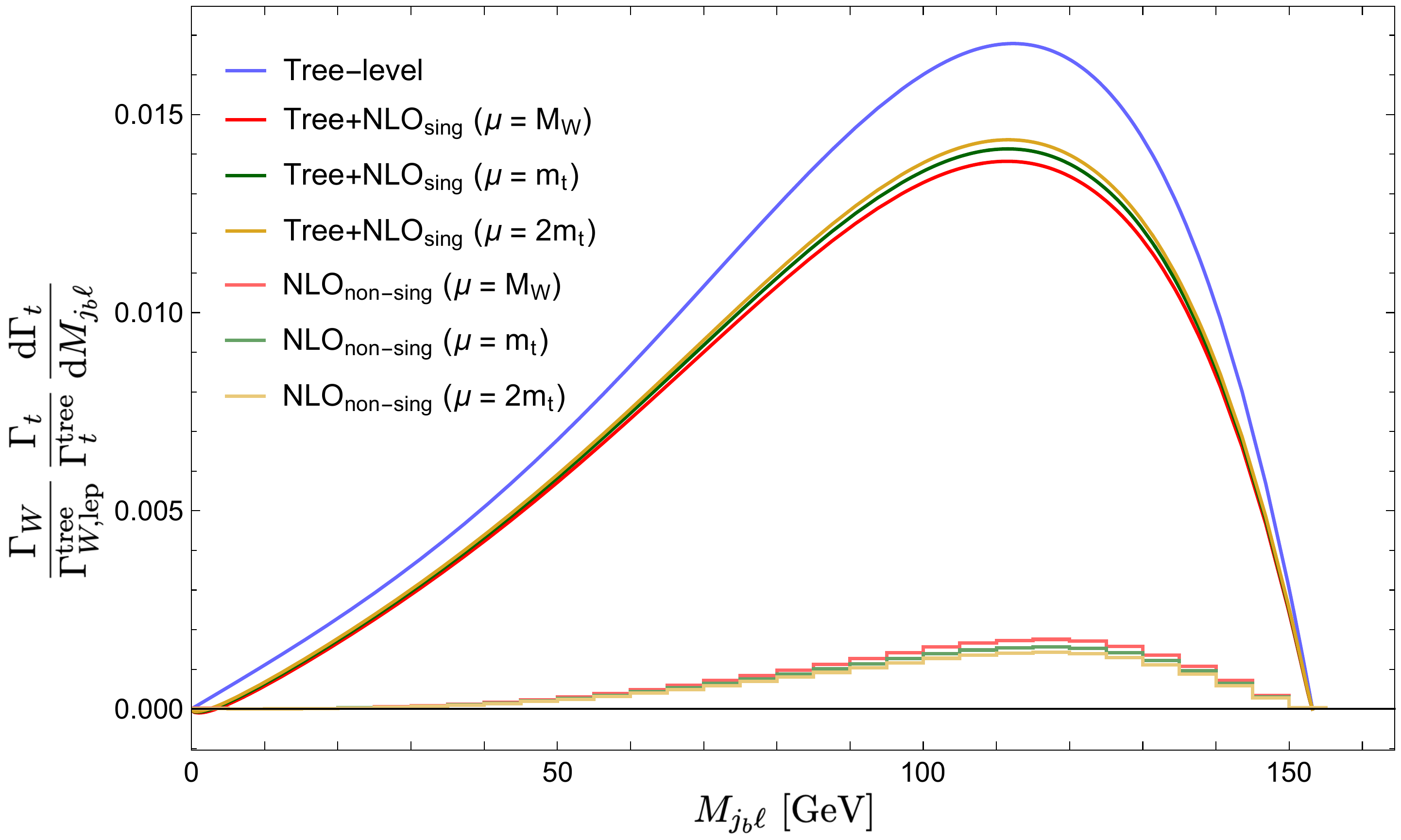}
	\hspace{5pt}
	\includegraphics[width=0.454\linewidth]{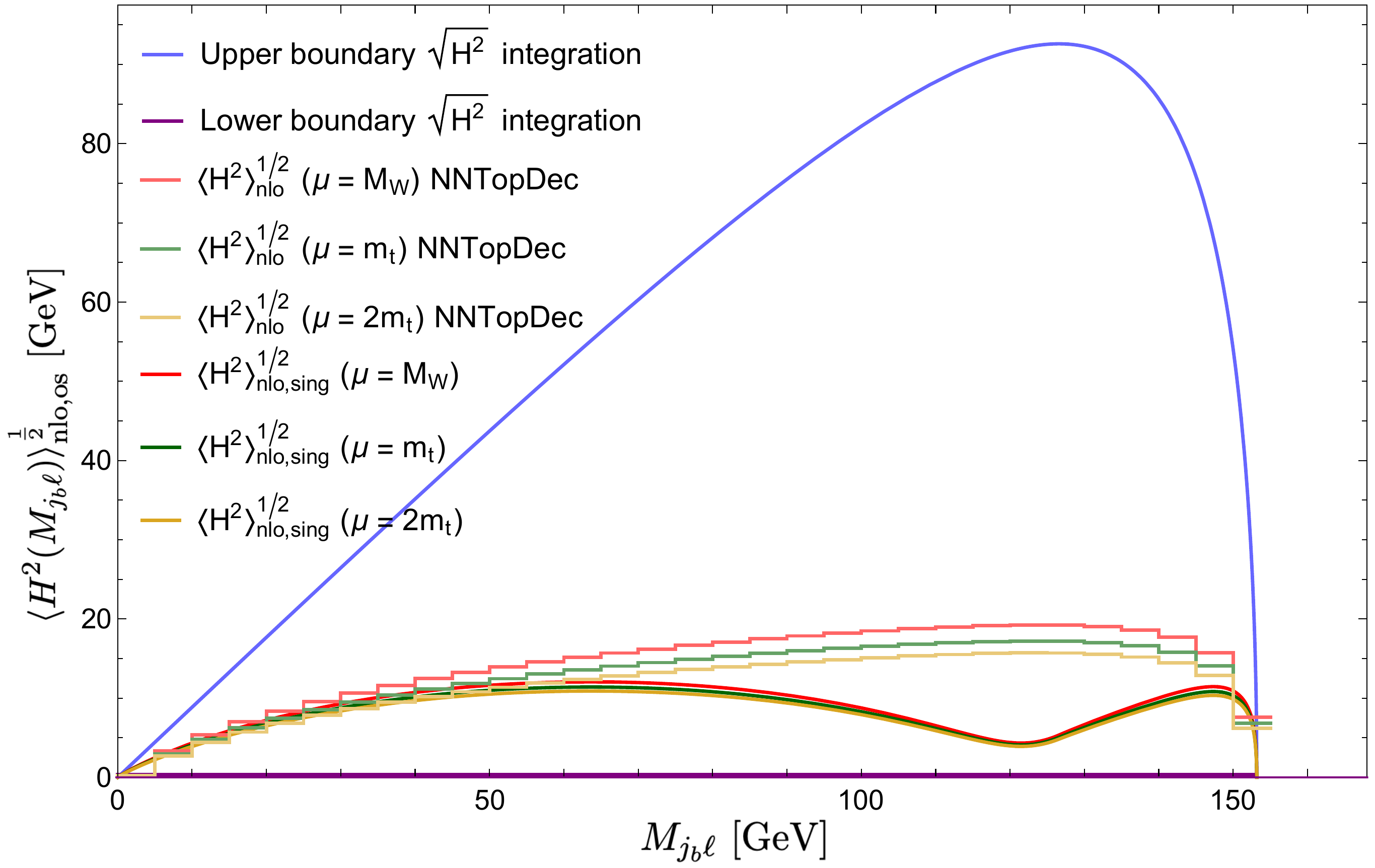}
	\caption{
	Left panel: Tree-level (blue) and ${\cal O}(\alpha_s)$ fixed order evaluation of the on-shell top quark endpoint factorization theorem for the $M_{j_b\ell}$ distribution of Eq.~(\ref{eq:ontopNLOfactorization}) with $m_t=173$~GeV in the pole mass scheme and without gap subtraction. The $W$-boson is treated in the narrow width limit $\Gamma_W\to 0$, and we used the normalization as for the left panel in Fig.~\ref{fig:Mbldist}. At ${\cal O}(\alpha_s)$ we used the renormalization scales $\mu=M_W$ (red), $m_t$ (green) and $2m_t$ (yellow). The light colored histograms show the ${\cal O}(\alpha_s)$ non-singular corrections for the same renormalization scales. Right panel:
	Average ${\cal O}(\alpha_s)$ invariant mass of the hadronic final state for the on-shell top decay as a function of $M_{j_b\ell}$ for the same renormalization scales. The light colored histograms show the corresponding full exact ${\cal O}(\alpha_s)$ results. The colored solid lines show the ${\cal O}(\alpha_s)$ singular results from the factorization theorem. The blue line is the maximal possible hadronic invariant mass as a function of $M_{j_b\ell}$. 
    }  
	\label{fig:onshellplots}
\end{figure}

In left panel of Fig.~\ref{fig:Mbldist} we show the distribution
\begin{eqnarray}
\left.\frac{\mathrm{d}\sigma}{\mathrm{d}M_{j_b\ell}}(\Delta M_t)\right|_{\rm NLO, sing}
\equiv
\int_{(m_t-\Delta M_t)^2}^{(m_t+\Delta M_t)^2}\mathrm{d} M_t^2
\int_{(m_t-\Delta M_t)^2}^{(m_t+\Delta M_t)^2}\mathrm{d} M_{\bar{t}}^2
 \left.\frac{\mathrm{d}^3\sigma}{\mathrm{d} M_t^2 \mathrm{d} M_{\bar{t}}^2 \mathrm{d} M_{j_b\ell}}\right|_{\rm NLO, sing} 
\end{eqnarray}
for $Q=700$~GeV, $m_t^{\rm pole}=173$~GeV, $\Gamma_t=1.42$~GeV, $M_W=80.4$~GeV and the invariant hemisphere mass cut $\Delta M_t=10$~GeV.  The cut on the hemisphere invariant masses ensures the validity of the bHQET power counting on which our factorization formula is based on. The blue line is the tree-level result. For the renormalization scales of the NLO results we adopt $\mu=M_W$ (red line), $m_t^{\rm pole}$ (green line) and $2m_t^{\rm pole}$ (yellow line). We remind the reader that for any possible choice of the renormalization scale there are large logarithms in the fixed-order expansion because of the wide range of physical scales entering the observable encoded by the different factorization functions. These scales are $\Gamma=\Gamma_t$, $m_t\Gamma/Q$, $\sqrt{m_t\Gamma}$, $m_t$ and $Q$, see Tab.~\ref{tab:modetable}.\footnote{Even if non-boosted top quark are considered, i.e.\ when $Q\gtrsim 2 m_t$, fixed-order predictions (in the resonance region) still suffer unavoidably from large logarithms due to the hierarchy $m_t\gg \Gamma_t$.}
Thus large logarithms affect the overall normalization of the fixed-order distribution (via the hard factors $H_Q$, $H_m$ and $H_{ij}$) and the shape of the distribution (via the momentum-dependent factorization functions). For comparison, in the left panel of Fig.~\ref{fig:onshellplots} we also display the analogous tree-level and the NLO fixed-order corrections for on-shell top quark decay (for $m_t^{\rm pole}=173$~GeV and $M_W=80.4$~GeV) using the same colors.

Let us first have a look at the tree-level result (blue line), which has the shape of the distribution familiar from previous fixed-order studies~\cite{Liebler:2015ipp,ChokoufeNejad:2016qux,Denner:2023grl}. The distribution starts at zero for $M_{j_b\ell}=0$, the kinematic limit for $m_b=0$ and drops again to very small values at 
$M_{j_b\ell}\approx [(m_t^{\rm pole})^2-M_W^2]^{1/2}=153.2$~GeV, with some smearing due to the finite top quark width. Apart from this smearing, the result looks very similar to the one for the on-shell top decay.
The strict kinematic upper boundary of the distribution is at $M_{j_b\ell,\rm max}=[M_{t,\rm max}^2-M_W^2]^{1/2}=164.4$~GeV, but there the distribution is already strongly suppressed as the bulk of the events are produced close to the top resonance peak, where $M_{t}$ is close to $m_t^{\rm pole}$.  As the visible endpoints are determined by kinematics, 
this implies that the behavior of the NLO QCD corrections close to the endpoints can be qualitatively understood from simple 3-body ($t\to j_b\bar\ell\nu_\ell$) kinematics where the corrections to the resonance peak location modify the mass of the initial top (plus coherent gluons) state and the average $b$-jet invariant mass $H^2$, which are increased due to the QCD radiation from production stage and decay state radiation. The impact of these QCD corrections cannot be captured fully correctly considering only the on-shell top quark decay.

\begin{figure}[!t]%!htbp
	\centering
	\includegraphics[width=0.95\linewidth]{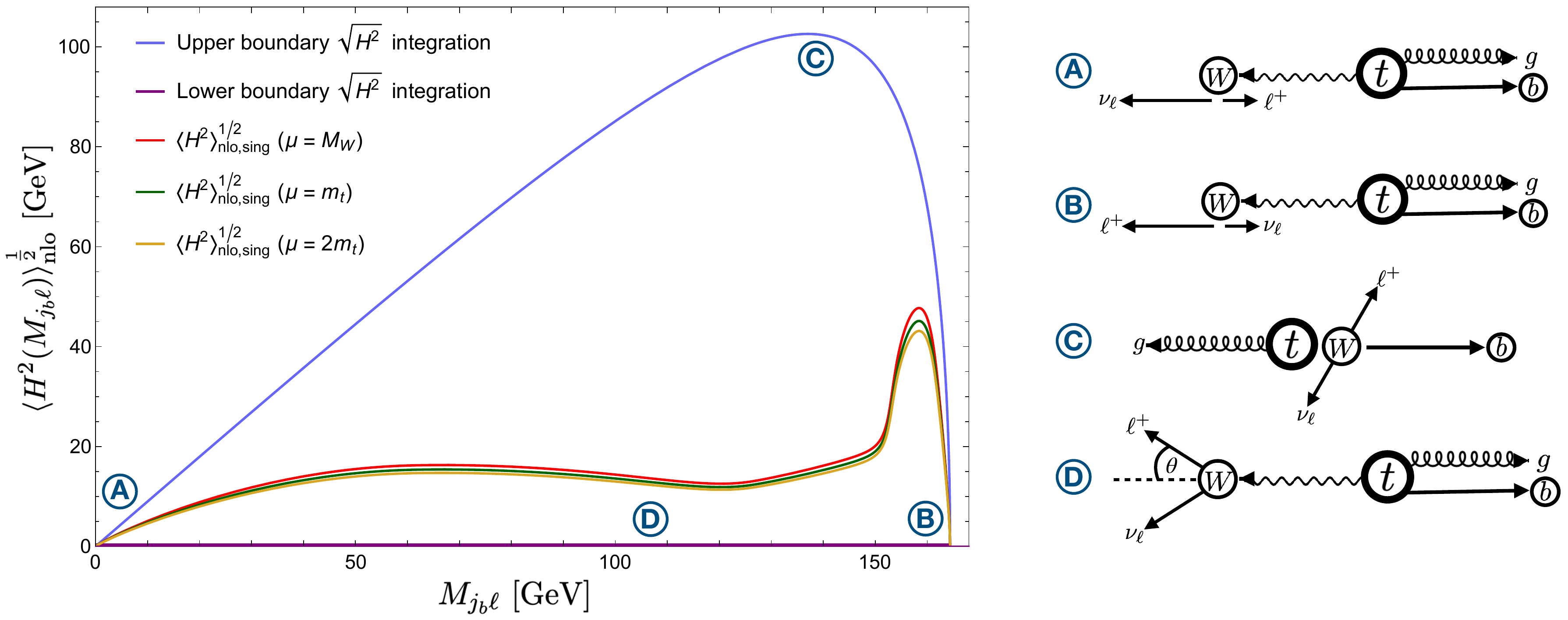}
	\caption{Average invariant mass of the hadronic $b$ plus gluons system $H^2=(M_t-p_\ell-p_\nu)^2$ in the top quark hemisphere as a function of $M_{j_b\ell}$ at ${\cal O}(\alpha_s)$ in the fixed-order expansion from the factorization theorem for the renormalization scales $\mu=M_W$ (red), $m_t$ (green) and $2m_t$ (yellow) and for the same  parameters as used in Fig.~\ref{fig:Mbldist}. The blue line is the maximal possible $\sqrt{H^2}$ as a function of $M_{j_b\ell}$ in the top hemisphere phase space integration. Indicated are the decay configurations for the lower (A) and upper (B) $M_{j_b\ell}$ endpoints and the phase space point where $H^2$ is maximal (C).}
	\label{fig:H2distribution}
\end{figure}

Before discussing the NLO results for the $M_{j_b\ell}$ distribution, it is useful to discuss the validity the small-$H^2$ ($b$-jet invariant mass) limit, on which our factorization result is based, and the possible impact of the ${\cal O}(\alpha_s)$ non-singular corrections which have not yet been calculated. In Fig.~\ref{fig:H2distribution} the blue line shows the upper boundary of the $\sqrt{H^2}$ phase space integration arising in the $M_{j_b\ell}$ distribution (see Tab.~\ref{table:mlb_h2_g2_plm_new}) as a function of $M_{j_b\ell}$. 
The analogous upper $\sqrt{H^2}$ boundary for the on-shell top decay is shown as the light blue line in the right panel of Fig.~\ref{fig:onshellplots}.
We see that the upper $H^2$ bound goes to zero at the lower and the upper endpoints. This is because $M_{j_b\ell,\rm min}=0$ (point (A)) or $M_{j_b\ell,\rm max}$ (point (B)) are reached when, in the rest frame of the decaying top quark, the $W$-boson is emitted with maximal energy back-to-back to the low invariant mass $b$-jet and the charged lepton is radiated in the $W$ or $b$ directions, respectively (see the illustrations on the right side in Fig.~\ref{fig:H2distribution}). Thus the small-$H^2$ limit in the description of the top quark decay is reliable in both endpoint regions,\footnote{For inclusive semileptonic $B$ decays, where the $W$ invariant mass is unrestricted the small-$H^2$ limit is only appropriate at the upper endpoint region.} and there the non-singular NLO corrections not captured by the factorization formula are vanishing. 
However, in the bulk of the $M_{j_b\ell}$ distribution, where also the maximum of the distribution is located, the kinematic phase space allows for values of $\sqrt{H^2}$ close to $100$~GeV, where the small-$H^2$ limit breaks down.\footnote{The maximal $b$-jet invariant mass $\sqrt{H^2}_{\rm max}=M_{t,{\rm max}}-M_W=102.6$~GeV (point (C)) arises when the $W$-boson is produced as rest, with $M_{j_b\ell}|_{H^2_{\rm max}}=[M_{t,{\rm max}} (M_{t,{\rm max}}-M_W)]^{1/2}=137.0$~GeV. For the on-shell top decay it is the top mass that determines the maximal $b$-jet invariant mass with $\sqrt{H^2}_{\rm max,os}=92.6$~GeV and $M_{j_b\ell}|_{H^2_{\rm max,os}}=126.6$~GeV. } 
It is therefore interesting to ask the question whether the large $H^2$ phase space region is populated by many events. To address this question we can calculate the average $H^2$ as a function of $M_{j_b\ell}$. At ${\cal O}(\alpha_s)$ in the fixed-order expansion this yields the expression 

\begin{align}
& \langle H^2(M_{j_b\ell}) \rangle_{\rm nlo, sing}	 \equiv  \frac{1}{\frac{\mathrm{d}\sigma}{\mathrm{d}M_{j_b\ell}}(\Delta M_t)|_{\rm tree}}  \\
& \qquad \times
	\int_{(m_t-\Delta M_t)^2}^{(m_t+\Delta M_t)^2}\mathrm{d} M_t^2
	\int_{(m_t-\Delta M_t)^2}^{(m_t+\Delta M_t)^2}\mathrm{d} M_{\bar{t}}^2
	\int_0^{H^2_{\rm max}(M_{j_b\ell})}\mathrm{d} H^2  
	\left.\frac{H^2 \times \mathrm{d}^4\sigma}{\mathrm{d} M_t^2 \mathrm{d} M_{\bar{t}}^2 \mathrm{d} H^2 \mathrm{d} M_{j_b\ell}}\right|_{\rm NLO, sing} \,, \notag
\end{align}  
where we note that the tree-level contribution in the numerator (which is proportional to $H^2\delta(H^2)$) vanishes. The results for $\langle H^2(M_{j_b\ell}) \rangle^{1/2}_{\rm nlo, sing}$ for the three renormalization scales are shown as the colored lines in Fig.~\ref{fig:H2distribution} and indicate that the average $b$-jet invariant mass stays well below  $18$~GeV except for hump at $M_{j_b\ell}>152$~GeV, where it reaches values around $45$~GeV. This hump is an artificial feature arising from the fact that the NLO fixed-order distribution is much larger than the tree-level distribution for $M_{j_b\ell}>152$ due to a shift of the endpoint region to larger $M_{j_b\ell}$ values caused by the QCD corrections (further discussed below). When we normalize the average to the entire NLO distribution the hump disappears and the average $b$-jet invariant mass stays below $30$~GeV for the entire spectrum. Using 3-body kinematics for decaying top quark with mass $173$~GeV, this corresponds to $(E_{j_b}-|\vec{p}_{j_b}|)/(E_{j_b}+|\vec{p}_{j_b}|)<0.05$, which indicates that the phase space region where the $b$ quark plus gluon system is not jet-like is suppressed. 
The yet unknown non-singular ${\cal O}(\alpha_s)$ corrections will of course modify the average $\langle H^2(M_{j_b\ell}) \rangle_{\rm nlo, sing}$, in the middle section of the $M_{j_b\ell}$ distribution, and need to be included for precise phenomenological applications, but they will not change this conclusion. It is thus
justified that we show the NLO fixed-order result generated from the factorization formula for the entire spectrum for our qualitative discussion.
	 
This can be cross checked for the on-shell top decay from the analogous ${\cal O}(\alpha_s)$ fixed-order results, where the non-singular corrections can be determined from the exact QCD fixed-order NLO calculations of Refs.~\cite{Czarnecki:1989bz,Gao:2012ja}. Here the average $H^2$ as a function of $M_{j_b\ell}$ is obtained from the expression
\begin{align}
\label{eq:ontopH2av}
& \langle H^2(M_{j_b\ell}) \rangle_{\rm nlo,os}	 \equiv  \frac{1}{\frac{\mathrm{d}\Gamma_t}{\mathrm{d}M_{j_b\ell}}|_{\rm tree}}  \times
\int_0^{H^2_{\rm max}(M_{j_b\ell})}\mathrm{d} H^2  
\left.\frac{H^2 \times \mathrm{d}^2\Gamma_t}{\mathrm{d} H^2 \mathrm{d} M_{j_b\ell}}\right|_{\rm NLO} \,.
\end{align}  
In the right panel of Fig.~\ref{fig:onshellplots}, again for $\mu=M_W$ (red line), $m_t^{\rm pole}$ (green line) and $2m_t^{\rm pole}$ (yellow line), the average $\langle H^2(M_{j_b\ell}) \rangle_{\rm nlo,os}^{1/2}$ is shown as obtained form the ${\cal O}(\alpha_s)$ singular contributions given in Eq.~(\ref{eq:ontopNLOfactorization}). The binned results in corresponding lighter colors show the exact ${\cal O}(\alpha_s)$ results obtained from a modified version of the NNTopDec library~\cite{Gao:2012ja}.
We see that the average is always below $17$~GeV and that the non-singular contributions, which are only sizeable in the middle section of the distribution do not affect the overall conclusion that the hadronic final state system is jet-like, such that the singular contributions already provide a reasonable description of the spectrum. 

Let us now have a closer look on the ${\cal O}(\alpha_s)$ fixed-order results for the $M_{j_b\ell}$ distribution for the renormalization scales $\mu=M_W$ (red line), $m_t^{\rm pole}$ (green line) and $2m_t^{\rm pole}$ (yellow line). The results for the double resonant $t\bar t$ production from Eq.~(\ref{eq:NLOfactorization}) are shown in the left panel of Fig.~\ref{fig:Mbldist}. We see that the NLO QCD corrections are negative and quite large. The major part of these large QCD corrections are caused by the fact that they yield a sizeable radiative tail to the hemisphere masses $M_{t,\bar t}$ above the resonance peak~\cite{Dehnadi:2023msm}\footnote{As discussed in Refs.~\cite{Liebler:2015ipp}, see also Ref.~\cite{ChokoufeNejad:2016qux}, for $b$-jets with a very small jet radius, the radiative tail arises below the resonance peak.},
which leads to a strong reduction of the fiducial inclusive cross section in the presence of the hemisphere invariant mass cut $\Delta M_t$ on the top and antitop hemisphere masses. For $\mu=(M_W,m_t^{\rm pole},2m_t^{\rm pole})$ we find that the ratio of fixed-order NLO versus tree-level fiducial cross sections is $(0.68,0.70,0.74)$. The same observation was also made in the NLO fixed-order analysis of Ref.~\cite{Liebler:2015ipp}, where is was also shown that this effect is only very weakly dependent on the size of the $b$-jet radius. These large QCD correction effects on the overall rate are an artifact of the fixed-order expansion and do not arise once the resummation of large logarithms is properly accounted for due to the improved convergence~\cite{Fleming:2007xt,Dehnadi:2016snl,Bachu:2020nqn}.  
The analogous ${\cal O}(\alpha_s)$ fixed-order corrections for the on-shell top decay obtained from Eq.~(\ref{eq:ontopNLOfactorization}) are shown in the left panel of Fig.~\ref{fig:onshellplots}. Here the corrections are negative as well, but overall much smaller, since the corrections cannot affect the top production rate. In the same figure we also show the size of non-singular corrections in corresponding lighter colors (as histograms) obtained from the NNTopDec library. (We refer to the figure caption for details.) As expected, they are small in the two endpoint regions, and comparable to the singular ${\cal O}(\alpha_s)$ corrections in the middle region of the spectrum, reconfirming the conclusions from our discussion above.
 
However, apart from the large NLO fixed-order effects on the overall norm of the $M_{j_b\ell}$ distribution for the double resonant $t\bar t$ production, there are also modifications to the shape of the distribution most notably in the endpoint regions which are particularly sensitive to the top quark mass and also the mass of the $b$-jet.  
This is better visible in the right panel of Fig.~\ref{fig:Mbldist} where we plotted the tree-level and NLO distributions both normalized to unity. 
We can now better see that the lower and upper endpoints are both shifted towards larger values. For the upper endpoint, the shift amounts to about $+1$~GeV, while at the lower endpoint the shift amounts to roughly $+10$~GeV. For the lower endpoint the fixed-order QCD corrections are so large that the distribution becomes even negative.  
We remind the reader again that a fully quantitative analysis requires the summation of large logarithms (and eventually also the inclusion of non-perturbative effects). Nevertheless, it is possible to provide
a qualitative explanation for both effects using the 
simple tree-level 3-body ($t\to j_b\bar\ell\nu_\ell$) kinematics mentioned above and accounting for the
QCD corrections to the (top) resonance mass and the $b$-jet invariant mass. 
The lower and upper endpoints for $M_{j_b\ell}$ distribution for the 3-body kinematics with a finite $b$ mass and the $W$ boson in the narrow width limit read
\begin{eqnarray}
M_{j_b\ell,\rm min}^{\rm 3-body} &=&
\bigg(\frac{1}{2} \big[ m_t^2 + m_b^2 - M_W^2 - \sqrt{\lambda (m_t^2, m_b^2, M_W^2)} \big]\bigg)^{1/2} \notag \\
& =& \frac{m_t\,  m_b}{\sqrt{m_t^2-M_W^2}}+\ldots 
\\
M_{j_b\ell,\rm max}^{\rm 3-body} &=&
\bigg(\frac{1}{2} \big[ m_t^2 + m_b^2 - M_W^2 + \sqrt{\lambda (m_t^2, m_b^2, M_W^2)} \big]\bigg)^{1/2} \notag \\
& =& \sqrt{m_t^2-M_W^2} \Big[1- \frac{M_W^2 m_b^2}{2\, (m_t^2-M_W^2)^2}+\ldots \Big]   \notag
\end{eqnarray}
where in the respective second equalities we only kept the leading $m_b$ correction. Taking $m_t$ and $m_b$ as a proxies for the top resonance mass and the $b$-jet invariant mass, respectively, generated by the QCD NLO corrections, we can track the impact of NLO corrections on the endpoints. We know from the inclusive double hemisphere invariant mass factorization formula discussed in Sec.~\ref{sec:hemispherefactorization_new} that the resonance mass acquires corrections from the large-angle soft and the production-stage ultra-collinear radiation and we can write the corresponding shift as $m_t\to m_t+\Delta m_t$. From previous numerical studies of the inclusive double hemisphere mass distribution in the pole mass scheme in Ref.~\cite{Fleming:2007xt} it is known that for $Q=700$~GeV the NLO QCD corrections yield 
a positive shift of the resonance peak mass of about half a GeV, $\Delta m_t\sim 0.5$~GeV, where the major part of this positive shift arises from the large-angle soft radiation. From the factorization of the $n'$-collinear ($b$-jet) and ultra-collinear radiation contributions discussed in Sec.~\ref{sec:factorizationtopdecay}
we can track the impact of this invariant mass shift through the dependence of the external $\hat{k}^+$ term and find (see the $\delta$-function shown in  Eq.~(\ref{eq:Wtopfactfin})) that is also leads to an increase of the $b$-jet mass. 
As the average value of the angle variable $1/\gamma^2$ is $1/2$, which corresponds to the situation where the $b$-jet is radiated orthogonal to the top jet axis $\vec n$ in the top rest frame, the average of the $\hat{k}^+$ variable is $(k^++k^-)/2=v_n\cdot k$ (see Eq.~(\ref{eq:tildekplusdef})). This implies that the squared $b$-jet mass is in average shifted by $m_t \, \Delta m_t$. Writing the additional effects on the squared $b$-jet mass from the $n'$-collinear and decay-stage ultra-collinear radiations as $m_t\,\Delta_b$, the total shift on the $b$-jet mass can be written as $m_b^2\to m_t (\Delta m_t+\Delta_b)$, and we find that the net effects on the upper $M_{j_b\ell}$ endpoint amounts to 
$\Delta M_{j_b\ell,\rm max}^{\rm on-shell}\sim 0.98 \Delta m_t - 0.15 \Delta_b$.
For the lower endpoint only the shift of the $b$-jet mass contributes which yields 
$\Delta M_{j_b\ell,\rm min}^{\rm on-shell}\sim \sqrt{m_t (\Delta m_t+\Delta_b)}$, which is a much larger effect. As $\Delta_b$ is certainly positive and assuming that $\Delta m_t$ and $\Delta_b$ are of comparable size, we can qualitatively explain the observed effects. As the linear $m_b$ dependence of the lower endpoint yields a much larger non-analytic dependence on $\Delta m_t$ and $\Delta_b$ we can also understand that the associated fixed-order QCD corrections which need to generate the effects dynamically are particularly large. We expect that the basic mechanisms just discussed for the endpoint regions remain true for the fully resummed result and when non-perturbative effects are accounted for. 

It is also interesting to apply our qualitative endpoint analysis of the NLO corrections in the $M_{j_b\ell}$ distribution for the on-shell top decay. From the NLO curves shown in the left panel of Fig.~\ref{fig:onshellplots} we see that the upper endpoint is essentially unchanged, while the lower endpoint is shifted by only about $+3$~GeV. This is indeed consistent with our qualitative endpoint analysis as for the on-shell top decay the radiative corrections cannot modify the mass of the decaying top state so that we have $\Delta m_t=0$. This implies $\Delta M_{j_b\ell,\rm max}^{\rm on-shell}\sim - 0.15 \Delta_b$, which is negative and very small, and $\Delta M_{j_b\ell,\rm min}^{\rm on-shell}\sim \sqrt{m_t \Delta_b}$, which is positive, but smaller than for the double resonant $t\bar t$ production.

\section{Conclusions and Outlook}
\label{sec:conclusions}

In this article we have derived a factorization formula for boosted off-shell top-antitop pair production in $e^+e^-$ annihilation in the double resonant region where the top decays semileptonically and the $b$-jet invariant mass is small. The antitop decays hadronically and is treated inclusively. The factorization does not rely on on-shell and asymptotic intermediate top quarks or the approximation of the narrow width limit. The double resonant region is defined by cuts on the invariant masses $M_{t,\bar t}$ of the top and antitop hemispheres around the top quark mass, $|M_{t,\bar t}-m_t|<\Delta M_{t\bar t}$, and is thus determined from a measurement. The hemispheres are defined with respect to the events' thrust axis and include the lepton momenta, and the $b$-jet is defined by clustering all hadrons in the top hemisphere yielding a jet with a large radius. The factorization formula, which we derived using Soft-Collinear-Effective-Theory and boosted Heavy-Quark-Effective-Theory, allows to describe all differential observables that can be constructed from the leptons and the $b$-jet 4-momenta, and to sum all large logarithms that arise from the hierachy of scales $Q=E_{cm}\gg m_t\gg \sqrt{m_t(M_{t,\bar t}-m_t)}\sim M_{j_b}\gg M_{t,\bar t}-m_t\gtrsim \Gamma_t > \Lambda_{\rm QCD}$. These scales represent all possible dynamical scales governing the process and the observables one can construct from the momenta of the $b$-jet and the leptons coming from the top quark decay.
The factorization is based on the separation of four distinct QCD radiation modes which cannot interfere in the kinematic regime we consider: the soft radiation in either one of the top and the antitop rest frames (called top-ultra-collinear and antitop-ultra-collinear, respectively), large-angle-soft radiation in the $e^+e^-$ c.m.\ frame and hard-collinear radiation associated to the $b$-jet. Our factorization formula represents a combination of previously established factorizations for the inclusive double hemisphere invariant mass distributions in the resonance regions and hadron inclusive semileptonic heavy meson decays in the endpoint region. For the consistent combination, a particular emphasis has to be put on the frame-dependence of the momentum variables involved in the factorization such that the top decay phase space can be treated in the $e^+e^-$ c.m.\ or the top hemisphere rest frame. 

Our factorization formula has two important new features.
First, the ultra-collinear radiation for the decaying top quark,  yields a novel factorization function, which we call the ultra-collinear-soft (ucs) function. The ucs function describes the light-cone momentum distribution of the coherent soft radiation in the rest frame of the boosted top quark and is sensitive to the hard processes of $t\bar t$ production and top quark decay as well as to the propagation of the resonant (off-shell) top quark. The ucs function, which is formulated in terms of a matrix element of the time-ordered products of top production and top decay operators, encodes all the non-factorizable corrections that arise for the process, and thus allows to study the physics of non-factorizable effects at the analytic and the operator level. 
Furthermore, due to the large top quark width, the ucs function can be computed in perturbation theory, and we have determined the QCD corrections at ${\cal O}(\alpha_s)$. 
Second, our factorization provides a first principles description of the leading power (and in particular leading ${\cal O}(\Lambda_{\rm QCD})$) non-perturbative corrections to top production and decay. At leading order in the power-counting the only source of hadronization effects is the large-angle soft radiation, which is sensitive to the hemisphere boundary, that also defines the large-radius $b$-jets we employ. 
As the large-angle soft radiation cannot distinguish between the boosted top quark and the $b$-jet, which is boosted in the top direction, these hadronization effects in the resonance region are the same as for the inclusive double hemisphere invariant mass distribution and can thus be determined from $e^+e^-$ invariant mass based event-shapes such as thrust, $C$-parameter or heavy jet mass in the dijet peak region. Thus the knowledge on the hadronization effects for these $e^+e^-$ event-shapes can be used to carry out first principles studies of non-perturbative effects on the top mass sensitive features of the distributions arising from the semileptonic top decay. For top quark decay sensitive distributions that are generated already at tree-level, the non-singular corrections not captured by the factorization are ${\cal O}(\alpha_s)$-suppressed, such that our factorization formula and the summation of large logarithms are applicable for the entire spectrum.

In this article, apart from the factorization derivation and the computation of the ucs function at ${\cal O}(\alpha_s)$, we have carried out a brief NLO fixed-order analysis of the $b$-jet lepton invariant mass distribution $M_{j_b\ell}$ and discussed the qualitative impact of the NLO QCD corrections on the top mass sensitive endpoints of the $M_{j_b\ell}$ distributions. We stress that the resummation of large logarithms, which we have not carried out here, is essential for a rigorous phenomenological discussions of the perturbative predictions. Large logarithms are unavoidable for predictions of spectra arising from the top quark decay (also for non-boosted top production) due to the large hierarchy of relevant scales. With the ${\cal O}(\alpha_s)$ corrections to the ucs function we have calculated, and given that all other factorization and matching functions that arise in the factorization formula are known to at least ${\cal O}(\alpha_s)$,  the summation can be carried out at NNLL or NLL' order. In this article we have not carried out the resummation due to the complexity of the dependence of the ucs function on two light-cone convolution variables, the top quark width and the angle between the top quark and the $b$-jet directions. The complicated analytic form of the ucs function is a reflection of non-factorizable effects that are encoded in it, and which arise from the coherent treatment of the production and the decay stage radiation.  Upcoming work will address the summation of the large logarithms, and focus on detailed studies of the non-factorizable corrections as well as the non-perturbative and infrared sensitive effects, and their impact on the top mass determination. The results of this work represent an important stepping stone toward a thorough analytical theoretical understanding of the observables that are used for direct top quark mass measurements carried out at the LHC. Another essential conceptual step to get even closer to these observables is the consideration of narrower $b$-jets.

\begin{acknowledgments}
We acknowledge support by the FWF Austrian Science Fund under the Project No.~P32383-N27 and under the FWF Doctoral Program ``Particles and Interactions'' No.~W1252-N27 and the COST Action No.\ CA16201 PARTICLEFACE. We are very grateful to the Erwin-Schr\"odinger International Institute for Mathematics and Physics for partial support during the Thematic Programme ``Quantum Field Theory at the Frontiers of the Strong Interactions'', July 31 - September 1, 2023, as well as the workshop "Physics at TeV Colliders and Beyond the Standard Model 2025" at Les Houches, June 16 - 25, 2025. We thank Alessandro Broggio for discussions and providing us results obtained based on a modified version of the NNTopDec~1.0 library~\cite{Gao:2012ja}. We thank Bernd Carman for designing and creating Fig.~\ref{fig:topdecay_modes}.  

\end{acknowledgments}

\begin{appendix}

\section{Wilson Lines}
\label{app:Wilson} 

In the factorization formulae derived in this article many different Wilson lines are employed. We generally distinguish between jet field $W$-Wilson lines and $Y$-Wilson lines, that arise from BPS decoupling~\cite{Bauer:2000yr,Bauer:2001yt}. In the following we collect all Wilson lines that appear in matrix elements for the processes that arise. Wilson lines for the conjugated matrix elements are obtained by taking the Hermitian adjoint. We start with the $Y$-Wilson lines for the large-angle (ultra-)soft radiation,
\begin{align} 
&Y_{n,+}^\dagger(x) = P \, \mathrm{exp} \bigg( i g_s \int_0^\infty \mathrm{d} s \, n \cdott A_s (n s +x) \bigg) \ , \\
&Y_{\bar{n},-}(x) = \overline{P} \, \mathrm{exp} \bigg( -i g_s \int_0^\infty \mathrm{d} s \, \bar{n} \cdott A_s (\bar{n} s +x) \bigg) \,, \notag 
\end{align}
which are generated by final state quarks and antiquarks in directions $n$ and $\bar n$, respectively. The Hermitian adjoint expressions read $(Y_{n,\pm})^\dagger = Y_{n,\mp}^\dagger$. The analogous relations are valid for all other Wilson lines. Overwriting path-ordering through time-ordering yields
\begin{align} 
\label{eq:overwriteP}
&T \big[ Y_{\bar{n},-} \big]^{ab} =  \big[ \overline{Y}_{\bar{n},+}^\dagger \big]^{ba} = \bigg[  P \, \mathrm{exp} \bigg( i g_s \int_0^\infty \mathrm{d}s \, \bar{n} \cdot \overline{A}_s (\bar{n} s + x) \bigg) \bigg]^{ba} = \big[ (\overline{Y}_{\bar{n},+}^\dagger)^\intercal \big]^{ab}\ , \\
&\overline{T} \big[ Y_{\bar{n},+}^\dagger \big]^{ab}= \big[ \overline{Y}_{\bar{n},-} \big]^{ba} = \bigg[ \overline{P} \, \mathrm{exp} \bigg(- i g_s \int_0^\infty \mathrm{d}s \, \bar{n} \cdot \overline{A}_s (\bar{n} s + x) \bigg) \bigg]^{ba} = \big[ (\overline{Y}_{\bar{n},-} )^\intercal \big]^{ab} \ , \notag
\end{align}
where $\overline{A}_s = A_s^A \overline{T}^A$ with $\overline{T}^A = - (T^A)^\intercal$ being the ultra-soft gluon field in the $\bar{3}$ representation of SU(3). In contrast, $P$($\overline{P}$) ordered Wilson lines are already $T$($\overline{T}$)-ordered, e.g.\ $T [Y_{n,+}^\dagger] =Y_{n,+}^\dagger$ and $\overline{T} [Y_{n,-}] =Y_{n,-}$. For $n$-ultra-collinear gluon fields $A_n^{\rm uc}$ we employ the Wilson lines
\begin{align} 
\label{eq:ucWilsonlines}
&Y_{n',+}^{{\rm uc},\dagger}(x) = P \, \mathrm{exp} \bigg( i g_s \int_0^\infty \mathrm{d} s \, n' \cdott A_n^{\rm uc} (n' s +x) \bigg) \ , \\
&Y_{v_n,+}^{{\rm uc},\dagger}(x) = P \, \mathrm{exp} \bigg( i g_s \int_0^\infty \mathrm{d} s \, v_n \cdott A_n^{\rm uc} (v_n s +x) \bigg) \ , \notag \\
&W_{n,-}^{\mathrm{uc}}(x) = \overline{P} \, \mathrm{exp} \bigg( -i g_s \int_0^\infty \mathrm{d} s \, \bar{n} \cdott A_n^{\mathrm{uc}} (\bar{n} s +x) \bigg) \,. \notag  
\end{align}
The $n'$-(hard-)collinear Wilson line employed for the light quark jet field $\bar{\chi}_{n'} = \bar{\xi}_{n'}  W_{n'}$ reads
\begin{align} 
&W_{n'}(x) = \sum_{\rm perms} \bigg[
\exp \bigg( 
-\frac{g_s}{\bar{n}'\cdott {\cal P}} \, \bar{n}'\cdott A_{n'}(x)
\bigg)\bigg]\,,
\end{align}
where ${\cal P}$ is the SCET label momentum operator~\cite{Bauer:2001yt}. For calculations of the inclusive jet function one can use either one of the QCD definitions 
\begin{align} 
&W_{n',+}(x) = P \, \mathrm{exp} \bigg( i g_s \int^0_{-\infty} \mathrm{d} s \, \bar{n}' \cdott A_{n'} (\bar{n}' s +x) \bigg) \ , \\
&W_{n',-}(x) = \overline{P} \, \mathrm{exp} \bigg( -i g_s \int_0^\infty \mathrm{d} s \, \bar{n}' \cdott A_{n'} (\bar{n}' s +x) \bigg) \,, \notag 
\end{align} 
associated to $n'$-collinear radiation generated by initial state quarks or final state antiquarks, respectively. 
For the light quark jet field ${\chi}_{n'} =  W_{n'}^\dagger {\xi}_{n'}$ the Wilson line reads
\begin{align} 
&W^\dagger_{n'}(x) = \sum_{\rm perms} \bigg[
	\exp \bigg( 
	- g_s \, \bar{n}'\cdott A_{n'}(x) \, \frac{1}{\bar{n}'\cdott {\cal P^\dagger}} 
	\bigg)\bigg]\,, 
\end{align}
and one can then employ one of the QCD definitions,
\begin{align} 
	&W_{n',-}^{\dagger}(x) = 	
	\overline{P} \, \mathrm{exp} \bigg(- i g_s \int^0_{-\infty} \mathrm{d} s \, \bar{n}' \cdott A_{n'} (\bar{n}' s +x)  \bigg) \ , \\
	&W_{n',+}^\dagger(x) =  P \, \mathrm{exp} \bigg( i g_s \int_0^\infty \mathrm{d} s \, \bar{n}' \cdott A_{n'} (\bar{n}' s +x) \bigg) 
 \,. \notag 
\end{align}

\section{ Notation for Renormalization and Evolution}
\label{app:apprenor} 

In this appendix we present our notations concerning the $\overline{\rm MS}$ renormalization and renormalization group evolution (RGE) of all quantities occurring in the factorization theorems in the main body of this article.
We generically use $m$ for the heavy quark mass, and $Q$ stands for the hard scattering scale of massive quark-antiquark production. 

We first discuss the renormalization of all local matching factors $H_{Q}=|C_Q|^2$, $H_m=|C_m|^2$ and $H_{jj'}=C_j C_{j'}$ associated to the hard scattering and mass scales. They are associated to the renormalization of the respective $t\bar t$ production in SCET and bHQET and the $t\to b$ decay currents in HQET and their $C$ Wilson coefficients, where the relation between the bare and renormalized Wilson coefficients generically reads $C^{\text{bare}} = Z_C C$. The relations between the bare and renormalized matching factors read
\begin{align}
\label{eq:renormalization_H}
H_Q^{\mathrm{bare}}(Q) &= |Z_{C_Q} (Q,\mu)|^2 \, H_Q (Q,\mu) \,, \\
H_m^{\mathrm{bare}} \Big( m, \frac{Q}{m} \Big) &= \Big|Z_{C_m}\Big( m_t, \frac{Q}{m}, \mu \Big) \Big|^2 \, H_m \Big( m, \frac{Q}{m}, \mu \Big)\,, \notag \\
H_{jj'}^{\mathrm{bare}} (m, \hat{v}^+ \hat{H}^-) &= Z_{C_j}^2 (m_b, \hat{v}^+ \hat{H}^-,\mu) \, H_{jj'} (m, \hat{v}^+ \hat{H}^-,\mu) 
 \,. \notag 
\end{align}
The RGEs for the matching factors read
\begin{align} 
\label{eq:RGEs_Wilson_coefficients_appendix}
&\mu \frac{\mathrm{d}}{\mathrm{d} \mu}  H_Q (Q,\mu) = \gamma_{H_Q} (Q,\mu) \, H_Q(Q, \mu), &  & \gamma_{H_Q} = \gamma_{C_Q} \plus \gamma_{C_Q}^*, \\
&  \mu \frac{\mathrm{d}}{\mathrm{d} \mu}  H_m \bigg( m, \frac{Q}{m}, \mu \bigg) = \gamma_{H_m} \bigg( \frac{Q}{m}, \mu \bigg) \,  H_m \bigg( m, \frac{Q}{m}, \mu \bigg) , &    &\gamma_{H_m} = \gamma_{C_m} \plus \gamma_{C_m}^* ,,   \notag \\
&\mu \frac{\mathrm{d}}{\mathrm{d} \mu}  H_{jj'} (m, \hat{v}^+ \hat{H}^-,\mu) = \gamma_{H_d} ( \hat{v}^+ \hat{H}^-) \, H_{jj'} (m, \hat{v}^+ \hat{H}^-,\mu) , &  & \gamma_{H_d} = 2\, \gamma_{C_j}  \,. \notag
\end{align}
where all current anomalous dimensions are generically defined by
\begin{align}
\gamma_{C}(\mu) &= - Z_{C}(\mu) \, \mu \frac{\mathrm{d}}{\mathrm{d} \mu}  Z_{C}(\mu)  \,. \notag
\end{align}
The RGE evolution for the matching factors away from their respective natural scales then reads
\begin{align} \label{eq:RGEs_hard_function_appendix}
H_Q (Q, \mu) &= H_Q (Q, \mu_Q) \, U_{H_Q} (Q, \mu_Q, \mu), \\
H_m \bigg( m, \frac{Q}{m_t},  \mu  \bigg) &= H_m \bigg(m , \frac{Q}{m}, \mu_m\bigg) \, U_{H_m} \bigg( \frac{Q}{m_t}, \mu_m , \mu \bigg) , \notag \\
H_{jj'} (m,\hat{v}^+ \hat{H}^-,\mu) &= H_{jj'} (m,\hat{v}^+ \hat{H}^-,\mu_m) \, U_{H_d}(\hat{v}^+ \hat{H}^-,\mu_m,\mu) \,, \notag
\end{align}
with the generic evolution factors being defined as
\begin{align}
U_H(\mu_1,\mu_2)=\mathrm{exp} \bigg \{ \int_{\log \, \mu_1}^{\log \, \mu_2} \mathrm{d} \log \mu \, \gamma_{H} (\mu)\bigg \} \ .
\end{align}
For the mass mode matching factor $H_m$ we do not account for the summation of rapidity logarithms. They arise only at ${\cal O}(\alpha_s^2)$ and beyond from insertion of massive quark loops and their summation has been discussed in Ref.~\cite{Hoang:2015vua}. Their impact is known to be small numerically~\cite{Butenschoen:2016lpz,Bachu:2020nqn,Hoang:2024zwl}. 
For the factorization functions the relations between bare and renormalized versions have the form
\begin{align} 
\label{eq:renormalization_functions_appendix}
J_{n}^{\mathrm{bare}} (p^2) &= \int \mathrm{d} p'^2 \, Z_{J_{n}} (p^2 \minus p'^2, \mu) \, J_{n} (p'^2, \mu) \ , \\
J_{B}^{\mathrm{bare}} (\hat s) &= \int \mathrm{d} \hat s' \, Z_{J_{B}} (\hat s \minus \hat s', \mu) \, J_{B} (\hat s', \mu) \ , \notag \\
S_{\text{hemi}}^{\mathrm{bare}} (\ell^+ , \ell^- ) &= \int \mathrm{d} \ell'^+ \, \mathrm{d} \ell'^- \, Z_s (\ell^+ \minus \ell'^{+}, \mu) \, Z_s( \ell^- \minus \ell'^- , \mu) \, S_{\text{hemi}} (\ell'^+, \ell'^-, \mu) \ .  \notag \\
S_{\text{shape}}^{\mathrm{bare}} (\hat{v}^+,\hat{\ell}^+ ) &= \int \mathrm{d} \hat{\ell}'^+ \, Z_{\text{shape}} (\hat{v}^+,\hat{\ell}^+ \minus \hat{\ell}'^+, \mu) \, S_{\text{shape}} (\hat{v}^+,\hat{\ell}'^+, \mu) \ .  \notag \\
S_{ucs}^{\mathrm{bare}}(\hat{v}^+,v^-, \gamma^2, s^+, s^-, \hat{\ell}^+) & = \int \mathrm{d} s'^+   \int \mathrm{d} \hat{\ell}'^+ \, Z_{J_B} (s^+ \minus s'^+, \mu ) \, Z_{	{\mathrm{shape}}} (\hat{v}^+, \hat{\ell}^+ \minus \hat{\ell}'^+, \mu) \notag \\
&\times  S_{ucs} (\hat{v}^+,v^-, \gamma^2, s'^+, s^-, \hat{\ell}'^+, \mu) \,, \notag
\end{align}
where the dynamical variables of all factorization functions have dimensions of energy, except for the SCET jet function $J_n$, where $p^2$ has dimension of energy squared. The resulting RGEs have the form
\begin{align}
\mu \frac{\mathrm{d}}{\mathrm{d} \mu} J_{n} (p^2, \mu) &=  \int \mathrm{d} p'^2 \, \gamma_{J_{n}} (p^2 \minus p'^2, \mu) \, J_{n} (p'^2, \mu) \ , \\
\mu \frac{\mathrm{d}}{\mathrm{d} \mu} J_{B} (\hat s, \mu) &= \int \mathrm{d} \hat s' \, \gamma_{J_{B}} (\hat s \minus \hat s', \mu) \, J_{B} (\hat s', \mu) \ , \notag \\
\mu \frac{\mathrm{d}}{\mathrm{d} \mu} S_{\mathrm{hemi}} (\ell^+ , \ell^- , \mu) &= \int \mathrm{d} \ell'^+ \,    \gamma_s(\ell^+ \minus \ell'^+, \mu) S_{\mathrm{hemi}} (\ell'^+, \ell^-, \mu) \notag \\
& \quad  \plus \int \mathrm{d} \ell'^-  \gamma_s( \ell^- \minus \ell'^-, \mu) S_{\mathrm{hemi}} (\ell^+, \ell'^-, \mu) \ , \notag \\
\mu \frac{\mathrm{d}}{\mathrm{d} \mu} S_{\mathrm{shape}} (\hat{v}^+,\hat{\ell}^+, \mu) &= \int \mathrm{d} \hat{\ell}'^+ \, \gamma_{S_{\mathrm{shape}}} (\hat{v}^+,\hat{\ell}^+ \minus \hat{\ell}'^+, \mu) \, S_{\mathrm{shape}} (\hat{v}^+,\hat{\ell}'^+, \mu) \ , \notag \\
\mu \frac{\mathrm{d}}{\mathrm{d} \mu}  S_{ucs} (\hat{v}^+,v^-, \gamma^2,& s^+, s^-, \hat{\ell}^+, \mu)  =  \int \mathrm{d} s'^+ \gamma_{J_{B}} ( s^+ \minus  s'^+, \mu) \, S_{ucs} (\hat{v}^+,v^-, \gamma^2, s'^+, s^-, \hat{\ell}^+, \mu) \notag \\
& \quad  \plus \int \mathrm{d} \hat{\ell}'^+ \, \gamma_{S_{\mathrm{shape}}} (\hat{\ell}^+ \minus \hat{\ell}'^+, \mu) \, S_{ucs} (\hat{v}^+,v^-, \gamma^2, s^+, s^-, \hat{\ell}'^+, \mu) \,,\notag
\end{align}
where the anomalous dimensions for the matching function $F(t,\mu)$ with momentum variable $t$ are generically defined by
\begin{align}
\gamma_{F} (t , \mu) &= -\int \mathrm{d} t' \,  Z^{-1}_{F} (t \minus t', \mu) \, \mu \frac{\mathrm{d}}{\mathrm{d} \mu} Z_{F} (t', \mu) \ ,
\end{align}
The RGE evolution of the factorization function away from their respective natural scales then reads
\begin{align} 
\label{eq:general_solution_functions_RGE_appendix}
J_{n} (p^2, \mu) &=\int \mathrm{d} p'^2 \, U_{J_{n}} ( p^2 \minus p'^2, \mu, \mu_J) \, J_{n} (p'^2 , \mu_J) \ , \\
J_{B} (\hat s, \mu) &= \int \mathrm{d} \hat s' \, U_{J_{B}} (\hat s \minus \hat s', \mu, \mu_B) \, J_{B_{t}} (\hat s', \mu_B) \ , \notag \\
S_{\mathrm{hemi}} (\ell^+ , \ell^- \mu) &= \int \mathrm{d} \ell'^+ \, \mathrm{d} \ell'^- \, U_s (\ell^+ \minus \ell'^{+}, \mu, \mu_S) \, U_s( \ell^- \minus \ell'^- , \mu, \mu_S) \, S_{\text{hemi}} (\ell'^+, \ell'^-, \mu_S) \,, \notag \\
S_{\text{shape}} (\hat{v}^+,\hat{\ell}^+ ) &= \int \mathrm{d} \hat{\ell}'^+ \, U_{\text{shape}} (\hat{v}^+,\hat{\ell}^+ \minus \hat{\ell}'^+, \mu,\mu_B) \, S_{\text{shape}} (\hat{v}^+,\hat{\ell}'^+, \mu_B) \,,  \notag \\
S_{ucs}(\hat{v}^+,v^-, \gamma^2, & s^+, s^-, \hat{\ell}^+) = \int \mathrm{d} s'^+   \int \mathrm{d} \hat{\ell}'^+ \, U_{J_B} (s^+ \minus s'^+, \mu ,\mu_B) \, U_{	{\mathrm{shape}}} (\hat{v}^+, \hat{\ell}^+ \minus \hat{\ell}'^+, \mu, \mu_B) \notag \\
&\times  S_{ucs} (\hat{v}^+,v^-, \gamma^2, s'^+, s^-, \hat{\ell}'^+, \mu_B) \,, \notag
\end{align}
with the generic evolution factors defined as
\begin{align}
U_F(t,\mu_2,\mu_1)=\mathrm{exp} \bigg \{ \int_{\log \, \mu_1}^{\log \, \mu_2} \mathrm{d} \log \mu \, \gamma_{F} (t,\mu)\bigg \} \ .
\end{align}

\section{Collection of ${\cal O}(\alpha_s)$ Results}
\label{app:NLO_corrections}

Here we collect the ${\cal O}(\alpha_s)$ results of all perturbative ingredients for the factorization formulae we discuss in this article, covering the matching factors, the factorization functions, their $Z$-factors as well as their anomalous dimensions. All results are given in the $\MSbar$ scheme. We also provide the references to all available higher order results.

The results for the Wilson coefficients of the SCET and bHQET quark-antiquark production~\cite{Fleming:2007xt} 
and the HQET heavy-to-light decay currents~\cite{Bauer:2000yr,Chay:2002vy} read  ($Q^2 \equiv Q^2 +i0^+$, $y=\hat{v}^+ \hat{H}^-/m$)
\begin{align}
&C(Q,\mu)=1 \plus \frac{\alpha_s C_F}{4\pi} \bigg[   -\! \log^2 \bigg( \frac{\mu^2}{-Q^2} \bigg) \minus 3\, \log \bigg( \frac{\mu^2}{-Q^2} \bigg)  \minus 8 \plus \frac{\pi^2}{6} \, \bigg] \ , \\
&C_m \Big (m, \frac{Q}{m}  , \mu \Big) = 1 \plus \frac{\alpha_s C_F}{4\pi} \bigg[ \log^2 \bigg( \frac{\mu^2}{m^2} \bigg) \plus \log \bigg( \frac{\mu^2}{m^2} \bigg) \plus 4 \plus \frac{\pi^2}{6} \bigg] \ , \notag \\
\begin{split}
&C_{1}(m, \hat{v}^+ \hat{H}^-, \mu)=1 \plus\frac{\alpha_s C_F}{4 \pi} \, \bigg[ \minus 2\,  \log^2 \bigg( \frac{y\,  m}{\mu} \bigg) \plus  5\, \log \bigg( \frac{y \,m}{\mu} \bigg) \minus 2 \,\log  y \minus \frac{ \log  y}{1 \minus y} ,\notag \\
&~~~~~~~~~~~~~~~~~~~~ \minus 2\, \text{Li}_2 (1 \minus y) \minus  \frac{\pi^2}{12} \minus 6 \bigg] ,\notag \\
&C_{2}(m, \hat{v}^+ \hat{H}^- ,\mu)=\frac{\alpha_s C_F}{4 \pi} \, \frac{2}{1 \minus y} \, \bigg[ \frac{y \, \log  y}{1 \minus y} \plus 1\bigg] , \notag \\
&C_{3}(m, \hat{v}^+ \hat{H}^-, \mu)=\frac{\alpha_s C_F}{4 \pi} \, \frac{y}{1 \minus y} \, \bigg[ \frac{1 \minus 2 y  }{1 \minus y}  \, \log  y  \minus 1\bigg] .
\end{split}
\end{align}
This yields the following expressions for the local matching factors
\begin{align}
&H_Q (Q,\mu) = 1\plus \frac{\alpha_s C_F}{4\pi} \bigg[   -2\, \log^2 \bigg( \frac{Q^2}{\mu^2} \bigg)  \plus 6\, \log \bigg( \frac{Q^2}{\mu^2} \bigg) \minus 16 \plus \frac{7 \pi^2}{3} \bigg] \ , \\
&H_m \Big( m,  \frac{Q}{m},  \mu \Big) = 1 \plus \frac{\alpha_s C_F}{4\pi} \bigg[ 2\, \log^2 \bigg( \frac{\mu^2}{m^2} \bigg) \plus 2\, \log \bigg( \frac{\mu^2}{m^2} \bigg) \plus 8 \plus \frac{\pi^2}{3} \bigg] \ , \notag \\
&H_{11}(m, \hat{v}^+ \hat{H}^-, \mu)=1 \plus\frac{\alpha_s C_F}{4 \pi} \, \bigg[ \minus 4\,  \log^2 \bigg( \frac{y\,  m}{\mu} \bigg) \plus  10\, \log \bigg( \frac{y \,m_b}{\mu} \bigg) \minus 4 \,\log  y \minus \frac{ 2\log  y}{1 \minus y}  \notag \\
&~~~~~~~~~~~~~~~~~~~~ \minus 4 \, \text{Li}_2 (1 \minus y) \minus  \frac{\pi^2}{6} \minus 12 \bigg] ,\notag \\
&H_{12}(m, \hat{v}^+ \hat{H}^-, \mu)=\frac{\alpha_s C_F}{4 \pi} \, \frac{2}{1 \minus y} \, \bigg[ \frac{y \, \log  y}{1 \minus y} \plus 1\bigg] , \notag \\
&H_{13}(m, \hat{v}^+ \hat{H}^-, \mu)=\frac{\alpha_s C_F}{4 \pi} \, \frac{y}{1 \minus y} \, \bigg[ \frac{1 \minus 2 y  }{1 \minus y}  \, \log  y  \minus 1\bigg] \,. \notag
\end{align}
Since only Wilson coefficient $C_1$ is non-zero at tree-level, the factors $H_{22}$, $H_{23}$ and $H_{33}$ do not contribute at ${\cal O}(\alpha_s)$.
The QCD corrections to the hard function $H_Q$ are known at the two-loop \cite{Gehrmann:2005pd, Matsuura:1988sm, Matsuura:1987wt} and the three-loop level \cite{Baikov:2009bg, Lee:2010cga} and for $H_m$ the $\mathcal{O}(\alpha_s^2)$ and $\mathcal{O}(\alpha_s^3)$ corrections were computed in Refs.~\cite{Hoang:2015vua} and \cite{Fael:2022miw}, respectively. 
The matching factor $H_m$ contains powers of the rapidity logarithm $\log(Q/m)$ starting at ${\cal O}(\alpha_s^2)$ from massive quark-antiquark fermion loops. The resummation of these rapidity logarithms has been derived in Ref.~\cite{Hoang:2015vua}, their numerical effect is, however, quite small~\cite{Dehnadi:2016snl,Dehnadi:2023msm}.  
The hard matching factors $H_{jj'}$ for the heavy-to-light currents in $\bar{B}$ meson decays were determined to $\mathcal{O}(\alpha_s^2)$ in Refs.~\cite{Bonciani:2008wf, Asatrian:2008uk, Beneke:2008ei, Bell:2008ws} and for $\bar{B} \to X_s \gamma$ the hard function was extracted at three loops in Ref.~\cite{Fael:2024vko}.

For the one-loop expressions of the hard matching factors the current anomalous dimensions read
\begin{align}
&\gamma_{C_Q}(\mu)=-\frac{\alpha_s C_F}{4\pi}\, \bigg[ 4\, \log \bigg( \frac{\mu^2}{-Q^2} \bigg) \plus 6 \bigg] \ , \\
&\gamma_{C_m} (\mu) = \frac{\alpha_s C_F}{4\pi} \bigg[ 4 \, \log \bigg( \frac{-Q^2}{m_t^2} \bigg) \minus 4 \bigg] \ , \notag\\
&\gamma_{C_j} (\hat{v}^+ \hat{H}^-, \mu)= \frac{\alpha_s C_F}{4\pi} \bigg[ 4\, \log \bigg( \frac{\hat{v}^+ \hat{H}^-}{\mu} \bigg) \minus 5 \bigg] \ . \notag  
\end{align}
The corresponding current $Z$-factors are
\begin{align}
&Z_{C_Q} (Q,\mu)=1 \minus \frac{\alpha_s C_F}{4\pi}\, \bigg[ \frac{2}{\epsilon^2} \plus\frac{1}{\epsilon} \bigg( 3 \plus 2\, \log \bigg( \frac{\mu^2}{-Q^2} \bigg)\bigg) \bigg] \ , \\
&Z_{C_m} \Big( m, \frac{Q}{m} \mu \Big) = 1 \minus \frac{\alpha_s C_F}{4\pi} \bigg[ \frac{2}{\epsilon} \, \log \bigg( \frac{m^2}{-Q^2} \bigg) \plus \frac{2}{\epsilon} \bigg] \ . \notag \\
&Z_{C_j} (m_b, \hat{v}^+ \hat{H}^-, \mu) = 1 \plus \frac{\alpha_s C_F}{4\pi} \bigg[ \minus \frac{1}{\epsilon^2} \plus \frac{2}{\epsilon} \log \bigg( \frac{\hat{v}^+ \hat{H}^-}{\mu} \bigg) \minus \frac{5}{2\epsilon} \bigg] \ . \notag 
\end{align}
The renormalized SCET jet function for massless quarks has the form~\cite{Bauer:2003pi, Bosch:2004th}
\begin{align}
J_{n} (p^2, \mu) = \delta (p^2) \plus \frac{\alpha_s C_F}{4\pi} \bigg \{ &\delta (p^2) \bigg[ 7 \minus \pi^2 \bigg] \minus  \frac{3}{\mu^2} \mathcal{L}_0 \Big( \frac{p^2}{\mu^2} \Big)  \plus \frac{4}{\mu^2} \mathcal{L}_1 \Big( \frac{p^2}{\mu^2} \Big) \bigg \} \,,
\end{align}
where we use the notation in Eq.~(\ref{eq:plus_distributions}) for the plus distributions.
The ${\cal O}(\alpha_s^2)$ and ${\cal O}(\alpha_s^3)$ corrections were determined in Refs.~\cite{Becher:2006qw} and \cite{Bruser:2018rad} respectively. The bHQET massive quark jet function for a finite width $\Gamma>0$ reads~\cite{Fleming:2007xt} 
\begin{align} 
\label{eq:bHQET_jet_function_app_1}
J_{B}^\Gamma (\shat<0,\mu)=&\, \frac{1}{\pi m}\, \frac{\Gamma}{\shat^2 \plus \Gamma^2} \bigg \{ 1 \plus \frac{\alpha_s C_F}{4 \pi} \bigg[ \log^2 \bigg( \frac{\mu^2}{\shat^2 \plus \Gamma^2} \bigg) \plus 2\,  \log \bigg( \frac{\mu^2}{\shat^2 \plus \Gamma^2} \bigg)  \minus 4\,  \mathrm{arctan}^2 \bigg( \frac{\Gamma}{\shat} \bigg) \notag \\
&+\! 4\, \frac{\shat}{\Gamma} \, \mathrm{arctan} \bigg( \frac{\Gamma}{\shat} \bigg) \bigg[ \log \bigg( \frac{\mu^2}{\shat^2 \plus \Gamma^2} \bigg)  \plus1 \bigg]  \plus 4 \plus \frac{5\pi^2}{6} \bigg] \bigg \} \plus \frac{1}{\pi m}\, \frac{(4 \shat \Gamma)\, \delta m(R)}{(\shat^2 \plus \Gamma_t^2)^2} \ , \notag \\
J_{B}^\Gamma (\shat>0,\mu)=&\, \frac{1}{\pi m}\, \frac{\Gamma}{\shat^2 \plus \Gamma^2} \bigg \{ 1 \plus \frac{\alpha_s C_F}{4 \pi} \bigg[ \log^2 \bigg( \frac{\mu^2}{\shat^2 \plus \Gamma^2} \bigg) \plus 2\,  \log \bigg( \frac{\mu^2}{\shat^2 \plus \Gamma^2} \bigg)  \minus 4 \, \mathrm{arctan}^2 \bigg( \frac{\Gamma}{\shat} \bigg) \notag \\
& \minus 4 \pi^2 \plus 8\pi \, \mathrm{arctan} \bigg( \frac{\Gamma}{\shat} \bigg)  \plus  4\, \frac{\shat}{\Gamma} \, \mathrm{arctan} \bigg( \frac{\Gamma}{\shat} \bigg) \bigg( \log \bigg( \frac{\mu^2}{\shat^2 \plus \Gamma^2} \bigg)  \plus1 \bigg)  \notag \\
&\minus 4\pi \,\frac{\shat}{\Gamma} \bigg( \log \bigg( \frac{\mu^2}{\shat^2 \plus \Gamma^2} \bigg) \plus 1 \bigg)\plus 4 \plus \frac{5\pi^2}{6} \bigg] \bigg \}  \plus \frac{1}{\pi m}\, \frac{(4 \shat \Gamma)\, \delta m(R)}{(\shat^2 \plus \Gamma^2)^2} \ ,
\end{align}
where we use $\shat = (p^2 - m^2)/m$ and $\delta m(R) = m^{\mathrm{pole}} - m(R)$ denotes the residual mass term parametrizing the difference of the ($R$-dependent short-distance) mass scheme $m(R)$ with respect to the pole mass.
The case distinction can be avoided by writing the result as the imaginary part of the bHQET jet function correlator
\begin{align} 
\label{eq:bHQET_jet_function_app_2}
J_{B}^\Gamma (\shat,\mu)= - \frac{1}{\pi m} \,  \mathrm{Im} \bigg[ &\frac{1}{\shat \plus i \Gamma} \Big \{ 1 \plus \frac{\alpha_s C_F}{4\pi} \Big[ 4 \log^2 \Big( \frac{\mu}{\minus \shat \minus i \Gamma} \Big)   \plus 4 \log \Big( \frac{\mu}{\minus \shat \minus i \Gamma} \Big) \plus 4 \plus \frac{5\pi^2}{6} \Big] \Big \}  \\
& \plus \frac{2 \delta m(R)}{(\shat \plus i \Gamma)^2}   \bigg] \,.  \notag
\end{align}
The bHQET jet function for a stable massive quark reads  
\begin{align} 
\label{eq:bHQET_jet_function_app_3}
J_{B}^{\Gamma=0} (\shat,\mu)= \frac{1}{m} \bigg[& \delta(\shat) \plus \frac{\alpha_s C_F}{4\pi} \Big \{ \delta(\shat) \Big[ 4 \minus \frac{\pi^2}{2} \Big] \plus \frac{8}{ \mu}\mathcal{L}_1 \Big( \frac{\hat{s}}{\mu} \Big) \minus \frac{4}{ \mu} \mathcal{L}_0 \Big( \frac{\hat{s}}{\mu} \Big) \Big \} \\
& \minus 2 \delta m(R) \, \delta'(\shat) \bigg] \,. \notag
\end{align}
The ${\cal O}(\alpha_s^2)$ and ${\cal O}(\alpha_s^3)$ corrections were determined in Refs.~\cite{Jain:2008gb} and \cite{Clavero:2024yav}, respectively. 
The results for the partonic dihemisphere soft function~\cite{Fleming:2007xt} and meson decay shape function~\cite{Bauer:2003pi, Bosch:2004th} read
\begin{align}
\label{eq:renormalozed_shape_function_app}
& \hat S_{\mathrm{hemi}} (\ell^+ , \ell^- \mu) = \delta (\ell^+) \delta (\ell^-) \plus \frac{\alpha_s C_F}{4 \pi} \bigg \{ \delta (\ell^+) \delta (\ell^-) \, \frac{\pi^2}{3}   \\
& \qquad - \! 8\, \delta (\ell^-) \frac{1}{\mu}  \mathcal{L}_1 \Big( \frac{\ell^+}{\mu} \Big) -8\, \delta (\ell^+)\frac{1}{\mu}  \mathcal{L}_1 \Big( \frac{\ell^-}{\mu} \Big) \bigg \} 
 -  \Big[ \delta (\ell^+) \delta' (\ell^-) +\delta' (\ell^+) \delta (\ell^-)\Big] \delta_{\rm gap}(R)  \ ,\notag \\
& \hat S_{\text{shape}} (1,\hat{\ell}, \mu) =  \delta (\hat{\ell}) \plus \frac{\alpha_s C_F}{4\pi} \bigg \{  \minus \frac{\pi^2}{6} \delta (\hat{\ell}) \minus \frac{4}{\mu} \mathcal{L}_0 \Big( \frac{\hat{\ell}}{\mu} \Big) \minus  \frac{8}{\mu} \mathcal{L}_1 \Big( \frac{\hat{\ell}}{\mu} \Big)  \bigg \} 
 + \delta m(R) \delta' (\hat{\ell})  \ , \notag
\end{align}
where $\delta_{\rm gap}(R) = \Delta-\bar\Delta(R)$ is the gap subtraction~\cite{Hoang:2008fs}.
The $\mathcal{O}(\alpha_s^2)$ corrections to the hemisphere soft function have been computed in Ref.~\cite{Kelley:2011ng, Monni:2011gb, Hornig:2011iu} and for the shape function the $\mathcal{O}(\alpha_s^2)$ and $\mathcal{O}(\alpha_s^3)$ corrections were determined in Refs.~\cite{Becher:2005pd} and \cite{Bruser:2019yjk}, respectively. We remind the reader that the expression for the shape function (with the first argument being unity) applies for the heavy quark rest frame. The one in an arbitrary frame is recovered from Eq.~(\ref{eq:shapefctrel}), which also applies in an analogous fashion for $Z_{\rm shape}$ and $\gamma_{\rm shape}$. 
Note that up to $\mathcal{O}(1+\alpha_s)$ we can write the dihemisphere soft function in the factorized form $\hat S_{\mathrm{hemi}}^{(1+\alpha_s)} (\ell^+ , \ell^- \mu) = 
\hat S_{\mathrm{hemi}}^{(1+\alpha_s)} (\ell^+, \mu) \hat S_{\mathrm{hemi}}^{(1+\alpha_s)} (\ell^-, \mu)=
\delta (\ell^+) \delta (\ell^-) + \delta (\ell^+)  \hat S_{\mathrm{hemi}}^{(\alpha_s)} (\ell^-, \mu) + \delta (\ell^-)  \hat S_{\mathrm{hemi}}^{(\alpha_s)} (\ell^+, \mu)$ with
$\hat S_{\mathrm{hemi}}^{(1+\alpha_s)} (\ell, \mu)=\mbox{Im}[\hat{\cal S}_{\rm hemi}^{(1+\alpha_s)}(\ell,\mu) ]$, where
\begin{align}
\label{eq:softhemi_function_single}
\hat {\cal S}_{\mathrm{hemi}}^{(1+\alpha_s)} (\ell, \mu) = & 
-\frac{1}{\pi} \bigg[ \frac{1}{\ell+i0}
\bigg\{1 - \frac{\alpha_s C_F}{4\pi} \Big[ 4 \log^2 \Big( \frac{\mu}{\minus \ell \minus i 0^+} \Big) + \frac{7\pi^2}{6} \Big] \bigg\}
 + \frac{\delta_{\rm gap}(R)}{(\ell+i0)^2}\bigg] \,,
\end{align}
which is convenient for the fixed-order analysis in Sec.~\ref{sec:pheno}. 

The anomalous dimensions of the SCET jet function and the hemisphere soft function at ${\cal O}(\alpha_s)$ have the form
\begin{align} 
\label{eq:anomalous_dimension_functions_app}
& \gamma_{J_n} (p^2,\mu) = \frac{\alpha_s C_F}{4\pi} \bigg \{ 6\,  \delta (p^2) \minus \frac{8}{\mu^2} \mathcal{L}_0 \Big( \frac{p^2}{\mu^2} \Big)  \bigg \} \ , \\
& \gamma_s (\ell^\pm,\mu) = \frac{\alpha_s C_F}{4\pi} \, \frac{8}{\mu} \mathcal{L}_0 \Big( \frac{\ell^\pm}{\mu} \Big)   \,, \notag 
\end{align}
and the $Z$-factors are
\begin{align}
\label{eq:Z_factorsx_app}
& Z_{J_n}(s,\mu)=\delta (p^2) \plus \frac{\alpha_s C_F}{4\pi} \bigg \{ \delta (p^2) \bigg[ \frac{4}{\epsilon^2} \plus \frac{3}{\epsilon} \bigg] \minus \frac{4}{\epsilon} \, \frac{1}{\mu^2} \mathcal{L}_0 \Big( \frac{p^2}{\mu^2} \Big)  \bigg \} \ , \\
& Z_s(\ell^\pm,\mu) = \delta (\ell^\pm)  \plus \frac{\alpha_s C_F}{ 4\pi} \bigg \{ -\! \frac{2}{\epsilon^2}\, \delta (\ell^\pm)  \plus \frac{4}{\epsilon} \, \frac{1}{\mu}  \mathcal{L}_0 \Big( \frac{\ell^\pm}{\mu} \Big)  \bigg \}  \,. \notag 
\end{align}
As discussed in Sec.~\ref{sec:renormalizationtopdecay} the renormalization and RG evolution of the ucs function is already determined by the renormalization properties of the bHQET jet function and the shape function due to consistency relations. For the anomalous dimensions of the bHQET jet function and the shape function we define
\begin{align}
&\gamma_{J_B} (\hat{s}, \mu) = -2\, \Gamma^{\rm cusp} [\alpha_s] \, \frac{1}{\mu}  \mathcal{L}_0 \Big( \frac{\hat{s}}{\mu} \Big) + \gamma^{J_B} [\alpha_s] \, \delta(\hat{s}) \,, \\ 
&\gamma_{\rm shape}(1, \hat{\ell}, \mu) = 2\,  \Gamma^{\rm cusp} [\alpha_s] \, \frac{1}{\mu}  \mathcal{L}_0 \Big( \frac{\hat{\ell}}{\mu} \Big) + \gamma^{\rm shape} [\alpha_s] \, \delta(\hat{\ell}) \,, \notag 
\end{align}
where the cusp and non-cusp anomalous dimensions can be expanded in powers of $\alpha_s$
\begin{align}
&\Gamma^{\rm cusp} [\alpha_s] = \sum_{n=0}^{\infty} \Gamma^c_n \Big( \frac{\alpha_s}{4\pi} \Big)^{n+1} \,, \\
&  \gamma^{J_B}[\alpha_s] = \sum_{n=0}^{\infty} \gamma^{B}_n \Big( \frac{\alpha_s}{4\pi} \Big)^{n+1},  \qquad  \gamma^{\rm shape}[\alpha_s] = \sum_{n=0}^{\infty} \gamma^{\rm sh}_n \Big( \frac{\alpha_s}{4\pi} \Big)^{n+1}  \,. \notag
\end{align}
The coefficients of the cusp anomalous dimension up to the three-loops level were computed in Refs.~\cite{Korchemsky:1987wg, Moch:2004pa} and read
\begin{align}
\Gamma_0^c =&  \, \, 4 C_F, \qquad \Gamma_1^c = \bigg( \frac{268}{9} - \frac{4\pi^2}{3} \bigg) C_F C_A - \frac{80}{9} \, C_F T_F n_\ell , \\
\Gamma_2^c=& \, \bigg( \frac{490}{3} - \frac{536 \pi^2}{27} + \frac{44\pi^4}{45}  + \frac{88\zeta_3}{3} \bigg) C_F C_A^2 + \bigg(  \frac{160 \pi^2}{27} - \frac{1672}{27}  - \frac{224\zeta_3}{3} \bigg) C_F C_A T_F n_\ell \notag \\
&+ \bigg( 64 \zeta_3 - \frac{220}{3}   \bigg) C_F^2 T_F n_\ell - \frac{64}{27} C_F T_F^2 n_\ell^2 \,, \notag
\end{align}
where $n_\ell$ denotes the number of light (massless) quark flavors and $T_F = 1/2$. The non-cusp anomalous dimensions $\gamma^{J_B} [\alpha_s]$ and $\gamma^{\rm shape} [\alpha_s]$ are currently known up to $\mathcal{O} (\alpha_s^3)$. Their coefficients are
\begin{align}
\gamma_0^{B} =&  \, \, 4 C_F, \qquad \gamma_1^{B} = \bigg(  \frac{1396}{27} - \frac{23\pi^2}{9} - 20\zeta_3 \bigg) C_F C_A +\bigg( \minus \frac{464}{27} + \frac{4\pi^2}{9} \bigg) \, C_F T_F n_\ell , \\
\gamma_2^{B}=& \,  \bigg( \frac{112\pi^2\zeta_3}{9} - \frac{2468 \zeta_3}{9} + 120 \zeta_5 +\frac{192347}{729} -\frac{11797 \pi^2}{243}  + \frac{44\pi^4}{15}   \bigg) C_F C_A^2 \notag \\
& + \bigg(  \minus \frac{1520 \zeta_3}{27} - \frac{42908}{729} + \frac{4268 \pi^2}{243} - \frac{16 \pi^4}{15}  \bigg) C_F C_A T_F n_\ell \notag \\
&+ \bigg(  \frac{1184\zeta_3}{9} - \frac{5402}{27} + \frac{4\pi^2}{3} + \frac{16\pi^4}{45}    \bigg) C_F^2 T_F n_\ell \notag \\
& + \bigg(  \frac{448\zeta_3}{27} - \frac{10048}{729} - \frac{80\pi^2}{81}  \bigg) C_F T_F^2 n_\ell^2 \,, \notag
\end{align}
and
\begin{align}
\gamma_0^{\rm sh} =&  \, \, 4 C_F, \qquad \gamma_1^{\rm sh} = \bigg( 36\zeta_3 - \frac{220}{27} - \frac{\pi^2}{9} \bigg) C_F C_A -\bigg( \frac{16}{27} + \frac{4\pi^2}{9} \bigg) \, C_F T_F n_\ell , \\
\gamma_2^{\rm sh}=& \,  \bigg( \frac{5428\zeta_3}{9} - \frac{64\pi^2 \zeta_3}{9} - 264 \zeta_5 -\frac{81215}{729} +\frac{853 \pi^2}{243}  - \frac{44\pi^4}{45}   \bigg) C_F C_A^2 \notag \\
& + \bigg(  \minus \frac{4432 \zeta_3}{27} + \frac{4460}{729} - \frac{1388 \pi^2}{243} + \frac{16 \pi^4}{15}  \bigg) C_F C_A T_F n_\ell \notag \\
&+ \bigg( \minus \frac{32\zeta_3}{9} + \frac{1442}{27} - \frac{4\pi^2}{3} - \frac{16\pi^4}{45}   \bigg) C_F^2 T_F n_\ell \notag \\
& + \bigg( \minus \frac{448\zeta_3}{27} + \frac{6592}{729} + \frac{80\pi^2}{81}  \bigg) C_F T_F^2 n_\ell^2 \,. \notag
\end{align}
The two- and three-loops coefficients given above were computed in Refs.~\cite{Jain:2008gb, Clavero:2024yav} (bHQET jet function) and Ref.~\cite{Becher:2005pd, Bruser:2019yjk} (shape function), respectively.
The renormalization $Z$-factors at $\mathcal{O}(\alpha_s)$ read 
\begin{align}
\label{eq:Z_factor_functions_app}
& Z_{J_{B}} (\hat{s}, \mu) = \delta (\shat) \plus \frac{\alpha_s C_F}{4\pi} \bigg \{ \delta (\shat) \bigg[ \frac{2}{\epsilon^2} \plus \frac{2}{\epsilon} \bigg] \minus \frac{4}{\epsilon} \, \frac{1}{\mu}  \mathcal{L}_0 \Big( \frac{\hat{s}}{\mu} \Big)  \bigg \} \,, \\
& Z_{{\text{shape}}} (1,\hat{\ell}, \mu) = \delta (\hat{\ell} ) \plus \frac{\alpha_s C_F}{4\pi} \bigg \{ \delta (\hat{\ell}) \bigg[ \minus \frac{2}{\epsilon^2} \plus \frac{2}{\epsilon} \bigg] \plus \frac{4}{\epsilon} \,  \frac{1}{\mu} \mathcal{L}_0 \Big( \frac{\hat{\ell}}{\mu} \Big)  \bigg \} \,. \notag 
\end{align}

\section{NLO Corrections to the Ultracollinear-Soft Function}
\label{eq:udsfctIandII}

In this appendix we present the $\mathcal{O}(\alpha_s)$ unrenormalized results for the class~I and II contributions of the ucs function computed in Sec. \ref{sec:ucsoftcomputations}. The sum of the shape function type diagrams of class I (diagrams (a) to (d) in Fig.~\ref{fig:one_loop_diagrams_Sucs}) reads
\begin{align} \label{eq:NLO_corrections_Sucs_class1}
&\tilde{S}_{ucs}^{\mathrm{\RomanNumeralCaps{1}}} (\hat{v}^+,v^-, \gamma^2, s, \bara) \equiv \tilde{S}_{ucs}^{(a)-(d)}(\hat{v}^+,v^-, \gamma^2, s, \bara)=  \\
&=\frac{1}{4 \pi m_t \hat{v}^+}\frac{1}{|\Delta|^2} \, \frac{\alpha_s C_F}{4\pi}   \bigg\{ \delta (\bara) \bigg[ -\! \frac{2}{\epsilon^2} \! + \! \frac{4}{\epsilon} \! + \! 8  \plus \frac{\pi^2}{2} \minus 4 \log \bigg( \frac{-2\Delta}{\mu} \bigg) \minus 4 \log \bigg( \frac{-2\Delta^*}{\mu} \bigg)  \notag \\
&\quad  +\! \log^2 \bigg( \frac{-2\Delta}{\mu} \bigg)   \plus   \log^2 \bigg( \frac{-2\Delta^*}{\mu} \bigg) \bigg] \plus \frac{2}{\mu} \mathcal{L}_0 \Big( \frac{\bara}{\mu} \Big) \bigg[ \frac{2}{\epsilon} \minus \log \bigg( \frac{\bara \minus 2\Delta}{\mu} \bigg) \minus \log \bigg( \frac{\bara \minus 2\Delta^*}{\mu} \bigg) \bigg] \notag \\
&\quad    -\! \frac{4}{\mu} \mathcal{L}_1 \Big( \frac{\bara}{\mu} \Big) \plus \frac{2}{\Delta \minus \Delta^*} \bigg[ \log \bigg( \frac{\bara \minus 2\Delta}{\mu} \bigg)  \minus \log \bigg( \frac{\bara \minus 2\Delta^*}{\mu} \bigg)  \bigg] \bigg\} \notag \,,
\end{align}
and for the sum of the class II diagrams ((e) to (h)) we obtain
{\allowdisplaybreaks
\begin{align} \label{eq:NLO_corrections_Sucs_class2}
&\tilde{S}_{ucs}^{\mathrm{\RomanNumeralCaps{2}}} (\hat{v}^+,v^-, \gamma^2, s, \bara)\equiv \tilde{S}_{ucs}^{(e)-(h)} (\hat{v}^+,v^-, \gamma^2, s, \bara)= \frac{1}{4\pi m_t \hat{v}^+} \, \frac{1}{|\Delta|^2}\,  \frac{\alpha_s C_F}{4\pi}   \\
&\times \bigg\{ \delta (\bara) \bigg[  \frac{2}{\epsilon^2} \minus \frac{4}{\epsilon} \log \big( \tilde{\gamma}^2 \big) \minus \frac{7\pi^2}{6} \plus 4 \log \bigg( \frac{-2\Delta}{\mu} \bigg) \log \big( \tilde{\gamma}^2 \big) \notag \\
& \qquad +\! 4 \log \bigg( \frac{-2\Delta^*}{\mu} \bigg) \log \big( \tilde{\gamma}^2 \big) \minus \log^2 \bigg( \frac{-2\Delta}{\mu} \bigg) \minus \log^2 \bigg( \frac{-2\Delta^*}{\mu} \bigg) \notag \\
& \qquad -\! 2 \log^2 \big( \tilde{\gamma}^2 \minus 1 \big) \plus 2 \log^2 \bigg( 1-\frac{1}{\tilde{\gamma}^2} \bigg) \plus 4 \, \mathrm{Li}_2 \bigg( \frac{1}{\tilde{\gamma}^2} \bigg) \bigg] \notag \\
& \qquad +\! \frac{|\Delta|^2}{|\bara \tilde{\gamma}^2  \minus 2\Delta|^2} \, \frac{8}{\mu} \mathcal{L}_0 \Big( \frac{\bara}{\mu} \Big) \bigg[ -\! \frac{2}{\epsilon} \plus 2 \log \bigg( \frac{\bara \tilde{\gamma}^2 \minus 2\Delta}{\mu} \bigg)  \notag \\
& \qquad +\! 2 \log \bigg( \frac{\bara \tilde{\gamma}^2 \minus 2\Delta^*}{\mu} \bigg) \minus \log \bigg( \frac{\bara \minus 2\Delta}{\mu} \bigg) \minus \log \bigg( \frac{\bara \minus 2\Delta^*}{\mu} \bigg)  \bigg] \notag \\
& \qquad +\! \frac{|\Delta|^2}{|\bara \tilde{\gamma}^2  \minus 2\Delta|^2} \, \frac{16}{\mu} \mathcal{L}_1 \Big( \frac{\bara}{\mu} \Big) \plus  \frac{2\tilde{\gamma}^2}{|\bara \tilde{\gamma}^2  \minus 2\Delta|^2} \, \frac{\Delta \plus \Delta^*}{\Delta \minus \Delta^*} \notag \\
& \qquad \times \bigg[ (\bara \tilde{\gamma}^2 \minus 2\Delta^*)  \bigg( 2 \log \bigg( \frac{\bara \tilde{\gamma}^2 \minus 2\Delta}{\mu} \bigg) \minus \log \bigg( \frac{\bara \minus 2\Delta}{\mu} \bigg) \bigg) \notag \\
& \qquad  ~~~ -\! (\bara \tilde{\gamma}^2 \minus 2\Delta) \bigg( 2 \log \bigg( \frac{\bara \tilde{\gamma}^2 \minus 2\Delta^*}{\mu} \bigg) \minus \log \bigg( \frac{\bara \minus 2\Delta^*}{\mu} \bigg) \bigg) \bigg] \notag \\
&  \qquad +\!  \frac{4 \tilde{\gamma}^2}{|\bara \tilde{\gamma}^2  \minus 2\Delta|^2} \, 
\bigg[ \Delta \bigg( \log \bigg( \frac{\bara \minus 2\Delta}{\mu} \bigg) \minus 2 \log \bigg( \frac{\bara \tilde{\gamma}^2 \minus 2\Delta}{\mu} \bigg) \bigg) \notag \\
& \qquad  ~~~ +\!   \Delta^* \bigg( \log \bigg( \frac{\bara \minus 2\Delta^*}{\mu} \bigg) \minus 2 \log \bigg( \frac{\bara \tilde{\gamma}^2 \minus 2\Delta^*}{\mu} \bigg) \bigg) \bigg] \bigg\}     \notag \,,
\end{align}
}
where we remind the reader of the notation introduced in Eq.~(\ref{eq:gammatilde}).

\section{Phase Space Integrations and On-Shell Top Decay}
\label{app:phasespace}

For the evaluation of the 3-body phase space $\mathrm{d} H^2\mathrm{d} \Pi_3(p; p_\ell, \pn, H)$ in the rest frame where $p^\mu=(\sqrt{p^2},\vec 0)$ we employ for simplicity the narrow width limit for the $W$ boson, which sets $q^2=(p_\ell+p_{\nu_{\ell}})^2=M_W^2$.
Here, we consider two observables, the $b$-jet lepton invariant mass $M_{j_b \ell} = \sqrt{(H+p_\ell)^2}$ and $E_\ell = (p \cdot p_\ell)/ \sqrt{p^2}$, the energy of the charged lepton in the rest frame of the initial momentum $p$ (which is the hemisphere rest frame in our analysis).
We can integrate over the $b$-jet azimuth  and the charged lepton angle w.r.\ to the top jet hemisphere axis $\vec n$
as our observables $M_{j_b \ell}$ and $E_\ell$ do not depend them. This allows us to write the phase space integrals as
\begin{align} 
\int \mathrm{d} H^2\,   \mathrm{d}\Pi_3(p; p_\ell, p_{\nu_\ell} , H)  =  \int \mathrm{d} E_\ell \, \mathrm{d} H^2 \,  \mathrm{d} \gamma^2   \, \frac{1}{64\pi^3} \, \frac{1}{(\gamma^2)^2 \sqrt{p^2}}
\end{align}
and
\begin{align}
\int \mathrm{d} H^2 \,   \mathrm{d}\Pi_3(p; p_\ell, p_{\nu_\ell} , H)  =  \int \mathrm{d} M_{j_b \ell}  \, \mathrm{d} H^2\,  \mathrm{d} \gamma^2 \, \frac{1}{128\pi^3}    \, \frac{2 M_{j_b \ell}}{\sqrt{p^2}} \, \frac{1}{(\gamma^2)^2 \sqrt{p^2}}\,,
\end{align}
where the phase space boundaries are given in Tabs.~\ref{table:El_h2_g2_plm_new} and \ref{table:mlb_h2_g2_plm_new}.

\begin{table}[h]
	\renewcommand*{\arraystretch}{1.5}
	\centering
	\begin{tabular}{|c|c|c|c|c|c|}
		\hline
		Variable & Lower Boundary &Upper Boundary\\
		\hline
		$E_\ell$& $\frac{M_W^2}{2\sqrt{p^2}}$ &$\frac{\sqrt{p^2}}{2}$\\
		\hline
		$H^2$& $0$ &$ \big(\sqrt{p^2} - 2E_\ell \big) \bigg(\! \!  \sqrt{p^2} - \frac{M_W^2}{2E_\ell} \! \bigg) $ \\
		\hline
		$\gamma^2$& $1$  &$\infty $\\
		\hline
	\end{tabular}
	\caption{Phase space boundaries for the phase space parameterization in the order $\{ E_\ell, H^2 , \gamma^2\}$ in the narrow-width limit for the $W$.}
	\label{table:El_h2_g2_plm_new}
\end{table}

\begin{table}[h]
	\renewcommand*{\arraystretch}{1.5}
	\centering
	\begin{tabular}{|c|c|c|c|c|c|}
		\hline
		Variable & Lower Boundary &Upper Boundary\\
		\hline
		$M_{j_b \ell}$& $0$ &$\sqrt{p^2-M_W^2}$\\
		\hline
		$H^2$& $0$ &$\frac{M_{j_b \ell}^2 (p^2 - M_{j_b \ell}^2 - M_W^2)}{p^2 - M_{j_b \ell}^2}$ \\
		\hline
		$\gamma^2$& $1 $  &$\infty $\\
		\hline
	\end{tabular}
	\caption{Phase space boundaries for the phase space parameterization in the order $\{ M_{j_b \ell}, H^2 , \gamma^2\}$ in the narrow-width limit for the $W$.}
	\label{table:mlb_h2_g2_plm_new}
\end{table}

All kinematic variables appearing in the matrix element of Eq.~(\ref{eq:NLOfactorization}) can be expressed in terms of 
$p^2$, $H^2$, $\gamma^2$ and $E_\ell$ and read
\begin{align}
&\hat{H}^- = \bar{n}' \! \cdot \! H= \frac{1}{2\sqrt{p^2}} \left[ p^2 \plus H^2 \minus q^2 \plus \sqrt{\lambda (p^2, H^2, q^2)} \right]\,, \\
&\hat{p}_\ell^- = E_\ell \plus \frac{q^2 \sqrt{p^2} \minus E_\ell (p^2 \plus q^2 \minus H^2) }{\sqrt{\lambda (p^2, H^2, q^2)}}, \notag \\
& \hat{p}_\ell^+ = E_\ell \minus \frac{q^2 \sqrt{p^2} \minus E_\ell (p^2 \plus q^2 \minus H^2) }{\sqrt{\lambda (p^2, H^2, q^2)}},  \notag  \\
&\hat{p}_{\nu_\ell}^- =  \frac{1}{2\sqrt{p^2}} \left[ p^2 \minus H^2 \plus q^2 \minus \sqrt{\lambda (p^2, H^2, q^2)} \right]  \minus          E_\ell \minus \frac{q^2 \sqrt{p^2} \minus E_\ell (p^2 \plus q^2 \minus H^2) }{\sqrt{\lambda (p^2, H^2, q^2)}}, \notag \\
&\hat{p}_{\nu_\ell}^+ =  \frac{1}{2\sqrt{p^2}} \left[ p^2 \minus H^2 \plus q^2 \plus \sqrt{\lambda (p^2, H^2, q^2)} \right]  \minus          E_\ell \plus \frac{q^2 \sqrt{p^2} \minus E_\ell (p^2 \plus q^2 \minus H^2) }{\sqrt{\lambda (p^2, H^2, q^2)}}\,. \notag
\end{align}
For the  $M_{j_b \ell}$ distribution the corresponding expressions are obtained via the identity
\begin{align}
E_\ell = \frac{M_{j_b \ell}^2 \plus q^2 \minus H^2}{2\sqrt{p^2}} \,.
\end{align}
The factorization formula for the differential inclusive semileptonic on-shell top decay at ${\cal O}(\alpha_s)$ in the fixed-order expansion and pole mass scheme in the top quark rest frame with $p_t^\mu=m_t^{\rm pole}(1,\vec 0)$ has the form
{
\allowdisplaybreaks
\begin{eqnarray}
\label{eq:ontopNLOfactorization}
\lefteqn{ 
\left.\frac{\mathrm{d}\Gamma_t}{\mathrm{d} X}\right|_{\rm NLO} = 
  \int \mathrm{d} H^2 \, \mathrm{d} \Pi_3(p_t ; p_\ell , \pn, H) \, \delta(X-{\cal X}(M_t, p_\ell , \pn, H))}
\notag \\
&&\times \, \Big( \frac{e}{\sqrt{2} \,s_w} \Big)^4 |V_{tb}|^2 \, \frac{1}{|(p_\ell+p_{\nu_{\ell}})^2 \minus M_W^2 + i M_W \Gamma_W|^2} \notag \\
&&\times  \bigg \{  2\, \hat{H}^- \hat{p}_{\nu_\ell}^+   \, (p_\ell \cdott  v) \,  \delta ( H^2)  +  2\, \hat{H}^-  \hat{p}_{\nu_\ell}^+     (p_\ell \cdott  v) \,   \bigg[ \frac{1}{\hat{H}^-} \, \hat{S}_{\rm shape}^{(\alpha_s)}\Big(1, \frac{H^2}{\hat{H}^-} \Big) \plus J_{n'}^{(\alpha_s)} (H^2)   \bigg] \notag \\
&& \quad  +       \delta ( H^2) \,  \hat{H}^-     \bigg[  2 \hat{p}_{\nu_\ell}^+  (p_\ell \cdott  v)\,  H_{11}^{(\alpha_s)}(m_t,\hat{H}^-) \notag \\
&& \hspace{2.6cm}  + \Big( \hat{p}_{\nu_\ell}^+ (p_\ell \cdott v) + \hat{p}_\ell^+ (\pn \cdott v) \minus \frac{q^2}{2}   \Big) \, H_{12}^{(\alpha_s)}(m_t,\hat{H}^-) \notag \\
&& \hspace{2.6cm} + 2 \hat{p}_\ell^+ \hat{p}_{\nu_\ell}^+ \, H_{13}^{(\alpha_s)}(m_t,\hat{H}^-) \bigg] \bigg \}\,,
	\end{eqnarray}
}%
where the terms $F^{(\alpha_s)}$ represent the $\mathcal{O} (\alpha_s)$ corrections of the different factorization functions quoted in App.~\ref{app:NLO_corrections}.

\end{appendix}

\bibliography{./sources}
\bibliographystyle{JHEP}

\end{document}